\documentclass{article}
\usepackage{authblk}
\usepackage[utf8]{inputenc}
\usepackage[left=1.5cm,right=1.5cm,top=1cm,bottom=1.5cm]{geometry}
\usepackage{mathtools}
\usepackage{xcolor}
\usepackage{hyperref}
\hypersetup{
    colorlinks=true,
    linkcolor=blue,
    filecolor=magenta,      
    urlcolor=cyan,
    citecolor=teal
    }
\usepackage{amsmath}
\usepackage{graphicx}
\graphicspath{ {figs/} }
\usepackage{subfig}
\DeclareMathOperator{\sech}{sech}

\usepackage[style=numeric-comp,style=phys]{biblatex}
\title{An effective description of the impact of inhomogeneities on the movement of the kink front in 2+1 dimensions}
\author[1]{Jacek Gatlik}
\author[2]{Tomasz Dobrowolski}
\author[3]{Panayotis G. Kevrekidis}
\affil[1]{\textit{Doctoral School, University of the National Education Commission in Krakow, Podchor\c{a}\.zych  2, 30-084 Cracow, Poland}}
\affil[2]{\textit{Department of Computer Physics and Quantum Computing, University of the National Education Commission in Krakow, Podchor\c{a}\.zych  2, 30-084 Cracow, Poland}}
\affil[3]{\textit{Department of Mathematics and Statistics, University of Massachusetts, Amherst, Massachusetts 01003-4515, USA}}
\date{\today}

\begin{document}
\maketitle

\begin{abstract}
In the present work we explore the interaction of a 
one-dimensional kink-like front of the sine-Gordon equation
moving in 2-dimensional spatial domains. We develop an
effective equation describing the kink motion,
characterizing its center position dynamics
as a function of the transverse variable. The relevant
description is valid both in the Hamiltonian realm and in the
non-conservative one bearing gain and loss. We subsequently
examine a variety of different scenarios, without and
with a spatially-dependent heterogeneity. The latter
is considered both to be one-dimensional ($y$-independent)
and genuinely two-dimensional. The spectral features and
the dynamical interaction of the kink with the heterogeneity
are considered and comparison with the effective quasi-one-dimensional
description (characterizing the kink center as a function
of the transverse variable)  is also provided. Generally, good
agreement is found between the analytical predictions and the
computational findings in the different cases considered.
\end{abstract} \hspace{10pt}

\section{Introduction}
For years, nonlinear field theories have attracted the attention of many researchers. The reasons for this are twofold.  First, they appear in the description of physical \cite{Frenkel1939,Zharnitsky1998,Bugaychuk2002,Lam1992,Josephson1962}, biological \cite{Xie2000,Georgiev2004,Walet2005} as well as chemical \cite{Bednar1989} systems.
Secondly, unlike linear systems, regardless of the practical context, their behavior is far more interesting and challenging
to explore. Some of the best-known and well-studied nonlinear field models are the Korteweg–De Vries (KdV) equation \cite{Zabusky1965,drazin}, the nonlinear Schr{\"o}dinger equation \cite{AblowitzPrinariTrubatch,ablowitz2} and the sine-Gordon model \cite{Bour1862,MKW14}. As shown, these models in 1+1 dimensions are integrable by means of the Inverse Scattering Method \cite{Gardner1967,Zakharov1973,Zakharov1973,Ablowitz1973}. 
The latter allows one, for such integrable models, to obtain, based on appropriately behaving initial data at spatial infinity, the configuration of the fields at any later instant of time. In particular, for appropriately chosen initial data, the explicit analytical form of the soliton solutions can be obtained and
the dynamics of such fundamental nonlinear coherent structures
can be explored in time.

The interest of this paper is focused on the sine-Gordon model.  Often, in practical contexts, this model appears in somewhat modified (i.e., perturbed), potentially relevant experimentally versions. These modifications have their origin in the existence of external forcing, dissipation in realistic physical systems or various types of inhomogeneities \cite{McLaughlin1978,Golubov1988,Kivshar1989,Malomed1990,Dobrowolski2020,Demirkaya2013,Kevrekidis2014,Monaco2016,Dobrowolski2012,Gatlik2021,Dobrowolski2009}. These modifications, though, significantly affect the integrability property, however, they do not affect the existence of kink solutions. Such models are
often referred to as nearly integrable ones. The situation becomes even more complicated when passing from 1+1 to 2+1, as well as to a larger number of dimensions; see, e.g., the work of \cite{Kevrekidis2018} and references therein. In the case of the sine-Gordon model, even without any modifications, such
higher-dimensional settings are not integrable within the framework of the Inverse Scattering Method \cite{Ablowitz1991}, nor does
the model have the properties that should be satisfied for proving integrability based on the Painlev{\'e} test \cite{Ablowitz1980,Weiss1983,Weiss1984}.
Despite these difficulties, various solutions have been constructed, among others, in the form of a kink front. 
Indeed, it is relevant to recall here that the quasi-one-dimensional
kink (i.e., the kink homogeneous in the transverse direction)
is trivially still a solution in the higher-dimensional setting.

In higher dimensions, part of the challenge towards
describing the dynamics of the solitary waves concerns the 
fact that 
the position of the coherent structure is 
dependent both on the time variable and the ``transverse'' spatial variable. For a kink, e.g., along the $x$-direction,
its center will be $y$-dependent, while for a radial
kink, its center can be varying azimuthally; see, 
e.g., \cite{Kevrekidis2018}. 
Moreover, kink-antikink interactions have also been studied in the 2+1 dimensional model \cite{Gonzalez2022}. The behavior of a kink with radial symmetry 
has been intriguing to researchers since the 
early days of soliton theory \cite{Christiansen1979_2,Christiansen1981}. A fairly interesting phenomenon observed for radial configurations is their alternating expansion and contraction. However, it turns out that in two dimensions such configurations can be destroyed at the
origin \cite{Geicke1983}. Moreover, the evolution of long-lived configurations of breather form has also been studied in the context of the sine-Gordon model in 2+1 dimensions \cite{Piette1998}. Another interesting
potential byproduct of the radial dynamics 
can be the formation of breather as a result of collisions with edges as studied in \cite{Caputo2013}. Among other things, the influence of various types of inhomogeneities and modifications of the sine-Gordon model on the evolution of the kink front has continued to
attract the attention of researchers; see, e.g.,
the discussions of \cite{Kivshar1989,Malomed2014}.
 New studies devoted to the effect of inhomogeneities on kink dynamics in 2+1 dimensional systems can also be found in the articles \cite{Garcia-Nustes2017,Marin2018,Castro-Montes2020}.

In the present article, we consider the behavior of the deformed kink front in the presence of the inhomogeneities.
The way in which these inhomogeneities enter the equation of
motion is motivated by studies conducted in 
earlier works by some of the present authors \cite{Dobrowolski2009,Dobrowolski2012,Gatlik2021}, for the 1+1 case and the quasi-1+1 dimensional Josephson junction. In this study, we explore how the existence of the mentioned modifications of the sine-Gordon equation have its origin in the curvature of the junction.
Our goal, more concretely, is to investigate the stability of static kink fronts in the presence of spatial inhomogeneities in the more computationally 
demanding and theoretically richer $2+1$-dimensional
setting, extending significantly our recent results
of the $1+1$-dimensional case \cite{Gatlik2023arXiv}.
In order to do so, we obtain and test an effective
reduced model, leveraging the fundamental
non-conservative variational formalism presented in the work of \cite{Galley2013,galley2014principle}. This formalism enables the formulation of a Lagrangian description of systems with dissipation. An important part of this approach is the introduction of a non-conservative potential in addition to conservative ones giving the possibility of formulating a non-conservative Lagrangian. The Euler-Lagrange equations are then obtained just based on this Lagrangian. Here, our theoretical emphasis
is on utilizing this methodology to provide a
reduced ($1+1$-dimensional) description of the
center of the kink as a function of the transverse
variable in the spirit of the filament method,
utilized also earlier in \cite{Kevrekidis2018}.

The work is organized as follows. In the next section, we will define the problem under consideration, namely
the evolution sine-Gordon $2+1$-dimensional kinks
in the presence of heterogeneities in the medium. 
We will also construct the effective approximate model obtained based on the non-conservative Lagrangian approach. Section 3 is divided into four subsections.  In the first one, in order to check the obtained effective model and numerical procedures, we analyze the motion of the kink front in a homogeneous system, but with dissipation and external forcing. Subsection 2 of this part contains a study of the front propagation in the presence of inhomogeneities  
homogeneous along the transverse direction. In subsection 3, we include an analysis of the motion of the kink in a system whose equation has a form analogous to that describing a curved Josephson junction but with an inhomogeneity having a functional dependence on the variable normal to the direction of kink motion.
Section 4 contains an analysis of the stability of the kink in the presence of the spatial inhomogeneity in the form of potential well and barrier. 
In section 5, we summarize our findings and present
our conclusions, as well as some direction for further
research efforts.
Analytical results on this issue are located in Appendices A, B and C.
The last section contains remarks.

\section{Model and Theoretical Analysis}
\noindent

\subsection{System Description}
In the present article we study the perturbed sine-Gordon model 
 in 2+1 dimensions in the form:
\begin{equation}
\label{fullmodel}
\partial_t^2 \phi + \alpha\partial_t \phi - \partial_x (\mathcal{F}(x,y)\partial_x \phi) - \partial_y^2 \phi + \sin \phi = - \Gamma,
\end{equation}
where the function $\mathcal{F}(x,y)$ represents the inhomogeneity present in the system, $\alpha$ describes the dissipation caused by the quasi-particle currents and $\Gamma$ is the bias current in the Josephson junction
setup \cite{Malomed2014}. 
For the inhomogeneity, we will typically assume
$\mathcal{F}(x,y)=1+\varepsilon g(x,y)$, where $\varepsilon$ is a small control parameter, while 
$g(x,y)$ reflects the corresponding spatial variation. 
When considering the motion of a kink in this two-dimensional system, we assume periodic boundary conditions along the second dimension parametrized by the variable $y$
\begin{equation*}
	\begin{gathered}
		 \phi (x,y_{min},t)= \phi (x,y_{max},t),\\
		 \partial_t \phi (x,y_{min},t)= \partial_t \phi (x,y_{max},t).
	\end{gathered}
\end{equation*}
The initial velocity of the kink when  $\Gamma$ is equal to zero is selected arbitrarily. On the other hand, if both quantities $\alpha$ and $\Gamma$ are different from zero then the initial velocity is assumed equal to 
\begin{equation}
\label{stationary_velocity}
u_{s}=\frac{1}{\sqrt{1+\left(\frac{4\alpha}{\pi\Gamma}\right)^2}}.
\end{equation}
This value corresponds to the movement at the stationary speed obtained in the classic work of \cite{Scott1978}. 
We use this value because at the initial time the kink is sufficiently far away from the inhomogeneity. With such a large distance at the initial position of the front, the $\mathcal{F}$-function is approximately equal to one.
In this work, we will describe the movement of the kink front, the shape of which will have different forms at the initial instant and which will encounter different
types of heterogeneities during propagation..
We propose an effective description of this movement within a 1+1 dimensional model, characterizing the
center motion as a function of the transverse variable,
that we now expand
on. 
In our work, we compare the results of the original model and the effective model to determine the limits of applicability of the proposed simplified description.

\subsection{Nonconservative Lagrangian Model}
\noindent
Due to the existence of dissipation in the studied system, we will use the formalism described in the paper \cite{Galley2013,galley2014principle}. The proposed approach introduces a non-conservative Lagrangian in which the variables describing the system are duplicated and
an additional term is added to the Lagrangian to 
account for the non-conservative forces. The variational principle for this Lagrangian only specifies (and 
matches across acceptable trajectories) the initial data. On the other hand, in the final time, the coordinates and velocities of the two paths are not fixed but for both sets of variables are equal. Doubling the degrees of freedom has this consequence that in addition to the potential function $V$, one can include an arbitrary function, $\mathcal{R}$ (called nonconservative potential), that couples the two paths together. Nonconservative forces present in the system are determined from the potential ${\cal R}$. The ${\cal R}$ function is responsible for the energy lost by the system. This formalism, in the article \cite{Kevrekidis2014}, was applied to describe the ${\mathcal PT}$-symmetric variants of field theories (bearing balanced gain and loss). The referred modification introduced into the field models simultaneously preserves the parity symmetry ($P$, i.e. $x \rightarrow -x$) and the time-reversal symmetry ($T$, i.e. $t \rightarrow -t$ ). In particular, this approach has been applied to solitonic models such as $\phi^4$ and sine-Gordon.

In the current work, we consider the system described by equation \eqref{fullmodel}. For $\alpha=0$ and $\Gamma=0$, this equation can be obtained from the Lagrangian density
\begin{equation}
\label{ConLag}
\mathcal{L}(\phi, \partial_t\phi,\partial_x\phi, \partial_y\phi) = \frac{1}{2} (\partial_t\phi)^{2}-\frac{1}{2}\mathcal{F}(x,y) (\partial_x \phi)^{2}-\frac{1}{2} (\partial_y \phi)^{2}-V(\phi).
\end{equation}
The nonconservative Lagrangian density is  formed from the Lagrangian density \eqref{ConLag} by doubling the number of degrees of freedom
\begin{equation}
\label{NonconLag0}
\mathcal{L}_N =\mathcal{L}(\phi_1, \partial_t\phi_1,\partial_x\phi_1, \partial_y\phi_1)-\mathcal{L}(\phi_2, \partial_t\phi_2,\partial_x\phi_2, \partial_y\phi_2) +{\cal R}
\end{equation}
Much more convenient variables to describe our system with dissipation are the field variables $\phi_{+}$ and $\phi_{-}$.
The relationship between the variables
$\phi_i$, $(i=1,2)$ and $\phi_{+}$, $\phi_{-}$ is of the form $\phi_1=\phi_{+}+\frac{1}{2} \phi_{-}$ and $\phi_2=\phi_{+}-\frac{1}{2} \phi_{-}$. The main advantage of using new variables is that in the physical limit (indicated by the characters $PL$) the $\phi_{+}$ variable reduces to the original variable $\phi$ while the $\phi_{-}$ variable becomes equal to zero thereby disappears from the description.
In the new variables, the nonconservative Lagrangian density is of the form
\begin{equation}
\label{NonconLag}
\mathcal{L}_N = (\partial_t\phi_{+}) (\partial_t\phi_{-})-\mathcal{F}(x,y) (\partial_x \phi_{+}) (\partial_x \phi_{-})-(\partial_y \phi_{+}) (\partial_y \phi_{-})-V\left(\phi_{+}+\frac{1}{2} \phi_{-}\right)+V\left(\phi_{+}-\frac{1}{2} \phi_{-}\right)-\alpha \phi_{-} \partial_t \phi_{+}-\Gamma \phi_{-}.
\end{equation}
The variational scheme proposed in the paper \cite{Galley2013} leads to an Euler-Lagrange equation 
\begin{equation}
\label{n-eq-1}
 \left[
\partial_{\mu} \left( \frac{\partial \mathcal{L}_N}{\partial (\partial_{\mu} \phi_{-})}\right)
- \frac{\partial \mathcal{L}_N}{\partial \phi_{-}}
  \right]_{PL} = 0 ,
\end{equation}
where the subscript $\mu$  denotes the partial derivatives with respect to the variables $x^{\mu}=(t,x,y)$.   A particularly convenient form of the field equation is the one that separates the effect of the existence of a nonconservative potential from the rest of the equation
\begin{equation}
\label{n-eq-2}
\partial_{\mu} \left( \frac{\partial \mathcal{L}}{\partial (\partial_{\mu}\phi)}\right)
- \frac{\partial \mathcal{L}}{\partial \phi} =  \left[ \frac{\partial
\mathcal{R}}{\partial \phi_{-}} - \partial_{\mu} \left( \frac{\partial
\mathcal{R}}{\partial (\partial_{\mu}\phi_{-})}\right)
 \right]_{PL}
\end{equation}
Inserting the Lagrangian density \eqref{ConLag} into the above equation and using the form of the function $\mathcal{R}=-\alpha \phi_{-} \partial_t \phi_{+}-\Gamma \phi_{-}$, we reproduce equation \eqref{fullmodel}.

So far, our calculations are exact (i.e., no approximations
have been made). Hereafter, we will use a kink-like
ansatz in the field $\phi(x,y,t)$, so as
 to construct an effective (approximate) $1+1$ dimensional reduced model describing the dynamics of the kink center.
 This is a significant step in the vein of dimension
 reduction, however, it comes at the expense of 
 assuming that the entire field consists of a 
 fluctuating kink (i.e., small radiative wavepackets
 on top of the kink cannot be captured). Nevertheless,
 this perturbation in the spirit of soliton 
 perturbation theory \cite{Kivshar1989} has a
 time-honored history of being successful in 
 capturing coherent structure dynamics in such models.
 
 To implement our approach, we introduce a kink ansatz of the form $\phi_i(t,x,y) = K(x-X_i(t,y))=4 \arctan \left( e^{x-X_i}\right)$ into the Lagrangian \eqref{NonconLag} of the field model in $2+1$ dimensions, and then integrate over the spatial variable $x$. The 
 resulting effective nonconservative Lagrangian density is as follows
\begin{equation}
L = L_{1} - L_{2} + R, \hspace{0,5cm} R = R_{1} + R_{2},
\end{equation}
where the effective conservative Lagrangian densities are 
\begin{equation*}
\begin{gathered}
    L_{1}=\frac{1}{2} M (\partial_t X_{1})^{2}- \frac{1}{2}\int_{-\infty}^{+\infty}\mathcal{F}(x,y)(K^{'}(x-X_1)^2)dx - \frac{1}{2}M (\partial_y X_{1})^2,\\
     L_{2}=\frac{1}{2} M (\partial_t X_{2})^{2}- \frac{1}{2}\int_{-\infty}^{+\infty}\mathcal{F}(x,y)(K^{'}(x-X_2)^2)dx - \frac{1}{2}M (\partial_y X_{2})^2,
\end{gathered}
\end{equation*}
on the other hand, both parts of the nonconservative effective potential are equal to
\begin{equation*}
\begin{gathered}
    R_{1}=\frac{1}{2}\alpha\int_{-\infty}^{+\infty}\left(K(x-X_1)-K(x-X_2)\right)\left(K^{'}(x-X_1)\partial_t X_{1}+K^{'}(x-X_2)\partial_t X_{2}\right)dx,\\
    R_{2}=-\Gamma\int_{-\infty}^{+\infty}\left(K(x-X_1)-K(x-X_2)\right)dx .
\end{gathered}
\end{equation*}
By analogy with  equation \eqref{n-eq-2},  
the (approximate) effective  field-theoretic
equation for $X(y,t)$ is of the form 
\begin{equation}
\label{eq-eff1} \partial_t \left( \frac{\partial L}{\partial
(\partial_t X)}\right) +\partial_y  \left( \frac{\partial L}{\partial
(\partial_y X)}\right) - \frac{\partial L}{\partial X} = \left[
\frac{\partial R}{\partial X_{-}} - \partial_t \left( \frac{\partial R}{\partial
(\partial_t X_{-})}\right) -\partial_y \left( \frac{\partial R}{\partial
(\partial_y X_{-})}\right)\right]_{PL} ,
\end{equation}
where we use the variables $X_{+}=(X_1+X_2)/2$ and $X_{-}=X_1-X_2$  to write the nonconservative potential.
Note that the left side of the equation describes a situation in which there are no nonconservative forces, while the right side introduces dissipation and forcing into the system.  
In the equation \eqref{eq-eff1}, $L$ is a simple conservative Lagrangian density written in terms of the physical variable $X$ 
\begin{equation}
    \label{lag-con}
     L=\frac{1}{2} M (\partial_t X)^{2}- \frac{1}{2} \varepsilon \int_{-\infty}^{+\infty} g(x,y)(K^{'}(x-X))^2dx - \frac{1}{2}M (\partial_y X)^2.
\end{equation}
In this formula, we used the decomposition of the $\mathcal{F}$ function into a regular part and a small perturbation, i.e., $\mathcal{F}(x,y)=1+\varepsilon g(x,y).$ 
On the other hand, the function $R$ appearing on the right side of the equation is written in auxiliary variables $X_{+}$ and $X_{-}$. 
 Let us notice that the left-hand side of equation \eqref{eq-eff1} contains the full information about the inhomogeneities present in the system
\begin{equation}
\label{eq-eff2} M \partial_t^2 X - \varepsilon \int_{-\infty}^{+\infty}
g(x,y) K^{'}(x-X) K^{''}(x-X) dx - M \partial_y^2 X = \left[
\frac{\partial R}{\partial X_{-}} - \partial_t \left( \frac{\partial R}{\partial
(\partial_t X_{-})}\right) -\partial_y \left( \frac{\partial R}{\partial
(\partial_t X_{-})}\right)\right]_{PL}  .
\end{equation}
In order to calculate the right side of the effective field equation, we rewrite the nonconservative potential $R$ to the $X_{\pm}$ variables 
\begin{equation*}
\begin{gathered}
        R_1=\frac{1}{2}\alpha\int_{-\infty}^{+\infty}\left(K\left(x-X_{+}-\frac{1}{2} X_{-}\right)- K\left(x-X_{+}+\frac{1}{2}X_{-}\right)\right)\cdot\\\left[K^{'}\left(x-X_{+}-\frac{1}{2}X_{-}\right) \left(X_{+t}+\frac{1}{2}X_{-t}\right)+K^{'}\left(x-X_{+}+\frac{1}{2}X_{-}\right)\left(X_{+t}-\frac{1}{2}X_{-t}\right)\right]dx,\\  
    R_{2}=-\Gamma\int_{-\infty}^{+\infty} \left(K\left(x-X_{+}-\frac{1}{2}X_{-}\right)-K\left(x-X_{+}+\frac{1}{2}X_{-}\right)\right)dx.
\end{gathered}
\end{equation*}
We then determine the classical limit of the right-hand side of the equation \eqref{eq-eff2}. In the course of the calculations, we use the asymptotic values of the kink solution. The Euler-Lagrange equation defining the effective $1+1$ dimensional model is thus  identified as:
\begin{equation}
\label{eq-eff3} M \partial_t^2 X -  M \partial_y^2 X - \varepsilon \int_{-\infty}^{+\infty}
g(x,y) K^{'}(x-X) K^{''}(x-X) dx  = -  \alpha  M \partial_t X + 2 \pi \Gamma.
\end{equation}
Let us consider the function $g$ being the product of  $g(x,y)=p(x)q(y)$, where $p(x)$ corresponds to the inhomogeneity occurring across the direction of the kink motion, and $q(y)$ may represent the gaps occurring within this inhomogeneity along the transverse direction. The function $q(y)$ does not depend on $x$ therefore we can exclude it before the sign of the integral and perform the explicit integration of the expression containing the function $p(x)$.
In the first example, the $p$-function is the difference of the step functions 
$p(x) =\frac{1}{2}(\Theta(x+\frac{h}{2})-\Theta(x-\frac{h}{2}))$.
This form of the $p$-function makes the inhomogeneity exactly localized between the points $x=0$ and $x=h$. The Euler-Lagrange equation in this case is 
\begin{equation}
    \label{model}
    \partial_t^2 X +\alpha \partial_t X - \partial_y^2 X+\frac{1}{8}\varepsilon q(y)\left(\sech\left(\frac{h}{2}+X\right)^2-\sech\left(\frac{h}{2}-X\right)^2\right)=\frac{1}{4}\pi\Gamma.
\end{equation}
The second example concerns inhomogeneity described by a continuous function
\begin{equation}
    \label{pfun}
    p(x)=\frac{1}{2}\left(\tanh\left(x+\frac{h}{2}\right)-\tanh\left(x-\frac{h}{2}\right)\right).
\end{equation}
For large values of $h$, this function can be successfully approximated by a combination of step functions of the form $p(x)=\frac{1}{2}(\Theta(x+\frac{h}{2})-\Theta(x-\frac{h}{2}))$. 
 However, for smaller values of $h$, 
some differences are observed. The effective field equation in this case has a slightly more complex form 
\begin{equation}
    \label{e-model}
    \partial_t^2 X+\alpha \partial_t X- \partial_y^2 X+ \frac{1}{2}\varepsilon q(y) \left(\frac{(\frac{h}{2}+X) \coth(\frac{h}{2}+X) - 1}{\sinh^2(\frac{h}{2}+X)}  - \frac{(\frac{h}{2} - X) \coth(\frac{h}{2} - X) - 1}{\sinh^2(\frac{h}{2}-X)}
     \right)
    =\frac{1}{4}\pi\Gamma.
\end{equation}
This effective $1+1$ dimensional model is the basis for comparisons with predictions of the initial field equation \eqref{fullmodel} in $2+1$ dimensions.

\section{Numerical results}
This section will be devoted to the comparison of the predictions
resulting from the effective
$1+1$-dimensional model and the full $2+1$-dimensional field model.  Our goal
is to examine the compatibility of the two descriptions and
determine the range of applicability of the approximate model.
\subsection{Kink propagation in the absence of inhomogeneities}

\begin{figure}
    \centering
    \subfloat{{\includegraphics[height=5cm]{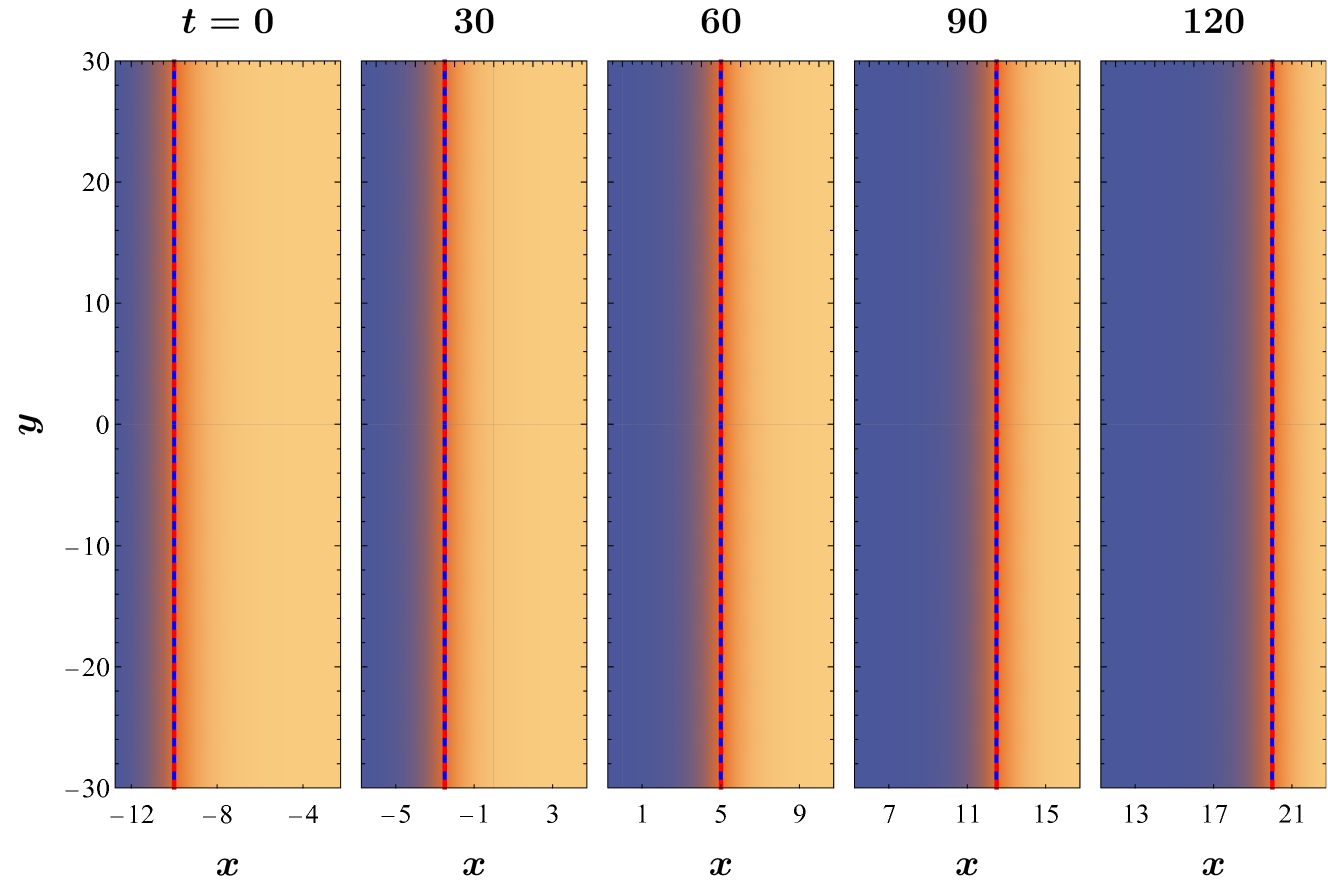}}}
    \qquad
    \subfloat{{\includegraphics[height=5cm]{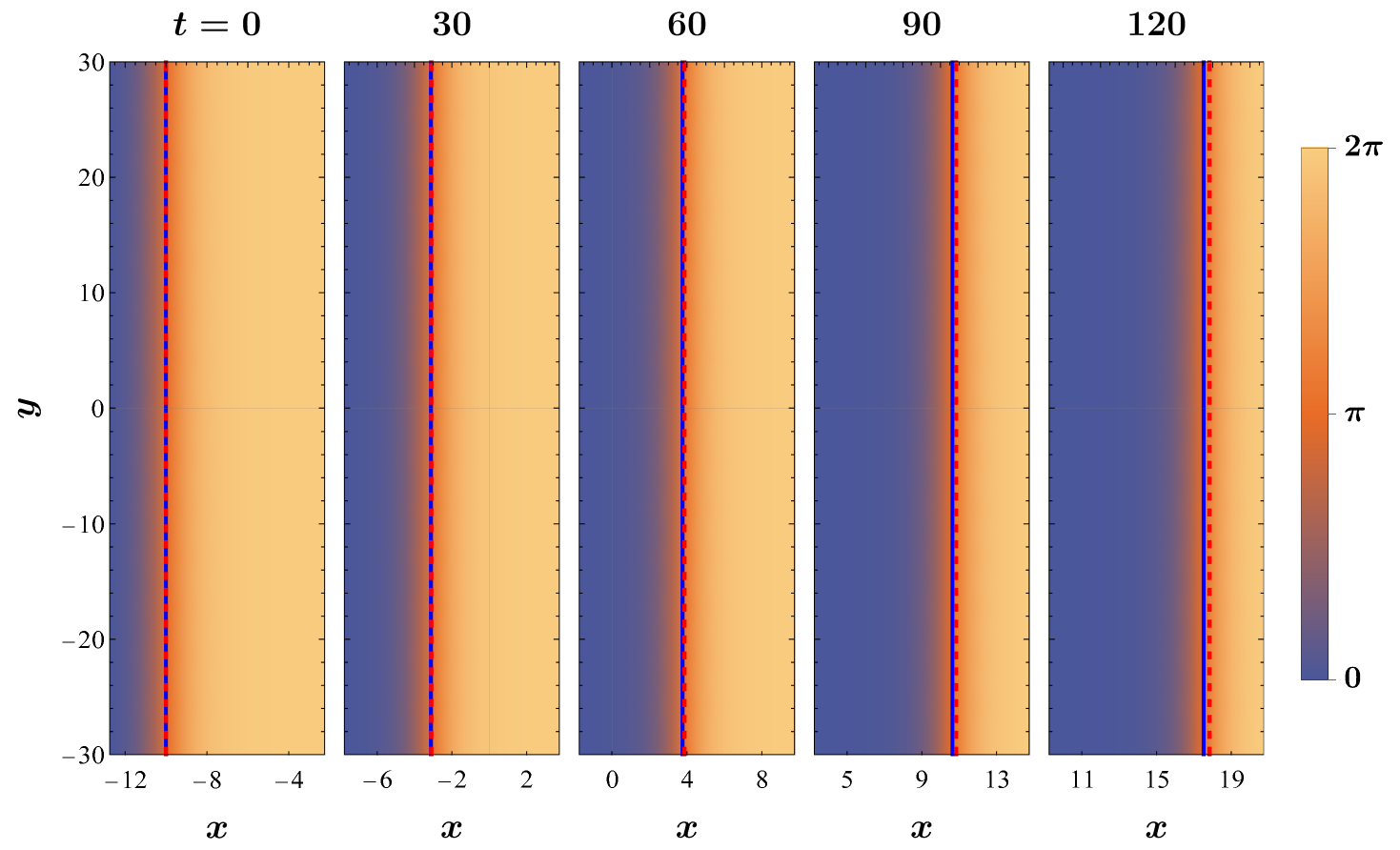}}}
    \caption{Comparison of the position of the center of the kink for the results obtained from the full field model and the approximate model. On the left the parameters in the figures shown have values $u=0.025$, $\Gamma=0$ and $\alpha=0$ and on the right $\Gamma=0.003$, $\alpha=0.01$ while velocity is equal to $u_s$ according to \eqref{stationary_velocity}. In both cases $\varepsilon=0$. The dashed red line is the position of the center of the kink according to the approximate model, while the blue line corresponds to the center of the kink from the solution of the full field model.}
    \label{fig_01}
\end{figure}

Initially, we performed tests to check the compatibility
of the two descriptions for a homogeneous system, i.e. for a
system for which the parameter representing the strength of
inhomogeneity $\varepsilon$ is equal to zero. The first check was
carried out for an initial condition with a kink of the form
of a straight line perpendicular to the $x$-direction,
i.e.,
direction of movement of the kink. The propagation of the kink
front is shown in Figure  \ref{fig_01}. The left panel shows the results
obtained from the field model of Eq.~\eqref{fullmodel}. The blue color represents the
area for which $\phi<\pi$, and the yellow color corresponds to
$\phi>\pi$. The areas are separated by the red line
$\phi(t,x,y)=\pi$. We identify this line with the kink front. This
panel shows the location of the front sequentially at moments
$t=0, 30, 60, 90, 120$. Each snapshot on the left
panel shows a sector of the system located in the interval $y \in
[-30,30]$, while $x\in [-25,15]$. It should be noted that the
simulations, nevertheless, were conducted on a much wider interval
$x$, i.e. $x \in [-70,70]$. At the ends of the interval (i.e. for $x=\pm 70$), Dirichlet boundary
conditions corresponding to a single-kink topological sector were
assumed. The right panel contains a comparison of the evolution of
the kink front obtained from the field equation (solid red line)
and that obtained from the approximate model (dotted blue line) given by the equation \eqref{e-model}.
The comparison was made at instants identical to those on the left
panel. Due to the very good agreement, the blue line is barely
visible. The simulation was performed for an initial velocity of the kink with $u_0=u_s=0.229339$.  It can be verified that this is the steady-state velocity resulting from equation \eqref{stationary_velocity} for the 
dissipation constant
$\alpha=0.01$ and bias current $\Gamma=0.003$. In this work, whenever $\Gamma \neq 0$ and $\alpha \neq 0$ we take the steady-state velocity resulting from equation \eqref{stationary_velocity} as the initial velocity. It is worth noting that, if we were to assume a velocity below the steady-state velocity during motion, this velocity will increase to the steady-state value due to the existence of an unbalanced driving force in the form of a bias current. On the other hand, if we assume an initial velocity above the stationary velocity then due to the unbalanced dissipation there will be a slowdown of the front to the stationary velocity. Finally, the initial position of the kink is taken equal to $X_0=-20$.

\begin{figure}
    \centering
    \subfloat{{\includegraphics[height=5cm]{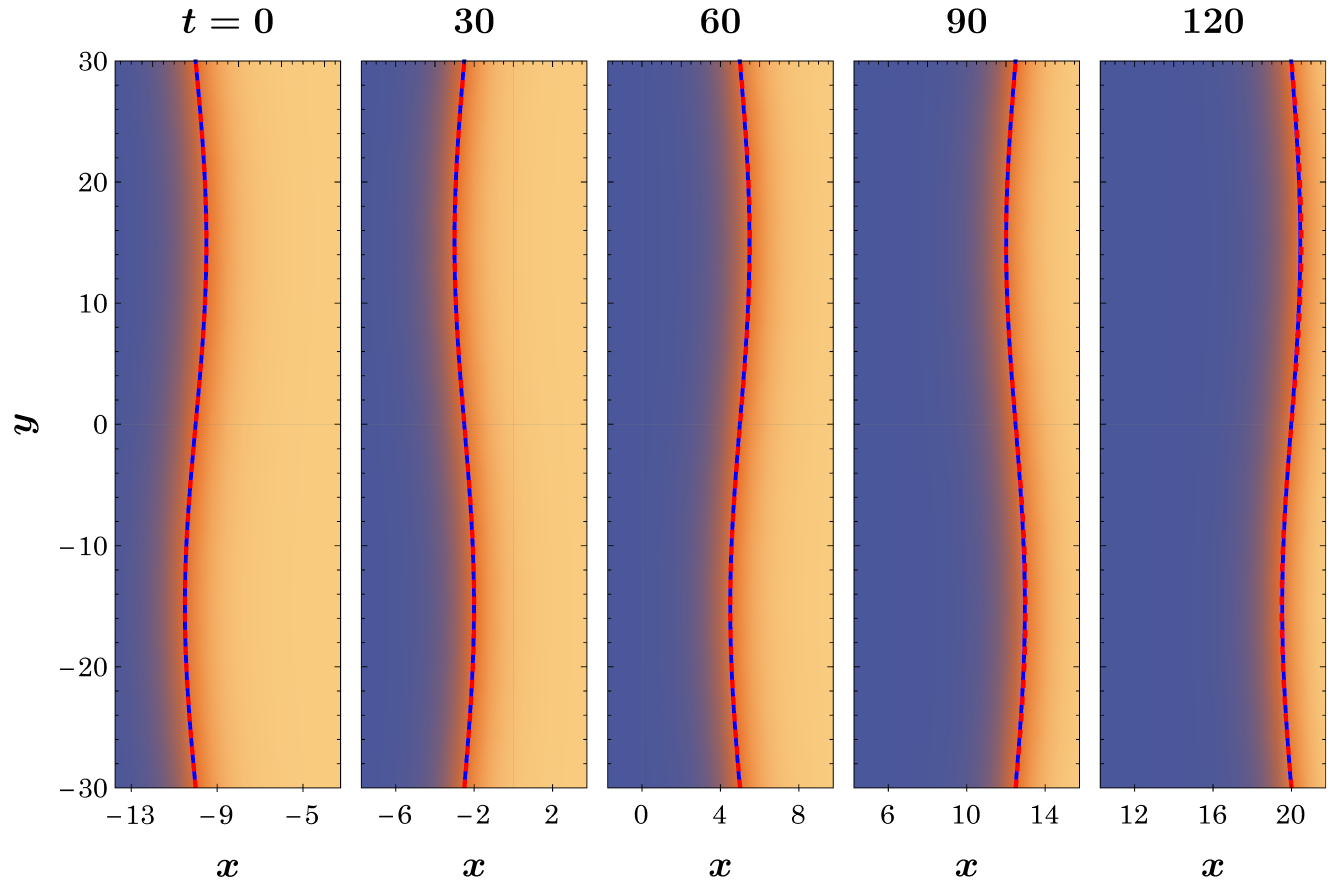}}}
    \qquad
    \subfloat{{\includegraphics[height=5cm]{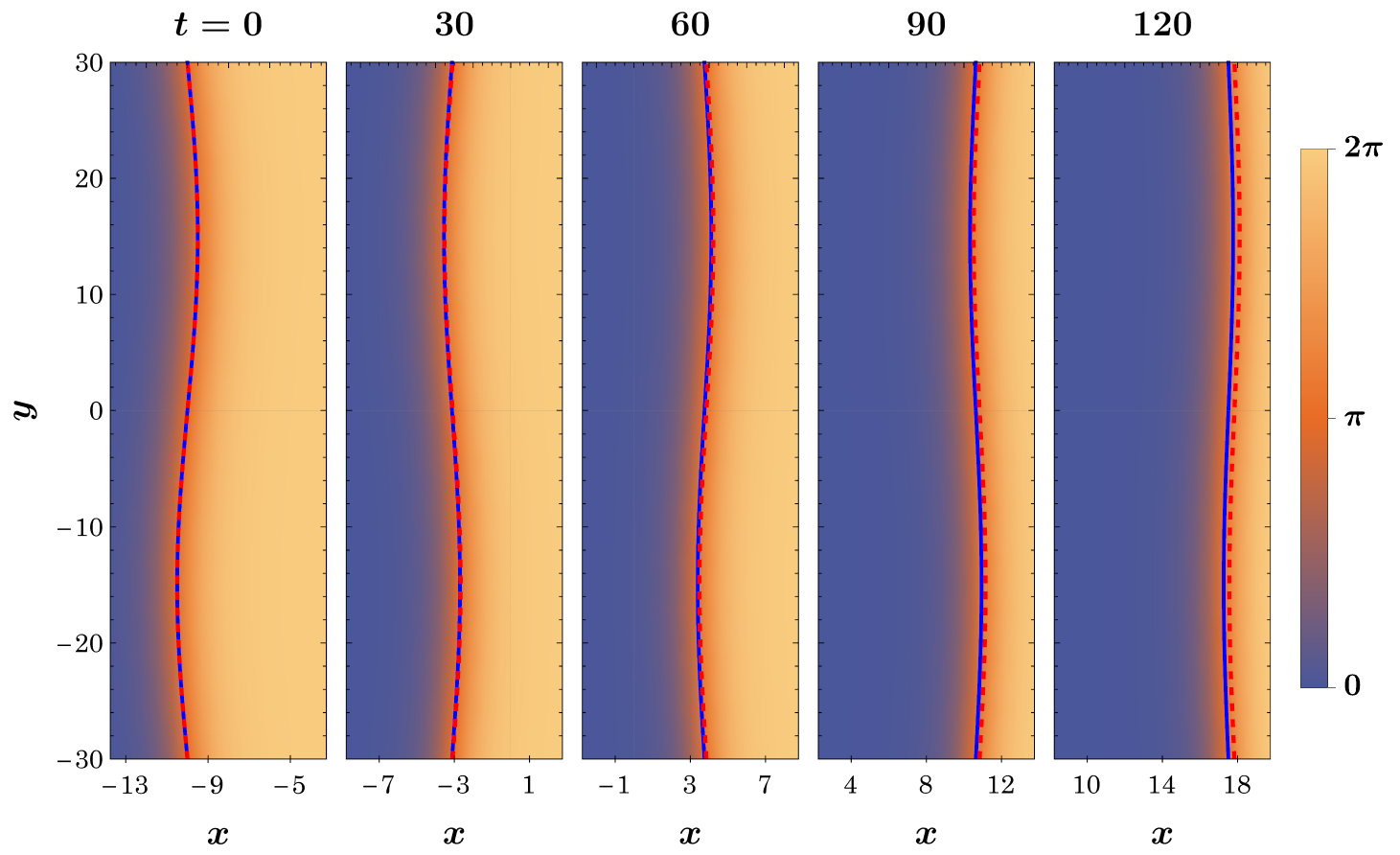}}}
    \caption{Comparison of the position of the center of the kink for the results obtained from the full field model and the approximate model with a modification of the initial position of the kink according to equation \eqref{modificationX0}. On the left the parameters in the figures shown have values $u=0.025$, $\Gamma=0$ and $\alpha=0$ and on the right $\Gamma=0.003$, $\alpha=0.01$ and $u_s$. In both cases $\varepsilon=0$ and $\lambda = 0.5$.}
    \label{fig_02}
\end{figure}

A slightly different situation is illustrated in Figure \ref{fig_02}. The first difference is that the bias current is zero $\Gamma=0$, and so instead of using equation (2) we can choose the initial velocity arbitrarily (here we take $u_0=0.2$). 
The second difference is that the shape of the front is deformed at the initial time. Here we assume the sinusoidal form of the deformation described by the formula
\begin{equation}
    X(y,t=0) =X_0 + \lambda\sin\left(\frac{2\pi y}{L_{y}}\right),
    \label{modificationX0}
\end{equation}
where $L_{y}=60$ is the width of the system along the direction of the $y$ variable.
This is selected with the mindset that the
any functional form of $X(y,t=0)$ should, in principle,
be decomposable in (such) Fourier modes.
The value of $X_0$ as before is $X_0=-20$, while the amplitude of the deformation is $\lambda=0.5$.
The value of the dissipation constant in the system is $\alpha = 0.001$. As before, there are no inhomogeneities in the system, i.e., $\varepsilon=0$.
The method of presenting the results is similar to that used in Figure \ref{fig_01}. 
The left panel illustrates the field configurations obtained from the equation \eqref{fullmodel},  sequentially at instants
$t=0, 30, 60, 90, 120$. The red solid line represents the kink front at the listed moments of time. On the right panel, the kink positions shown on the left panel (red lines) are compared with those obtained from the effective model \eqref{e-model}. The results of the effective model are represented by blue dashed lines. As can be seen, until $t=150$ there are no apparent differences between the results of the field model and the approximate model. 

\begin{figure}
    \centering
    \subfloat{{\includegraphics[height=5cm]{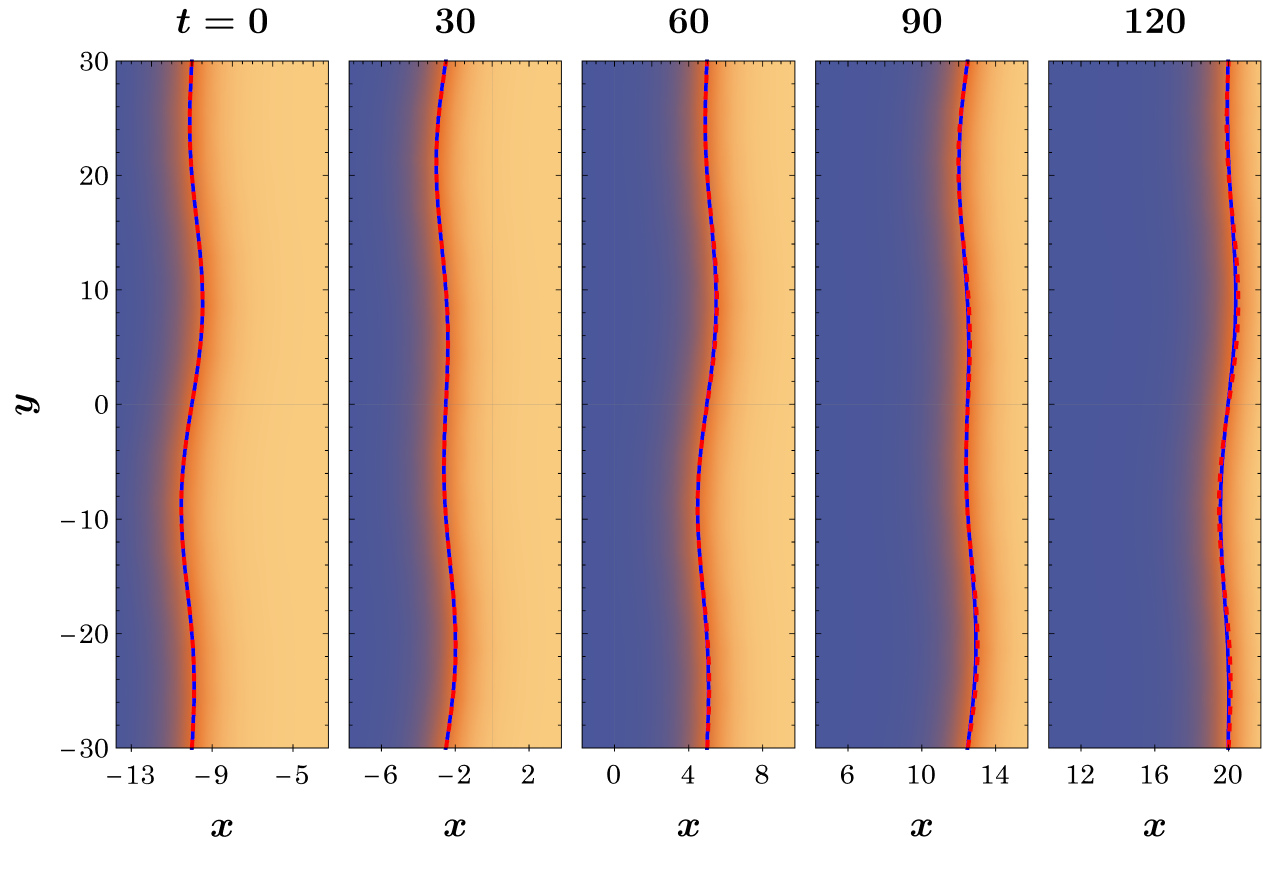}}}
    \qquad
    \subfloat{{\includegraphics[height=5cm]{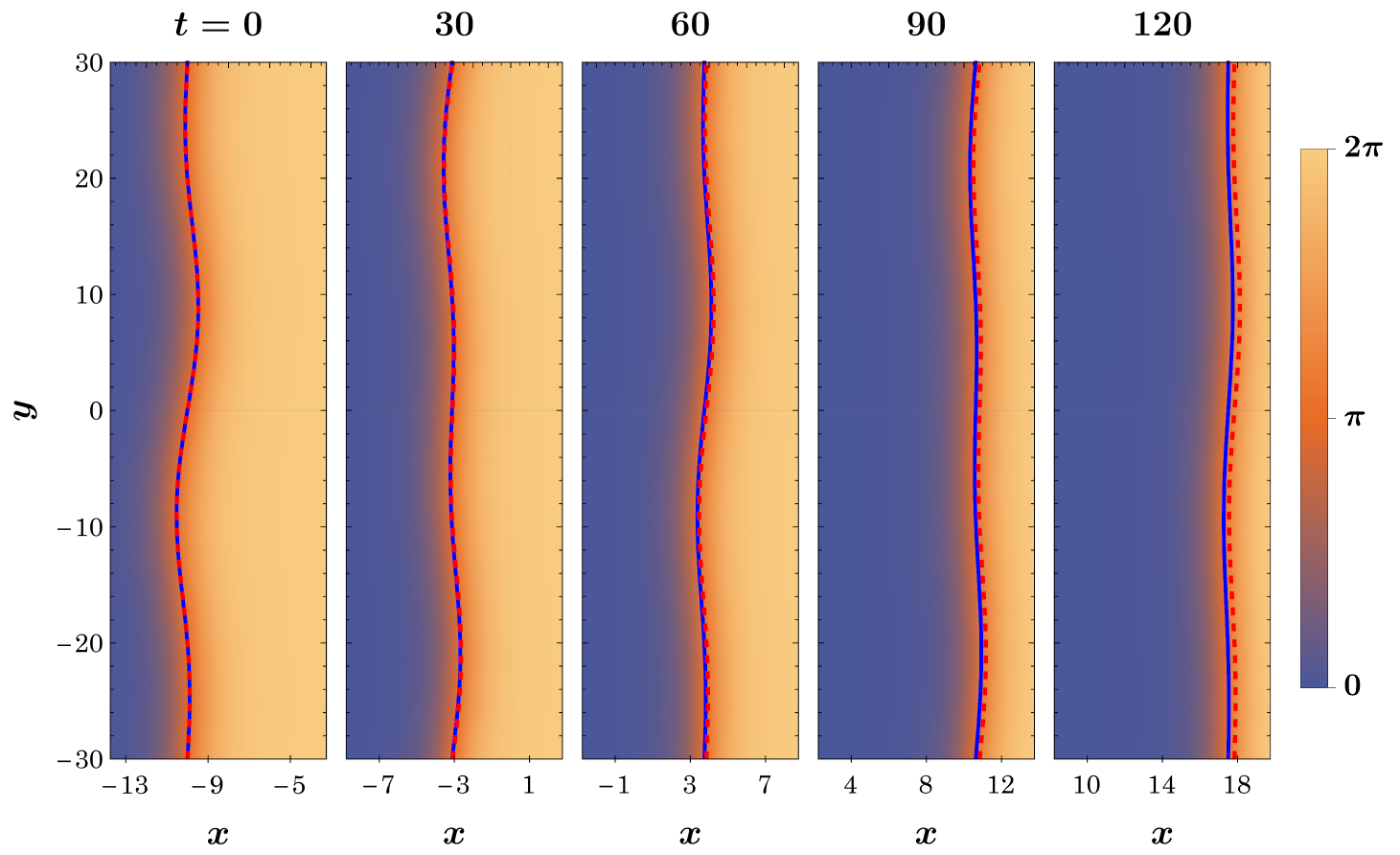}}}
    \caption{Comparison of the position of the center of the kink for the results obtained from the full field model and the approximate model with a modification of the initial position of the kink according to equation \eqref{modificationX0_full} for $N=2$. On the left the parameters in the figures shown have values $u=0.025$, $\Gamma=0$ and $\alpha=0$ and on the right $\Gamma=0.003$, $\alpha=0.01$ and $u_s$. In both cases $\varepsilon=0$ and $\lambda = 0.5$.}
    \label{fig_03}
\end{figure}

\begin{figure}
    \centering
    \subfloat{{\includegraphics[height=5cm]{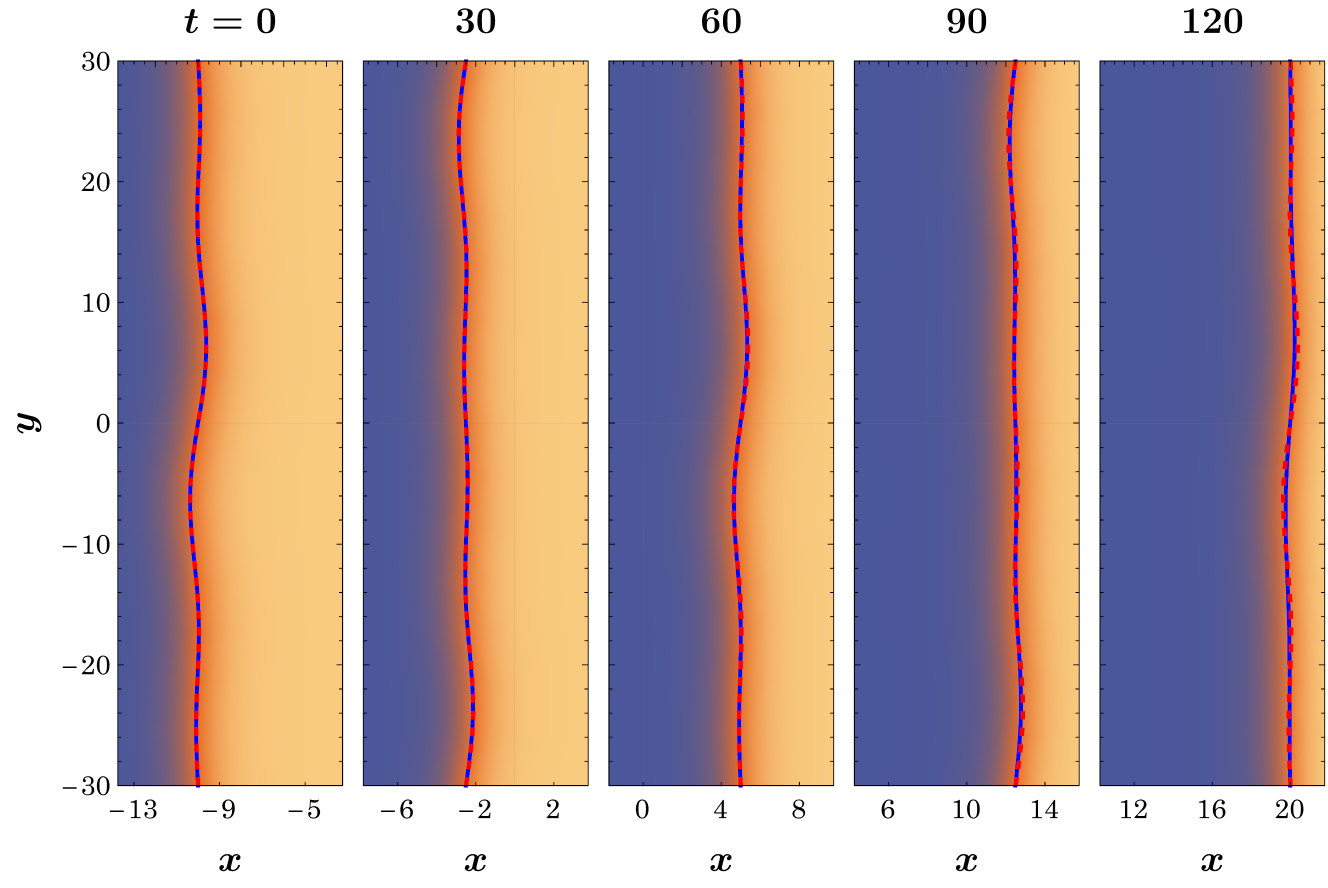}}}
    \qquad
    \subfloat{{\includegraphics[height=5cm]{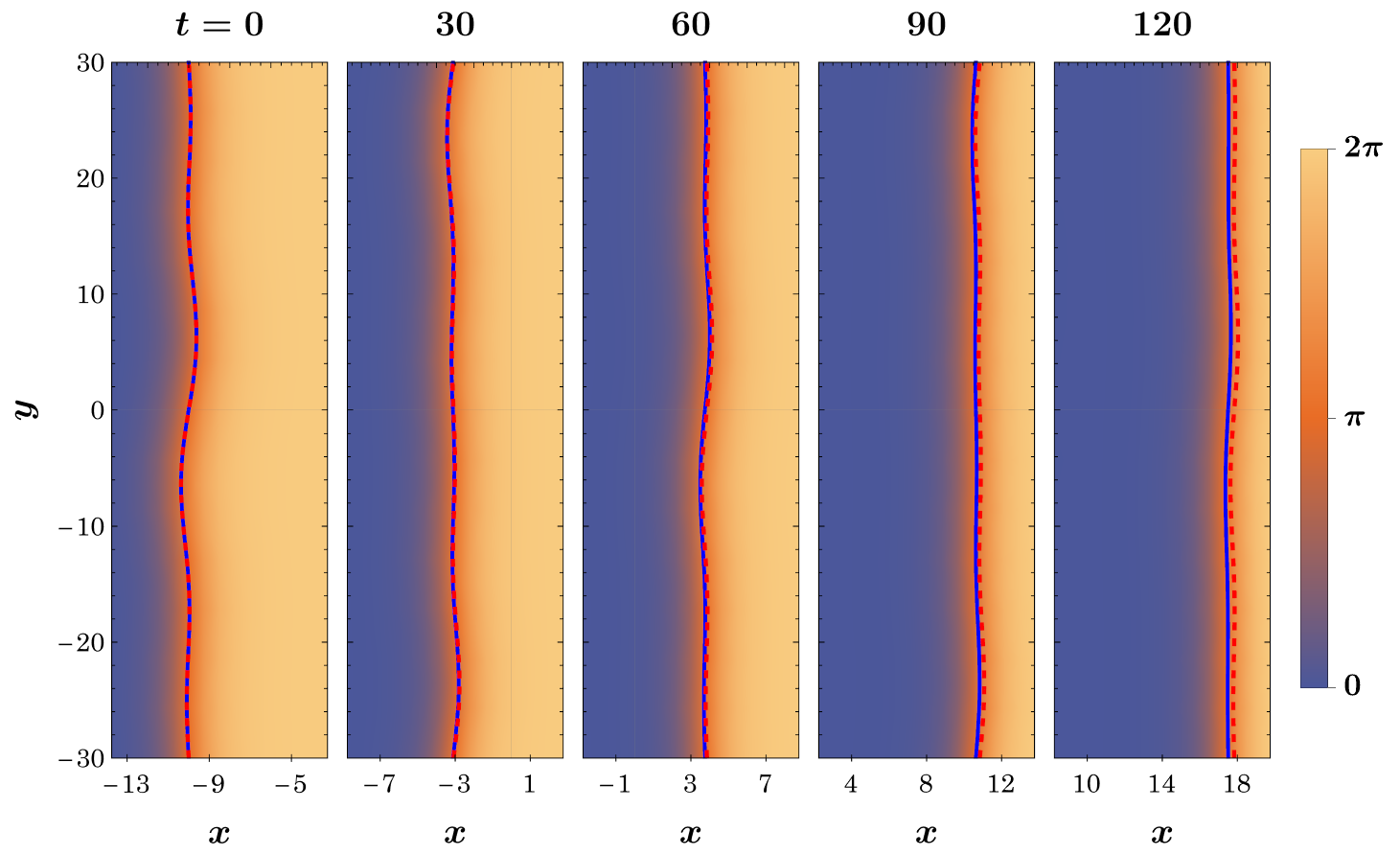}}}
    \caption{Comparison of the position of the center of the kink for the results obtained from the full field model and the approximate model with a modification of the initial position of the kink according to equation \eqref{modificationX0_full} for $N=3$. 
    On the left the parameters in the figures shown have values $u=0.025$, $\Gamma=0$ and $\alpha=0$ and on the right $\Gamma=0.003$, $\alpha=0.01$ and $u_s$. In both cases $\varepsilon=0$ and $\lambda = 0.5$.}
    \label{fig_04}
\end{figure}

A similar comparison to Figure \ref{fig_02} was made for a more complex shape of the kink initial front. In Figure \ref{fig_03}, we  studied the case of the initial kink front deformation containing more harmonics 
\begin{equation}
    X(y,t=0) = X_0 + \lambda\sum_{n=1}^{N}\sin\left(\frac{2\pi n y}{L_{y}}\right).
    \label{modificationX0_full}
\end{equation}
In this figure we have shown the evolution of the initial configuration with $N=2$ and $\lambda=0.5$. The other parameters for this case are exactly the same as for the process shown in Figure  \ref{fig_02}, i.e., among other things, the tested system is homogeneous $\varepsilon=0$ and the kink is not subjected to external force, i.e., $\Gamma=0$. As can be seen in the figure, the correspondence is very good even for $t=150$. An almost identical situation is shown in Figure \ref{fig_04}. In the case of this figure, the only difference from Figure \ref{fig_03} is the more complicated form of the kink front, which this time corresponds to $N=3$.  In this case, the first noticeable deviations appear for $t=120$. 

Summarizing, the simulations shown in the left panels of Figures \ref{fig_02}, \ref{fig_03} and \ref{fig_04} demonstrating the evolution of initially deformed kink fronts for $N=1, 2, 5$ and $\Gamma=0$ were repeated for non-zero bias current. 
The right panels of these figures  show the evolution of the kink front at a bias current equal to $\Gamma=0.003$ and a dissipation coefficient $\alpha=0.01$.
In these instances, the initial velocity calculated from equation \eqref{stationary_velocity} is $u_0=0.229339$. This velocity is the initial condition for the evolution of the kink fronts shown in right panels of Figures \ref{fig_02}, 
 \ref{fig_03},  \ref{fig_04}. 
Figure \ref{fig_02} according to the formula \eqref{modificationX0_full} shows the evolution of a deformed kink front with $N=1$, Figure  \ref{fig_03} corresponds to $N=2$, while Figure \ref{fig_04} describes the evolution of a front with $N=5$. In all cases, the front determined on the basis of the approximate equation \eqref{e-model} is slightly delayed compared to the front determined on the basis of the full field equation \eqref{fullmodel}. It turns out that in the first two cases ($N=1$, $N=2$) describing relatively slow deformation of the front (at the initial time), the approximate model gives even for $t=120$ the waveform of the front well reflecting the waveform of the front obtained from the full field model. The situation is 
slightly different for $N=5$. In this case, quite good agreement is obtained for $t=60$ and even $t=90$, while for $t=120$ we observe small differences.

\subsection{Propagation of the front in the presence of an $x$-axis directed inhomogeneity}
In this subsection, we will assume that the parameter $\varepsilon$ in equations \eqref{fullmodel} and \eqref{e-model} is non-zero. Such an assumption means that there is inhomogeneity in the system. In this work, we will describe the effect of inhomogeneity described by the function $g(x,y)=p(x)q(y)$, where $p(x)$ is given by equation \eqref{pfun}. In this first introduction
of the inhomogeneity, we will assume that $q(y)=1$, which means that the inhomogeneity is in the form of an elevation of height $\varepsilon$, orthogonal to the $x$-direction (which defines the direction of the kink movement). The spatial size of the inhomogeneity 
along the $x$-direction is approximated by the parameter $h$ appearing in equation \eqref{pfun}. In the simulations in this section, we assume $h=10$ and  $\varepsilon=0.01$. 
We study three types of kink dynamics. 

\begin{figure}
    \centering
    \subfloat{{\includegraphics[height=5cm]{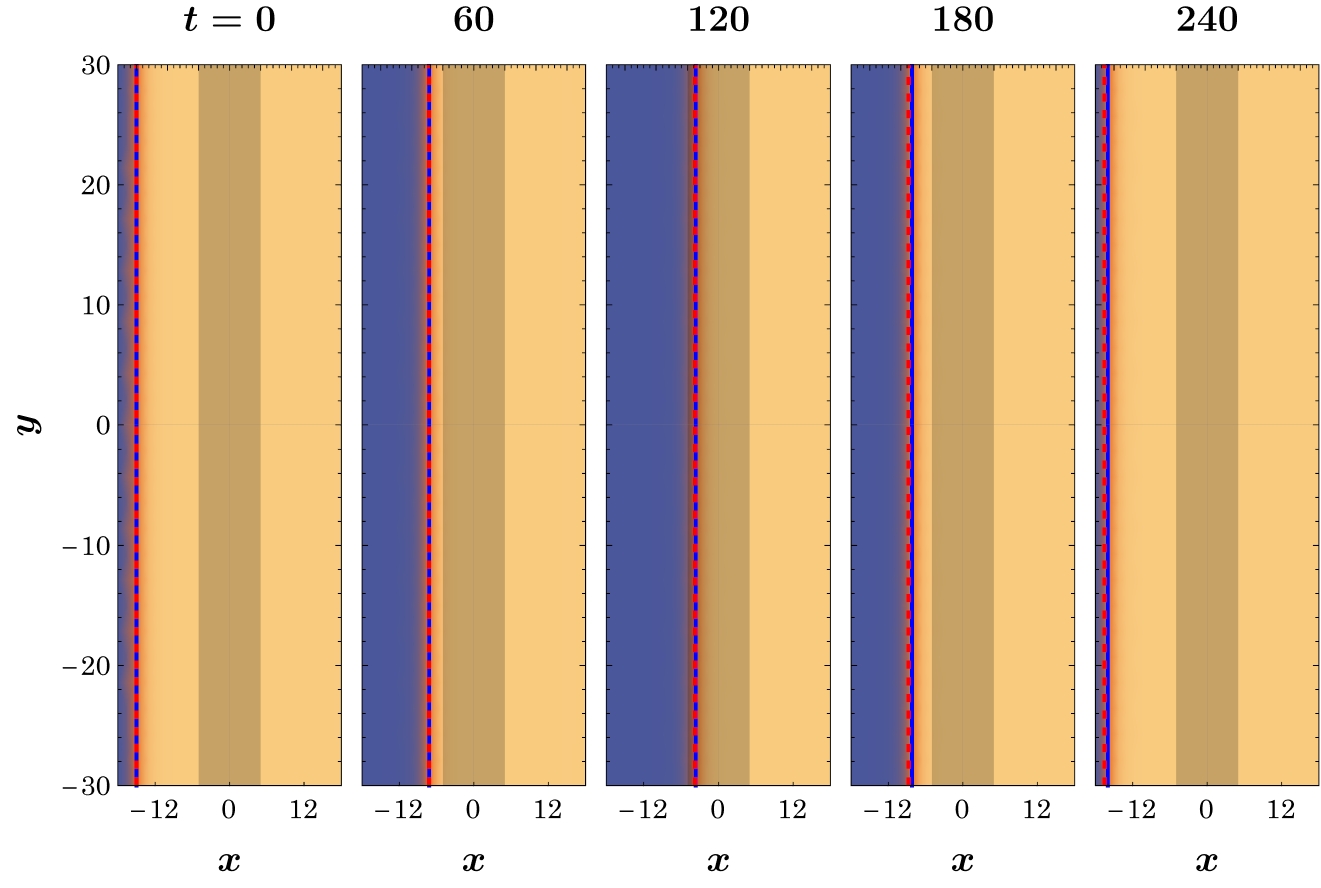}}}
    \qquad
    \subfloat{{\includegraphics[height=5cm]{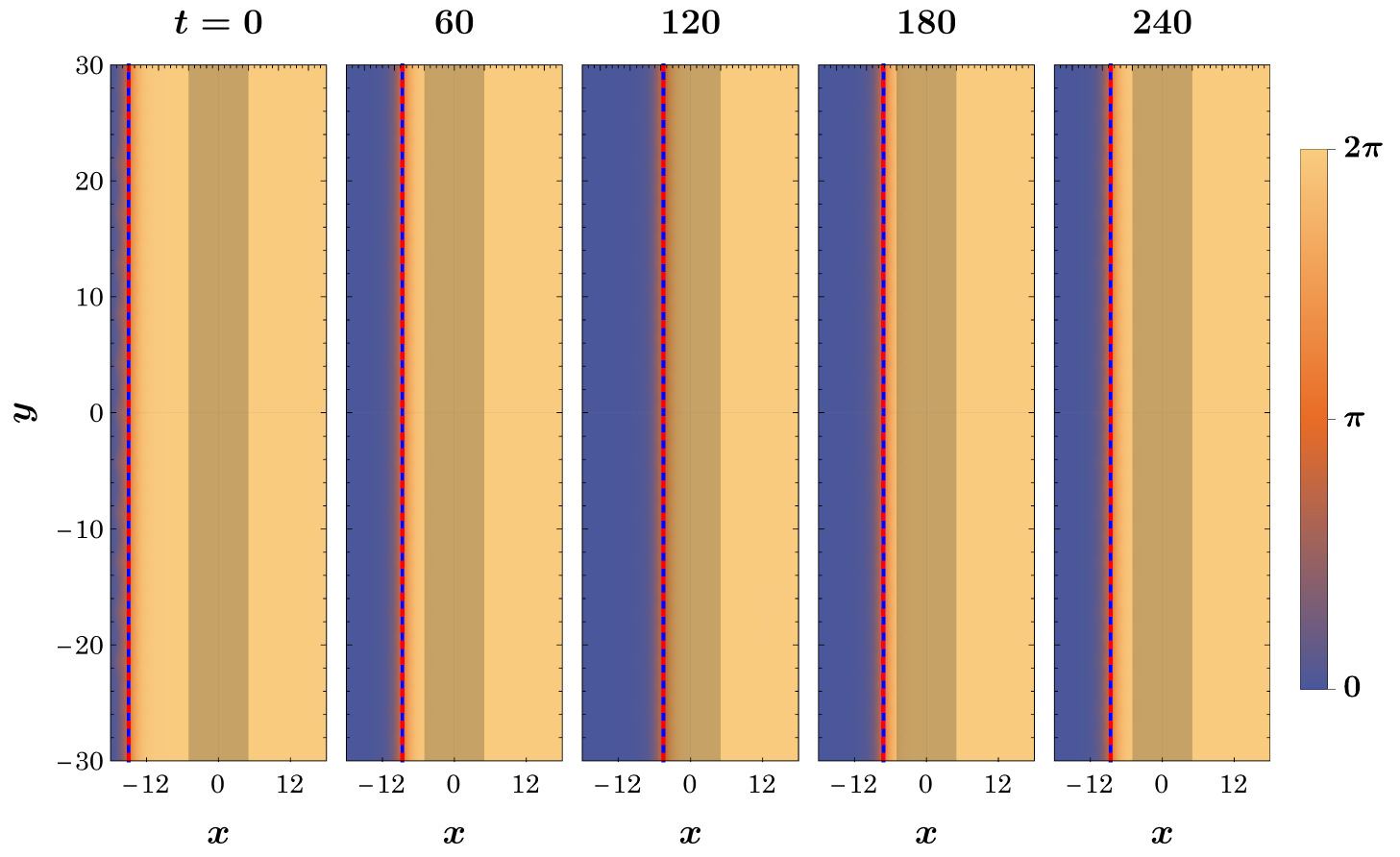}}}
    \caption{Comparison of the position of the center of the kink for the results obtained from the full field model and the approximate model. On the left the velocity has value $u=0.13$, and on the right $\Gamma=0.00135$, $\alpha=0.01$ and $u_s$. In both cases $\varepsilon=0.01$.}
    \label{fig_05}
\end{figure}

In the first case, we
consider the a reflection of the kink
from a barrier. The course of this
process is shown in Figure \ref{fig_05}. The case of reflection in the absence of external forcing ($\Gamma=0$) and dissipation
($\alpha=0$) is shown in the left figure. The initial condition in
this case is a straight kink front with a velocity $u=0.13$. As in
the previous section, the kink front is identified with the line
$\phi(t,x,y)=\pi$ (obtained from the field equation \eqref{fullmodel}). The front is
represented by the red line. Regions with $\phi(t,x,y)<\pi$ are once again
represented as blue areas, and $\phi(t,x,y)>0$ as yellow. On the
other hand, the position of the front determined from equation \eqref{e-model}
 is represented by the blue dashed line. The gray area
represents the position of inhomogeneity. The figure shows the
position of the front at instants $t=0, 60, 120, 180$ and $240$.
The kink at moments $t=0, 60, 120$ approaches the inhomogeneity while between moments $t=120$ and $t=180$ it is reflected and turns around, while
at instants between $t=180$ and $t=240$  it is already moving towards the initial position.
As can be seen, the correspondence of the two
descriptions, namely the ones based
on equation \eqref{fullmodel} and on equation \eqref{e-model} is very good, until t=120, while above this value we observe
slight deviations. The right figure
shows the same process in the case of occurrence of a dissipation
$\alpha=0.01$ and forcing $\Gamma=0.00135$ in the system. The course of the
front at the same moments as in the left figure also shows very
good agreement of the approximate model \eqref{e-model} with the initial
model \eqref{fullmodel}, also for t=240. In this figure, the initial velocity of the front is
chosen based on the formula \eqref{stationary_velocity}, i.e., as the stationary velocity.
It should be mentioned that the bouncing process in this case is
slightly more complex and has an identical 
(effectively one-dimensional) nature to that
described in the one-dimensional case in the paper \cite{Gatlik2023arXiv}. It
consists of multiple (damped) reflections from the barrier, which
eventually ends up stopping before the barrier.
As was shown in \cite{Gatlik2023arXiv}, this reflects
the presence of a stable spiral point at such a
location which asymptotically attracts the kink
towards the relevant fixed point.

\begin{figure}
    \centering
    \subfloat{{\includegraphics[height=5cm]{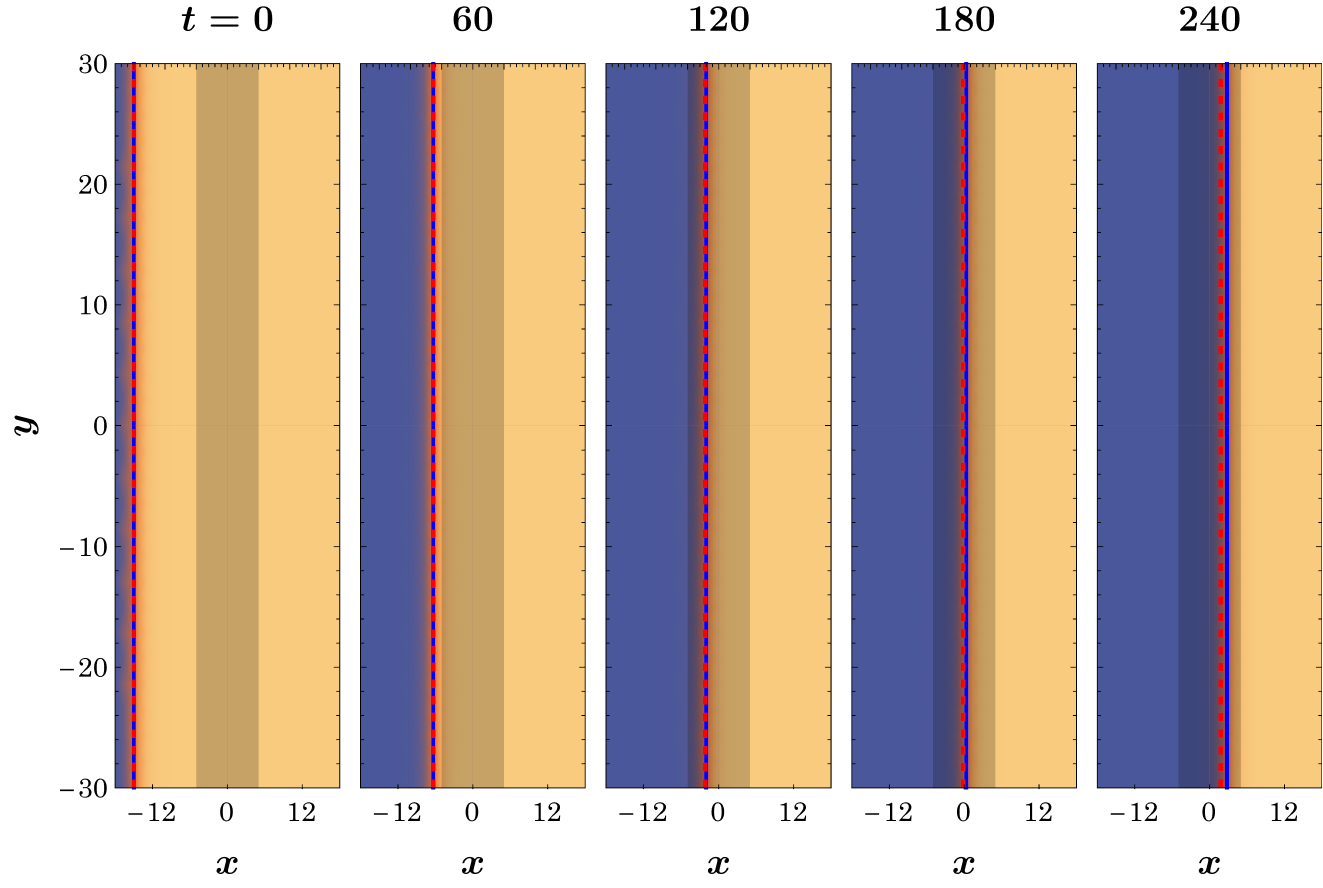}}}
    \qquad
    \subfloat{{\includegraphics[height=5cm]{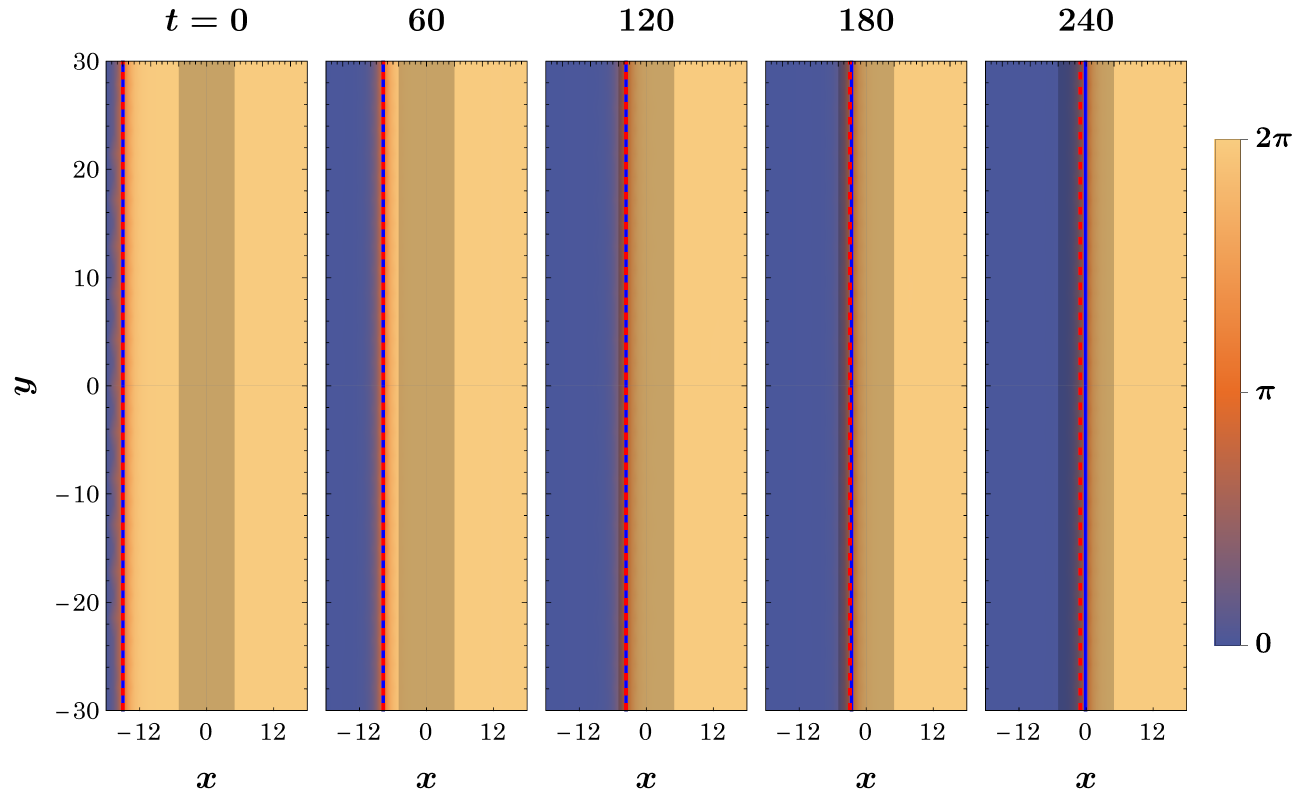}}}
    \caption{Comparison of the position of the center of the kink for the results obtained from the full field model and the approximate model. On the left the velocity has value $u=0.145$, and on the right $\Gamma=0.00155$, $\alpha=0.01$ and $u_s$. In both cases $\varepsilon=0.01$.}
    \label{fig_06}
\end{figure}

In the second case, we are dealing with the interaction of the
kink with the inhomogeneity for nearly critical parameter values.
This means an initial speed close to the critical velocity in
the absence of forcing and dissipation. When dissipation
in the system is present and when the forcing is non-zero, {then we assume that the forcing takes a value that leads through the formula \eqref{stationary_velocity} to a stationary speed approximately equal to the critical velocity. }
Figure  \ref{fig_06}, demonstrates this process
in detail. The left panel of this figure shows (with labeling
identical to this in Figure  \ref{fig_05}) the interaction of the kink with
the inhomogeneity at velocity $u=0.145$.
In this case, the kink
stays in the inhomogeneity region for a long time.  
Indeed, by the end of the
time frame monitored {in Figure \ref{fig_06},} the kink has not exited
the inhomogeneity. {Ultimately, if the time is extended even further then the movement of the kink front to the other side of the inhomogeneity can be observed.}
The
position of the front determined from the field equation  \eqref{fullmodel}, with
$\alpha=0$ and $\Gamma=0$ is in good coincidence with the position
of the kink obtained from the equation  \eqref{e-model}, up to the time t=120.
For longer times slight deviations are observed. On the other
hand, the right panel shows results for $\Gamma=0.00155$ and
$\alpha=0.01$. It can be seen that, this time as well, 
the agreement of
the position of the front determined from the original equation
and the effective one is very good up to the instant t=120. At
later moments we observe slight deviations. We would like to
underline that  Figures show only a part of the space
(i.e., from $-18$ to $18$) which in
the direction of the $x$-axis is contained in the range from
$-70$ to $70$, while in the $y$-axis direction it is contained in
the range from $-30$ to $30$.

\begin{figure}
    \centering
    \subfloat{{\includegraphics[height=5cm]{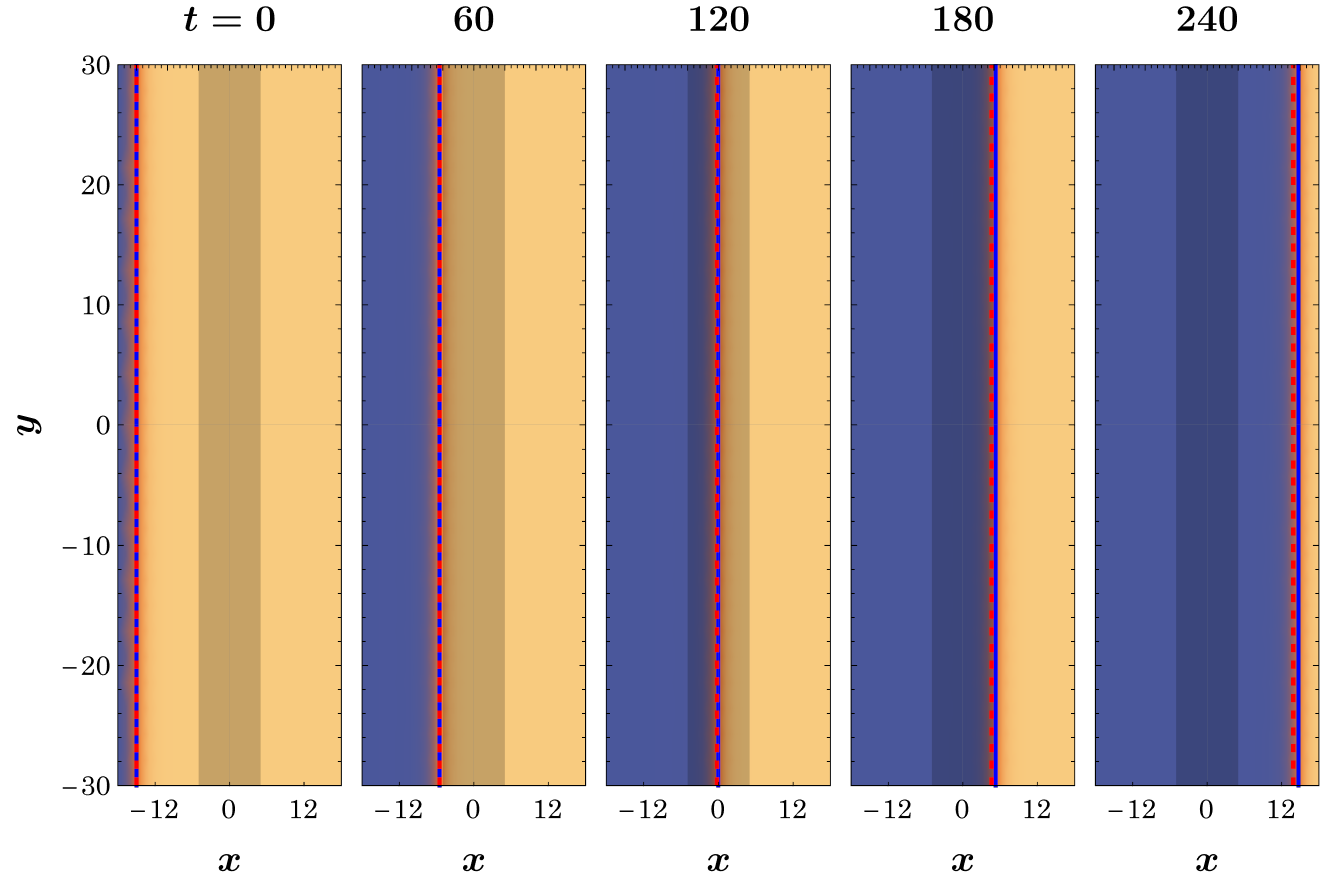}}}
    \qquad
    \subfloat{{\includegraphics[height=5cm]{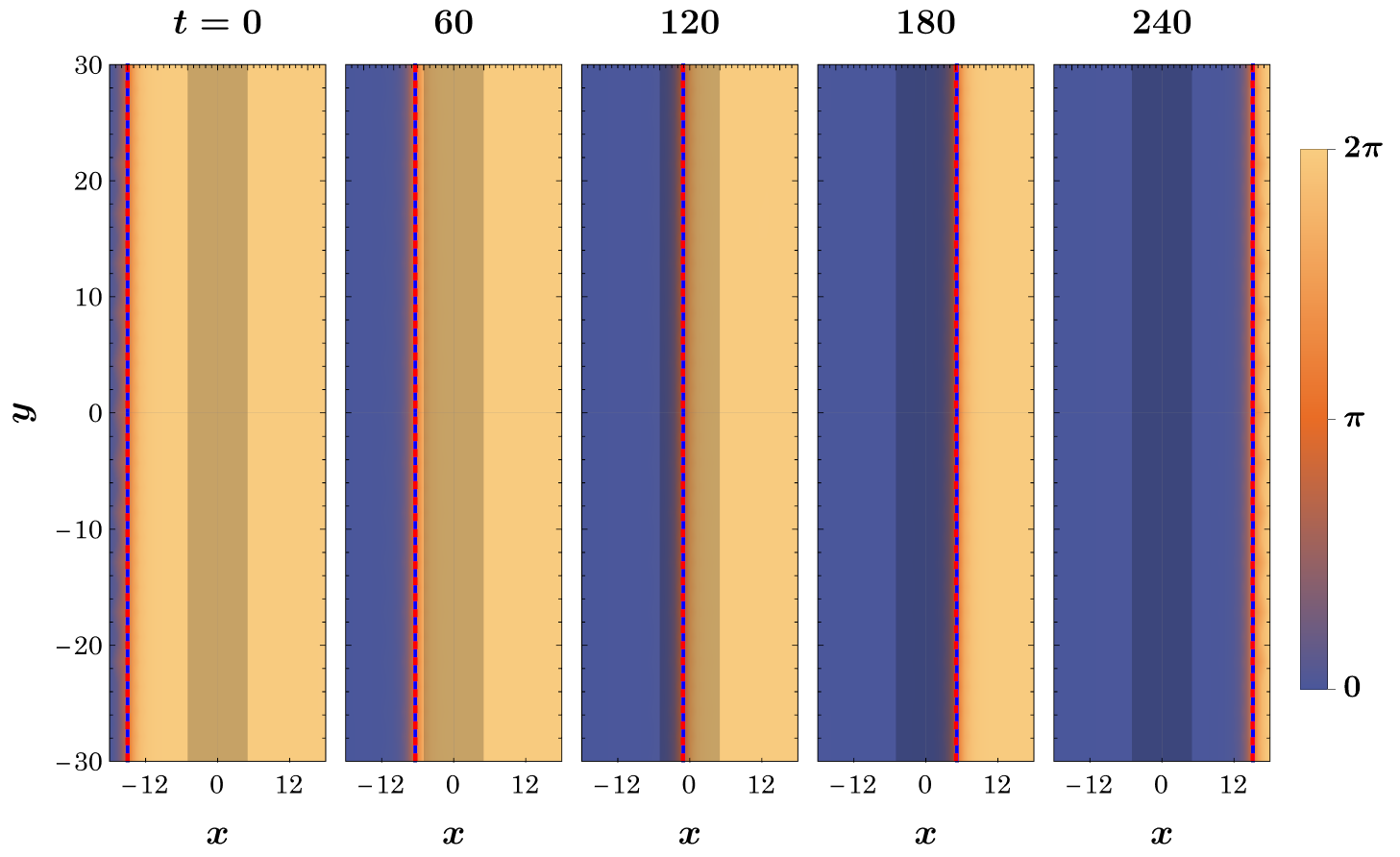}}}
    \caption{Comparison of the position of the center of the kink for the results obtained from the full field model and the approximate model. On the left the velocity has value $u=0.16$, and on the right $\Gamma=0.00185$, $\alpha=0.01$ and $u_s$. In both cases $\varepsilon=0.01$.}
    \label{fig_07}
\end{figure}

The last case is shown in Figure  \ref{fig_07}. The left panel shows the
movement of a kink  with an initial velocity $u=0.16$
significantly exceeding the critical speed. In this case, slight
deviations are already observed for $t=120$. On the other hand,
the case with dissipation is presented in the right panel. This
figure shows a kink front with an initial speed equal to the
stationary velocity determined for dissipation $\alpha=0.01$ and
forcing $\Gamma=0.00185$. In this case, the correspondence of the
description obtained from equation \eqref{fullmodel} and equation  \eqref{e-model} are
striking up to $t=240$.
The results obtained in this section are analogous to those
described in the paper \cite{Gatlik2023arXiv}, 
as the effective motion of the kink is practically
one-dimensional and the transverse modulation neither
plays a critical role to, nor destabilizes (as is, e.g.,
the case in nonlinear Schr{\"o}dinger type models \cite{kuznetsov1988instability}) the
longitudinal motion.


\subsection{Kink propagation for inhomogeneities dependent on both variables}

In this section we will consider some examples of heterogeneities
bearing a genuinely two-dimensional character, i.e., having a non-trivial
dependence not only on the $x$-variable initially 
aligned with the
direction of movement of the kink, but also on the $y$ variable, along which the front is initially homogeneous.

\subsubsection{Barrier-shaped inhomogeneity}
\begin{figure}
    \centering
    \subfloat{{\includegraphics[width=8cm]{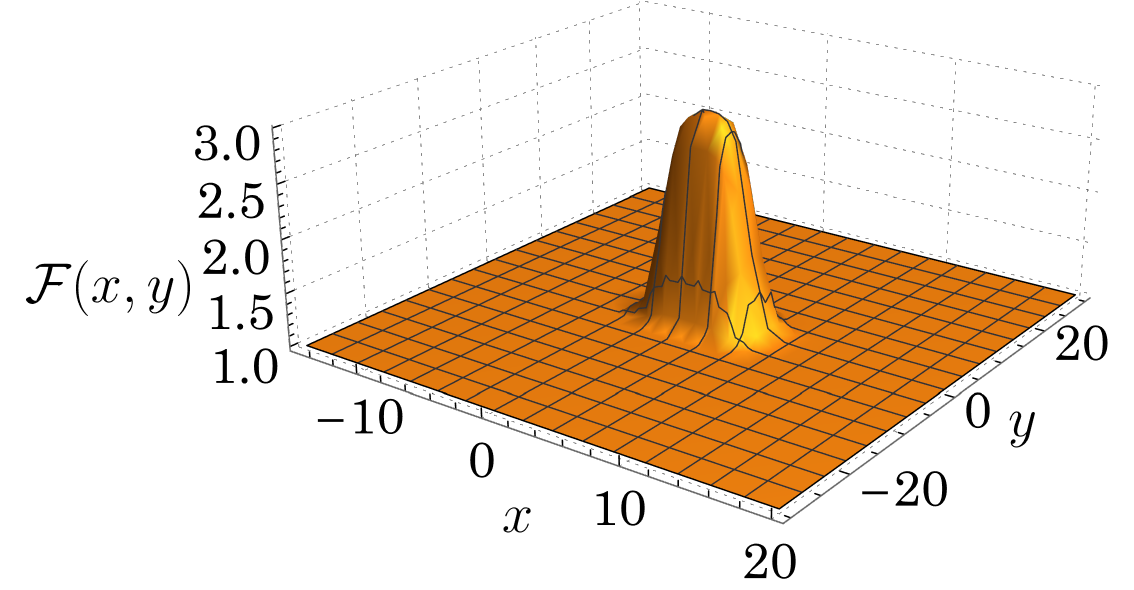}}}
    \qquad
    \subfloat{{\includegraphics[width=8cm]{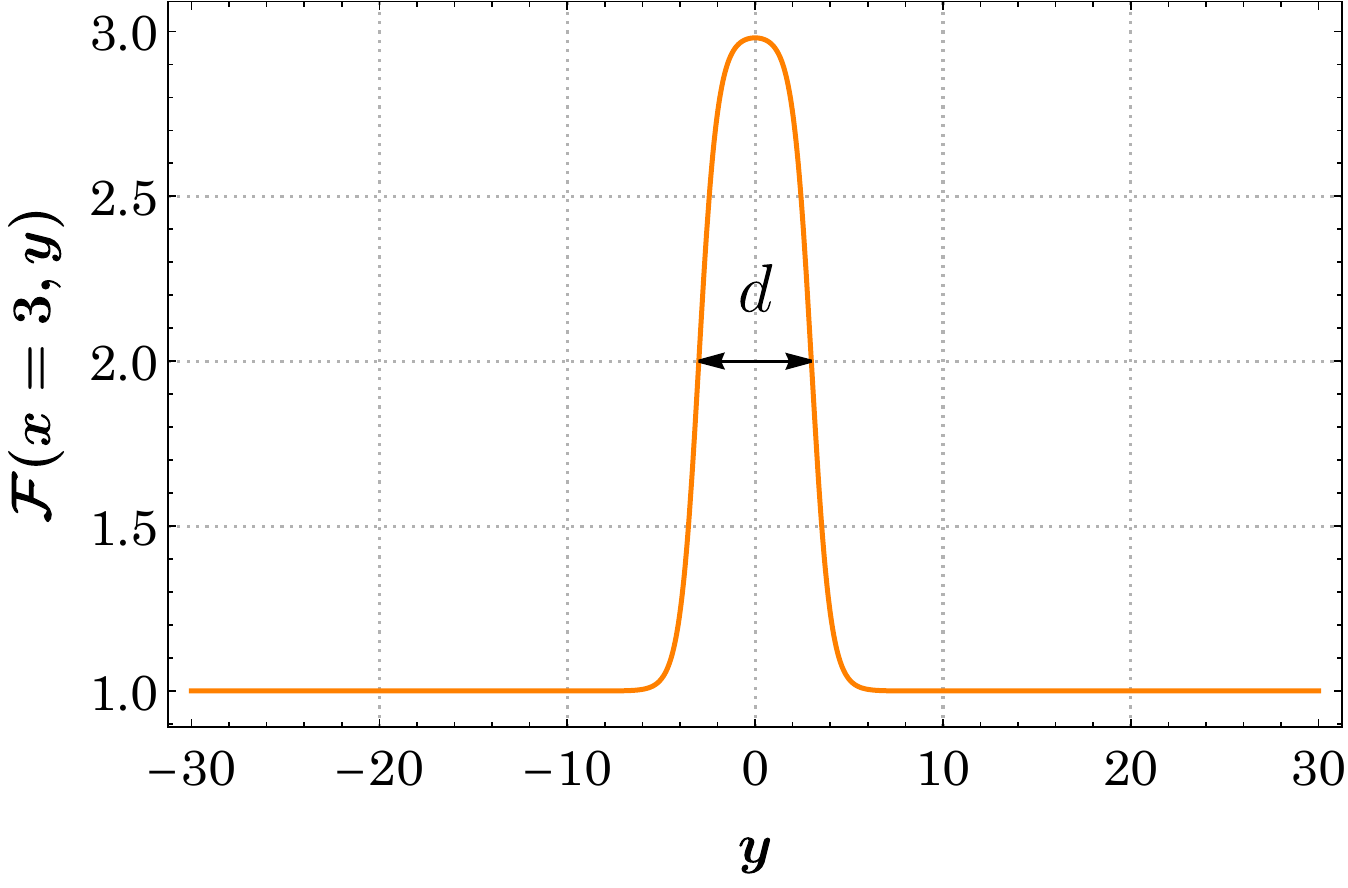}}}
    \caption{Left figure presents peak-shaped inhomogeneity $\mathcal{F}(x,y)$, while the right figure shows its section along a line $x=3$. Both parameters $h$ and $d$ are equall to $6$.}
    \label{fig_08}
\end{figure}

\begin{figure}
    \centering
    \subfloat{{\includegraphics[height=5cm]{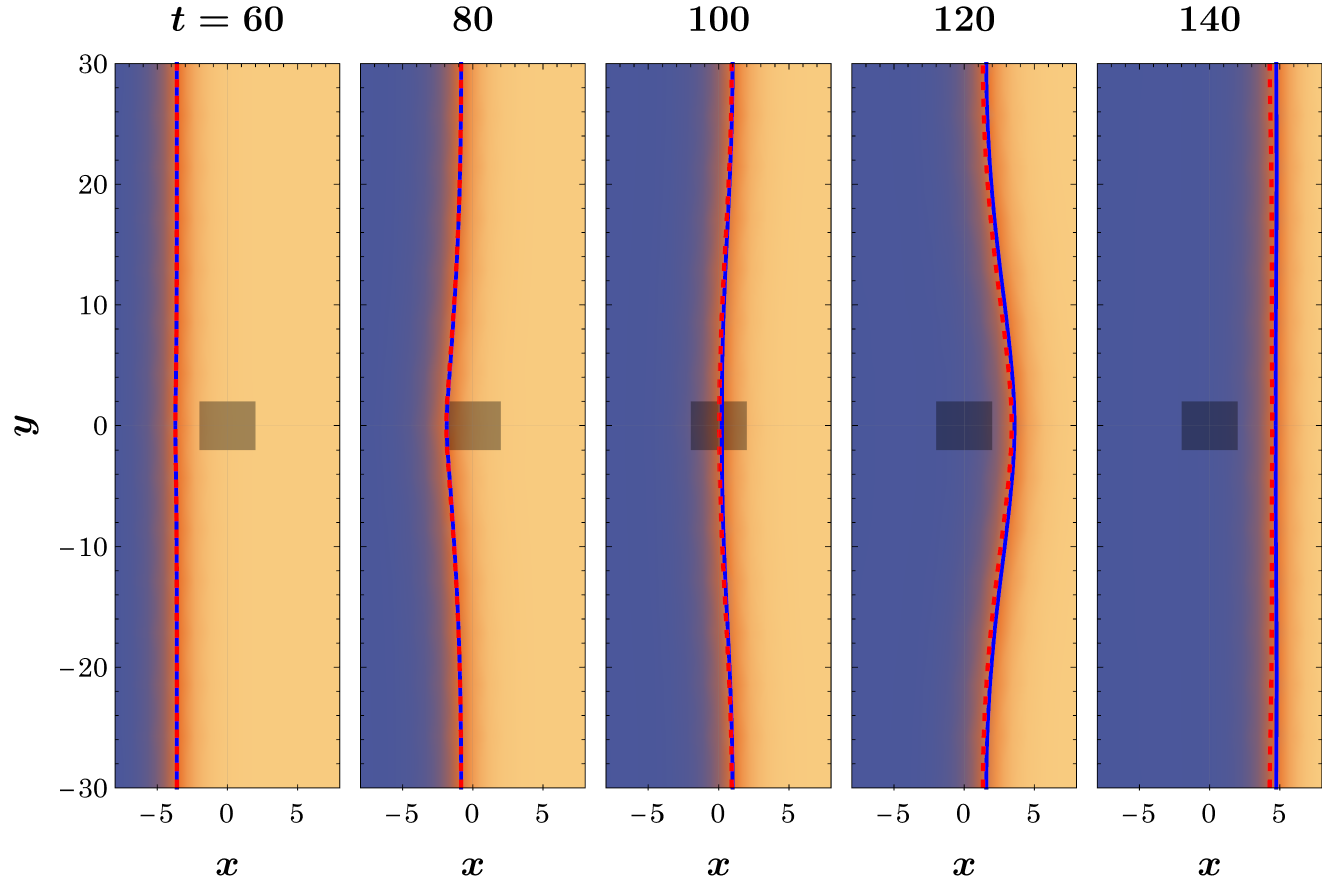}}}
    \qquad
    \subfloat{{\includegraphics[height=5cm]{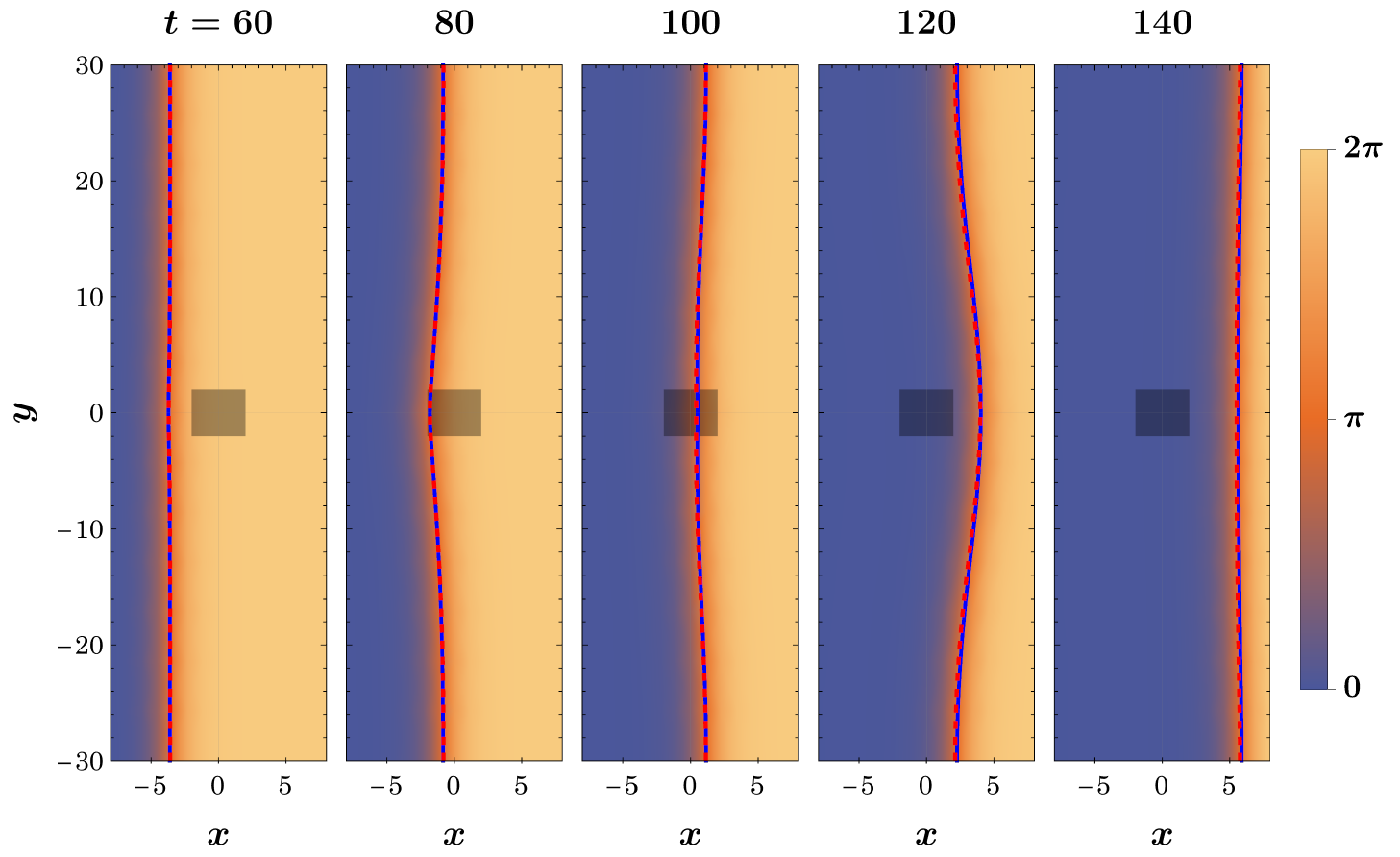}}}
    \caption{Passing over heterogeneity. In the left figure, the system without
forcing and dissipation. The kink front has an initial velocity
equal to $u=0.14$. The right figure shows an analogous process in
a system with forcing $\Gamma=0.0018$ and dissipation
$\alpha=0.01$. In both images the gray area represents
inhomogeneity. Here, we have that $\varepsilon=0.1$. The animations are available through the links \url{https://tinyurl.com/56wwcarr} for the left one and \url{https://tinyurl.com/3224zzf7} for the right one.}
    \label{fig_09}
\end{figure}

The first example is described by the function
$\mathcal{F}(x,y)=1+ \varepsilon g(x,y) = 1+ \varepsilon p(x)
q(y)$. The shape of this function is shown in Figure \ref{fig_08}.  In this case, the function $p(x)$ is given by formula  \eqref{pfun}
while $q(y)$ has the form:
\begin{equation}
\label{qfun}
    q(y) = \frac{1}{2}\left(\tanh(y + \frac{d}{2}) - \tanh(y - \frac{d}{2})\right).
\end{equation}
We will consider two cases. In the first case, the kink front passes over the inhomogeneity. In the second case, it is stopped by the inhomogeneity. To be more precise, the kink, in the absence of dissipation and forcing, bounces and returns towards its initial position, while when dissipation and forcing are non-zero the kink stops in front of the inhomogeneity due to the emergence
of a stable fixed point there.
The results of comparing the initial model \eqref{fullmodel} with the effective
model \eqref{eq-eff3} are very good, as can be seen in Figure \ref{fig_09}. 
In the
simulations, we assumed a parameter describing the strength of the
inhomogeneity equal to $\varepsilon=0.1$. The left panel shows the
interaction of the front with the inhomogeneity in the absence of
dissipation and forcing. The initial condition in this case is a
straight front with a velocity of $u=0.14$. It can be seen that in
the course of the evolution the front deforms (the 
kink bends around the inhomogeneity, which is represented in the figure as a gray
area) and then overcomes it. 
After crossing the 
 inhomogeneity, the tension of the string (the front of the kink) causes it to vibrate, i.e., it excites
 a transverse mode of the ``kink filament''. 
Obviously, we must remember
that local perturbations of the $\phi$-field profile can slightly
change the distribution of energy density along the kink front.
 As a consequence of the existence of tension,
 the string tends to straighten but excess kinetic energy causes it to vibrate in the direction of the motion of the
front, in the absence of dissipation and drive. This oscillation persists for a long time because the mechanism of energy reduction associated with its radiation is not very effective.
On the other hand, the right figure shows an analogous process in
the case where in the system we have a forcing of $\Gamma=0.0018$
and a dissipation characterized by the coefficient $\alpha=0.01$.
In this case, the initial speed is the stationary velocity
determined by the formula \eqref{stationary_velocity}. The course
of the process and the results are analogous to the case without
dissipation, i.e., we observe local changes in shape
that are similar to the left panel.
Nevertheless, after
passing over the inhomogeneity, we observe damped vibrations that ultimately lead to straightening of the front, as a result of this damped-driven system's
possessing of an attractor (and contrary to the 
scenario of the conservative Hamiltonian case). {The results shown in the figures have also been presented in the form of animations in the associated links. 
Since in the absence of forcing and dissipation, the mechanism of getting rid of excess energy through radiation is not sufficiently effective, extending the animation time in this case did not lead us to times at which the transverse oscillations of the kink front would disappear. The situation is different when there is dissipation in the system. The animation conducted for long times 
in the latter setting shows that the kink front straightens.}

\begin{figure}
    \centering
    \subfloat{{\includegraphics[height=5cm]{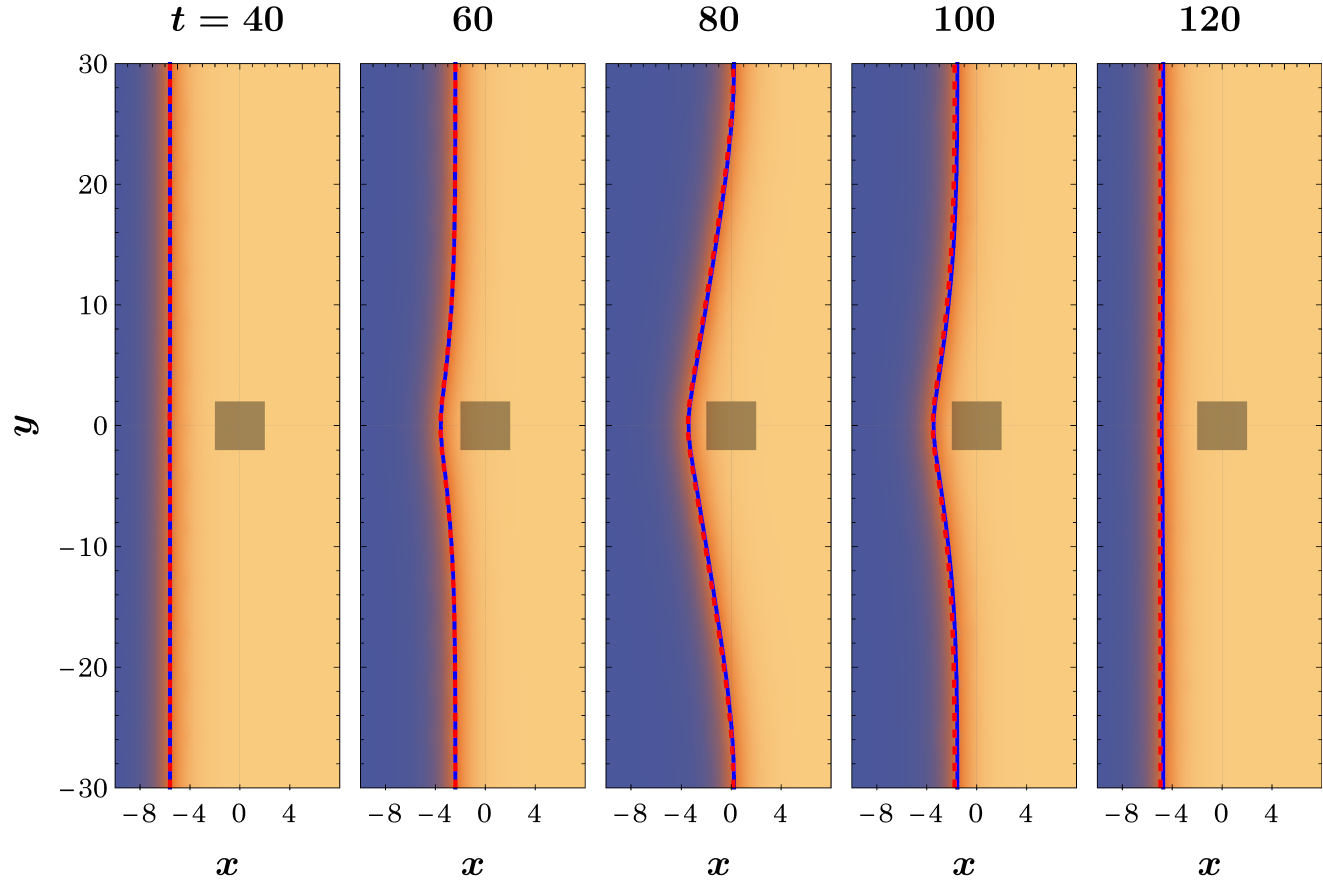}}}
    \qquad
    \subfloat{{\includegraphics[height=5cm]{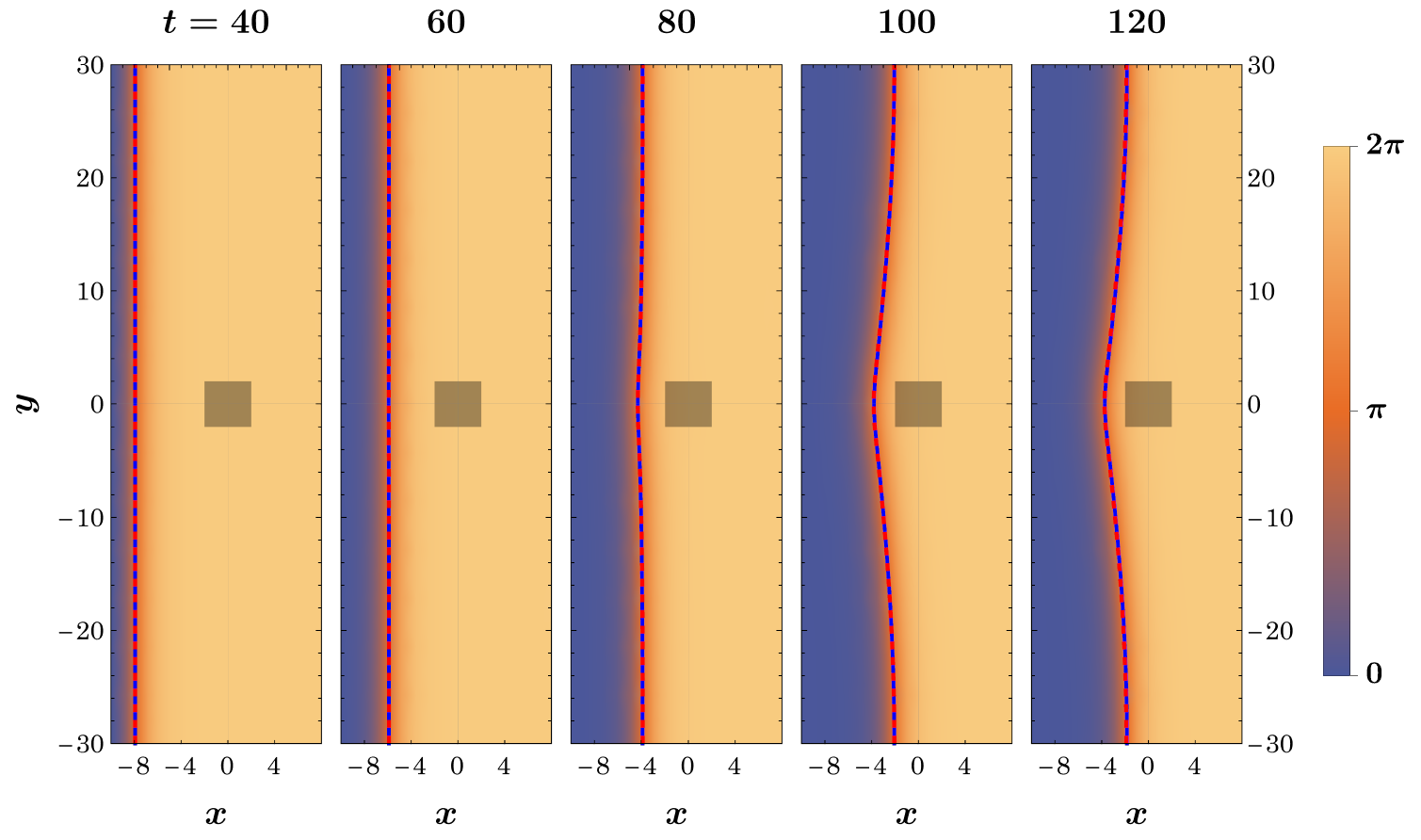}}}
    \caption{Reflection/stopping on inhomogeneity. In the left figure, the
system without forcing and dissipation is shown. The kink front has an
initial velocity equal to $u=0.16$. The right figure shows an
analogous process in a system with forcing $\Gamma=0.0013$ and
dissipation $\alpha=0.01$. In both pictures, the gray area
represents the inhomogeneities with $\varepsilon=0.5$. The animations are available through the links \url{https://tinyurl.com/54u972sb} for the left one and \url{https://tinyurl.com/mt6hd9pu} for the right one.}
    \label{fig_10}
\end{figure}

In the second case, shown in Figure \ref{fig_10}, we take a large value of the 
inhomogeneity strength
$\varepsilon=0.5$.

Accordingly, even a front with a velocity slightly greater than the
velocity reported in the previous figure is not sufficient to
overcome the inhomogeneity. The left panel shows the process of
interaction of a front with initial velocity $u=0.16$ with the
inhomogeneity represented by the gray area of the figure. As can
be seen during the interaction the front is attempting to pass over
the inhomogeneity, however, it finally bounces back towards
its initial
position.    Despite the large value of $\varepsilon$, 
and the substantial deformation of the kink filament,
the agreement between the original model \eqref{fullmodel} and the effective model \eqref{eq-eff3} remains 
very good. 
The right panel shows an even more interesting interaction of the kink front with the inhomogeneity.
In the figure, in addition to the value of the parameter
$\varepsilon=0.5$, a forcing of $\Gamma=0.0013$ and a dissipation
coefficient of $\alpha=0.01$ are assumed.
Initially the front moving towards the inhomogeneity experiences a deformation.
Then, a series of damped reflections of the front from the barrier occur.
During the reflections and returns, deformations of the entire front occur having the form of vibrations in the direction of motion. The subsequent turning of the front in the direction of the barrier is a consequence of the existence of an external forcing. Vibrations are damped due to the presence of dissipation in the system. What is interesting here is the final shape of the front, which is a consequence of multiple factors. The first factor is of course, the presence of a barrier that constrains the movement of the front and leads to an energetically
induced bending of the kink filament. The second is the presence of forcing, which in the middle is balanced by the presence of the barrier. The situation is different at the ends, where the front does not have ``feel'' the barrier (and hence is once again straightened). 
The combination of these factors with the geometric
distribution of our inhomogeneity leads to a
stable equilibrium analogous to the $1+1$-dimensional
case of \cite{Gatlik2023arXiv}. Yet, the 
present case also features
a spatial bending of the kink profile, given the
geometry of the heterogeneity and the
tendency to shorten the length of the kink, in a
way resembling the notion of string tension at
the front. 

\subsubsection{Heterogeneity in the  form of well.}

A slightly different type of inhomogeneity is a potential well. In
this section, the well is obtained by replacing $g(x,y)$ in the formula $\mathcal{F}(x,y)=1+ \varepsilon g(x,y) = 1+ \varepsilon p(x)
q(y)$ by $-g(x,y)$ and preserving the form of functions $p(x)$ and $q(y)$. 
In the relevant dip (rather than bump) of the heterogeneity,
 the parameters are taken as $h=6$ and
$d=6$. As in the previous section, we will consider two cases. In the first case, the front passes over the well, and in the second it is stopped by it.

\begin{figure}
    \centering
    \subfloat{{\includegraphics[height=5cm]{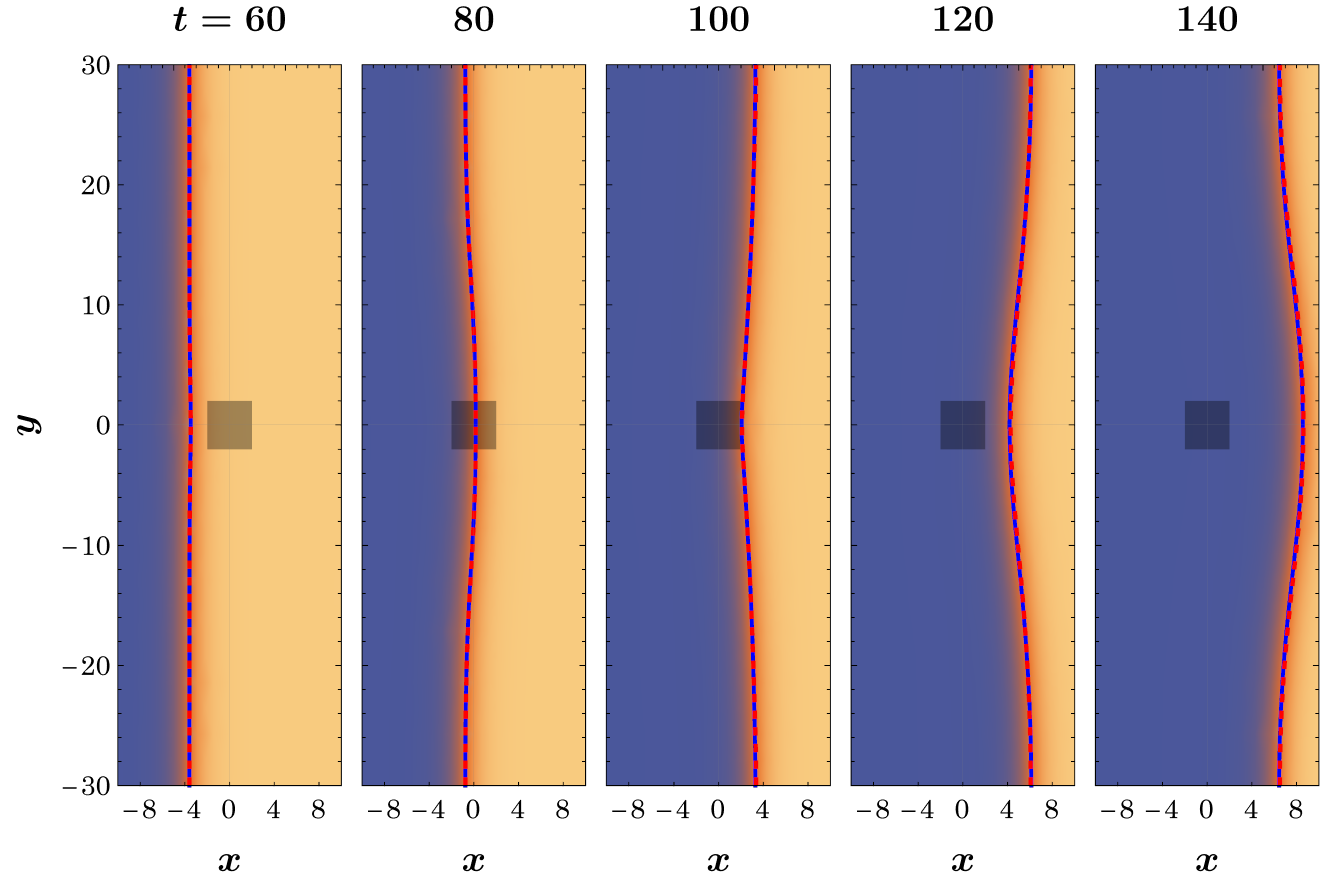}}}
    \qquad
    \subfloat{{\includegraphics[height=5cm]{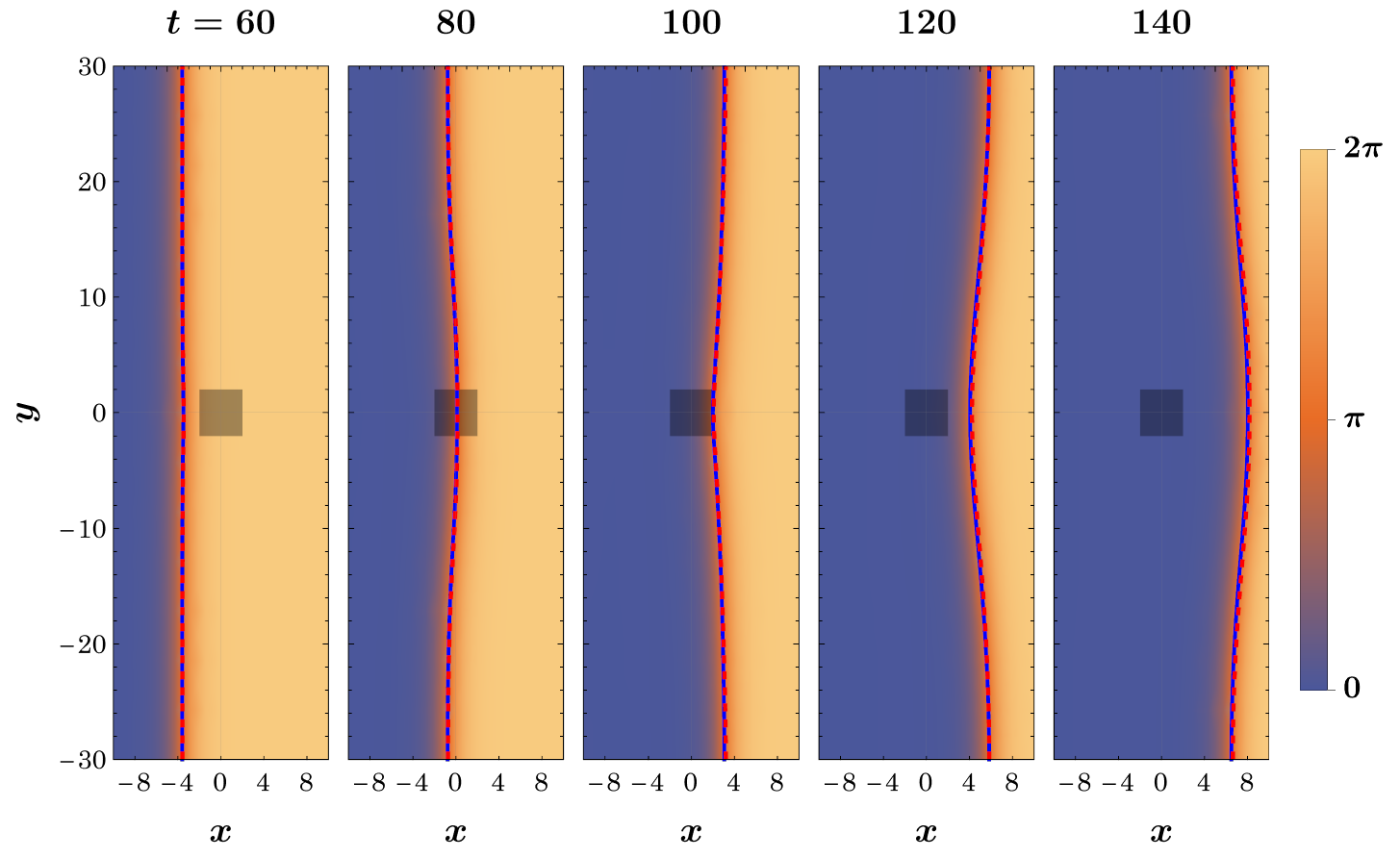}}}
    \caption{Passing over the well. In the left figure, the system without
forcing and dissipation. The kink front has an initial velocity
equal to $u = 0.14$. The right figure shows an analogous process
in a system with forcing $\Gamma = 0.0018$ and dissipation $\alpha
= 0.01$. In both images the gray area represents the region of inhomogeneity.
The parameter describing the depth of the well has a value of
$\varepsilon=0.1$. The animations are available through the links \url{https://tinyurl.com/2krhahj6} for the left one and \url{https://tinyurl.com/59p27xps} for the right one.}
    \label{fig_12}
\end{figure}

Figure \ref{fig_12} shows the case of a front passing over a well. The left
panel describes the case of no forcing and dissipation. The
parameter describing the depth of the well is $\varepsilon=0.1$.
The initial velocity of the front is $u=0.14$ in this case. A
straight front during its approach to the inhomogeneity deforms in the
middle part which is related to the attraction by the well (cf. with the opposite scenario of the barrier
case explored previously). In the
course of crossing the well the situation reverses. Due to the
attraction by the inhomogeneity,
the central part of the kink advances faster (than the
outer parts).
Then, we
observe the kink moving outside the well, which, in turn,
results in
vibrations along the direction of motion. These vibrations persist (in the Hamiltonian case)
for a very long time due to the lack of dissipation in the system. 
The right panel shows the same process, but when in the system
there is dissipation $\alpha=0.01$ and forcing $\Gamma=0.0018$.
The parameter describing the depth of the well is, as before,
$\varepsilon=0.1$. The course of the interaction is 
similar to
that in the left panel. The main difference is that the vibration
that the front performs after the impact visibly 
decays and eventually disappears due to
the existence of dissipation in the system.
Interestingly, in both cases, the agreement of the approximate model with the original one is very good even for long times.
 As before, we include animations showing the interaction process both in the case without dissipation and with dissipation.

\begin{figure}
    \centering
    \subfloat{{\includegraphics[height=5cm]{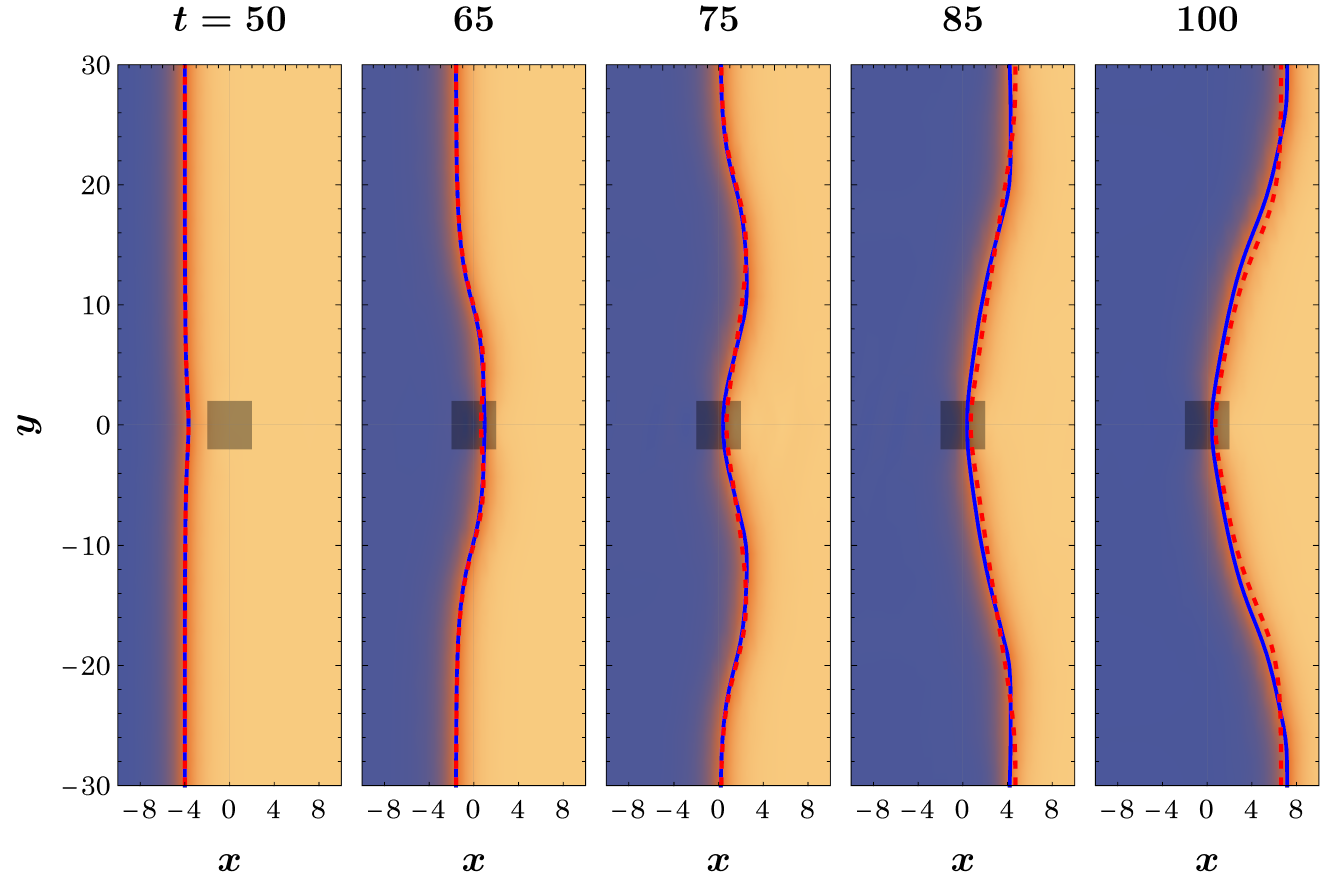}}}
    \qquad
    \subfloat{{\includegraphics[height=5cm]{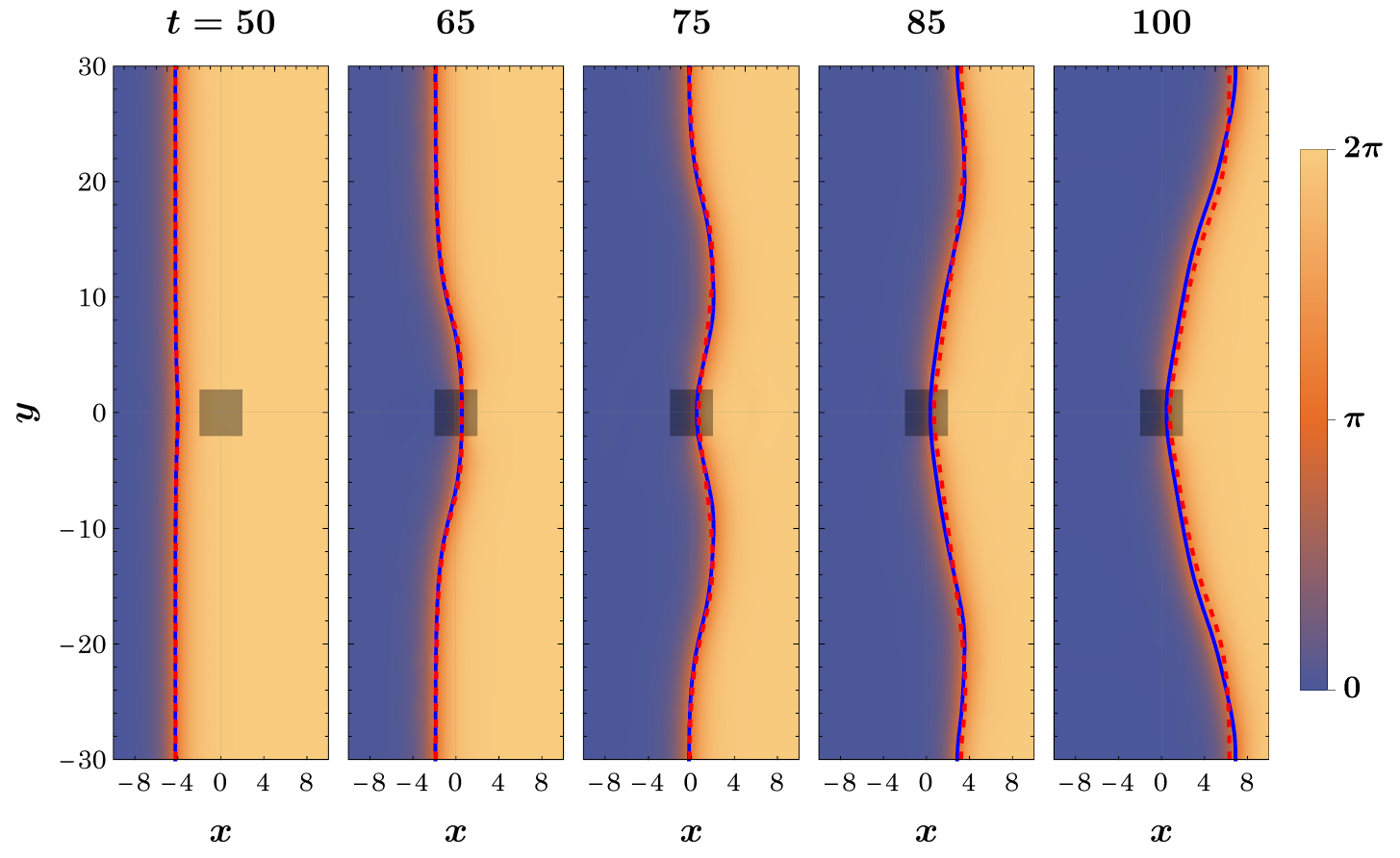}}}
    \caption{Intercepting of the kink front through a potential well. On the left,
the case without dissipation and forcing is shown. The initial velocity of
the front is $u=0.16$. On the right, the dissipation is
$\alpha=0.01$ while the forcing is $\Gamma=0.0013$. In both cases,
the parameter $\varepsilon$ is equal to $0.5$. The animations are available through the links \url{https://tinyurl.com/2p9h289t} for the left one and \url{https://tinyurl.com/y4jxb5sw} for the right one.}
    \label{fig_13}
\end{figure}

The situation becomes even more interesting in the case shown in Figure \ref{fig_13}. In this case, we observe the process of interception of the front by the potential well. The left panel of this figure shows the process of interaction in the absence of forcing and dissipation. The depth of the well here is quite large because it is determined by the parameter $\varepsilon=0.5$. The initial velocity of the kink front is $u=0.16$. As in the previous figure, initially, due to the attraction of heterogeneity, the front in its central part is pulled into the well. Then, there are long-lasting oscillations and deformations of the front, which is the result of interaction with the well. Due to the large value of the parameter $\varepsilon$, the approximation model is 
less accurate for long times, i.e., ones 
exceeding $t=100$. 

The right panel illustrates an identical process, i.e., interception of the front by the well but with both  dissipation ($\alpha=0.01$) and  external forcing ($\Gamma=0.0013$) in the system. As in the left panel, the front is initially, in the middle part, pulled into the well and then repeatedly deformed due to interaction with heterogeneity. The important change, once again, is that the deformations of the front, due to dissipation, become gradually smaller. Ultimately, the kink becomes static, adopting a shape different from a straight line,
due to the presence of (and attraction to) the heterogeneity. The final shape of the kink is a compromise between the forcing of
$\Gamma$ and the tension of the kink filament. 
Tension, as already
mentioned tends to minimize the length of the front while the
forcing pushes the free ends of the front to the right. Due to the large value of the $\varepsilon$ parameter, the approximate model 
has a more limited predictive power for sufficiently
long times, e.g., $t>1000$.
The discrepancies between the two descriptions seem to have a time shift nature.
However, the presence of dissipation leads to a gradual reduction in the kink's distortion, and thus to the differences between the initial model and the approximate one. It turns out that the final configuration is identical in both models.  We have put the course of the impact process in the form of an animations in the additional materials.

\section{Linear stability of the deformed kink front}
In this section we consider the model defined by the equation \eqref{fullmodel} with $\alpha=0$ and $\Gamma=0$
\begin{equation}
\label{s-0}
\partial_t^2 \phi - \partial_x (\mathcal{F}(x,y)\partial_x \phi) - \partial_y^2 \phi + \sin \phi = 0 .
\end{equation}
In the framework of this model we study the 
stability of the deformed static kink solution $\phi_0(x,y)$ satisfying the equation 
\begin{equation}
\label{s-1}
 - \partial_x (\mathcal{F}(x,y)\partial_x \phi_0) - \partial_y^2 \phi_0 + \sin \phi_0 = 0 .
\end{equation}
This study of the spectrum of the kink 
will help us further elucidate the internal vibrational
modes of the kink filament observed and discussed
in the previous
sections. 
 Indeed, whenever kink vibrations
are excited, they can be decomposed on the
basis of oscillations of the point spectrum
of the kink discussed below (while the
extended modes of the continuous spectrum
represent the small amplitude 
radiative wavepackets within the system). 
Moreover, this spectral analysis
can be leveraged to appreciate which configurations
are unstable (e.g., the ones where the
kink is sitting on top of a barrier)
vs. which ones are dynamically stable
(e.g., when the kink is trapped by a well).

\begin{figure}
    \centering
    \includegraphics[width=8cm]{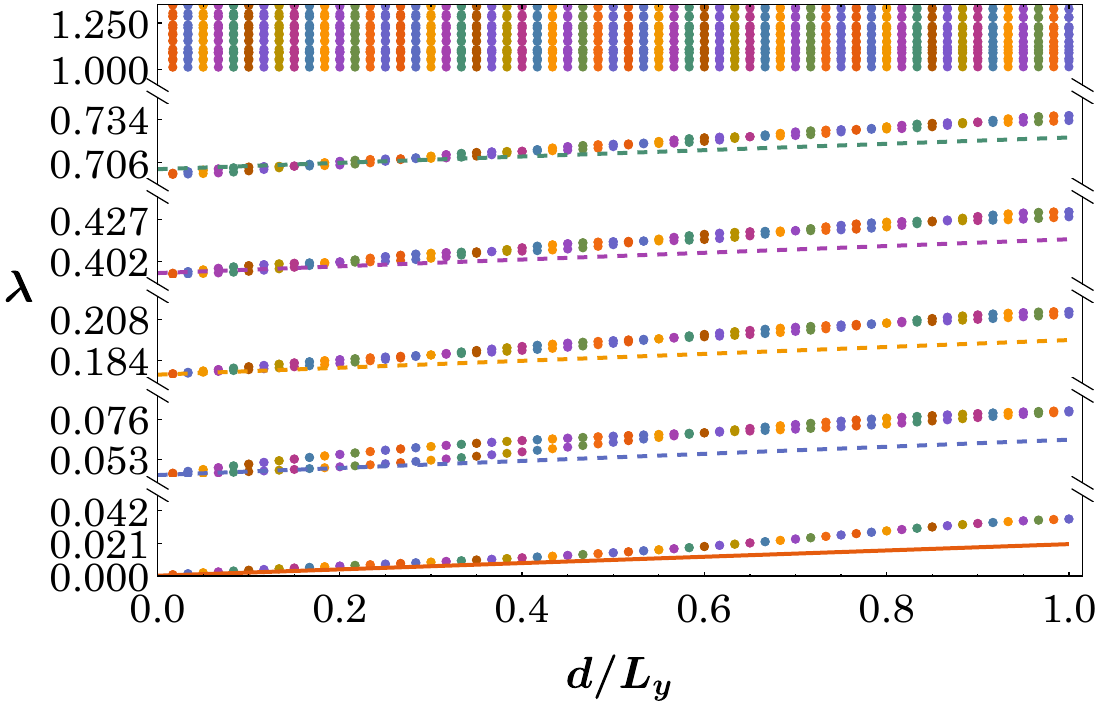}
    \caption{ Squared eigenfrequencies $\lambda=\omega^2$ calculated for the static kink front configuration (trapped by inhomogeneity with the form of a well) depending on the value of $d/L_y$ for $h=4$, $\varepsilon = 0.1$. The red line represent the eigenvalue of the ground state obtained from the approximate model described in Appendix C.}
    \label{fig_14}
\end{figure}

\begin{figure}
    \centering
    \subfloat{{\includegraphics[width=8cm]{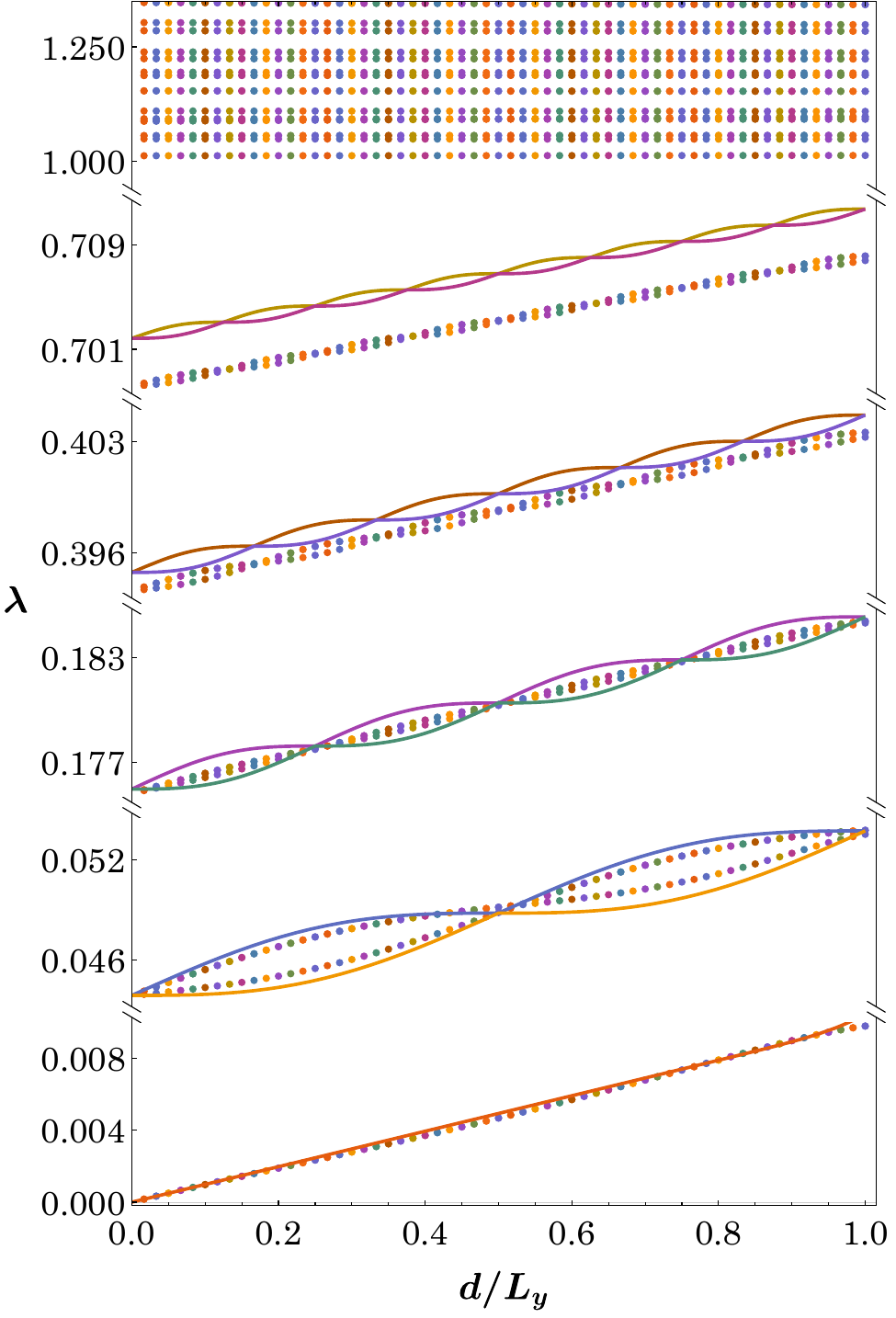}}}
    \qquad
    \subfloat{{\includegraphics[width=8cm]{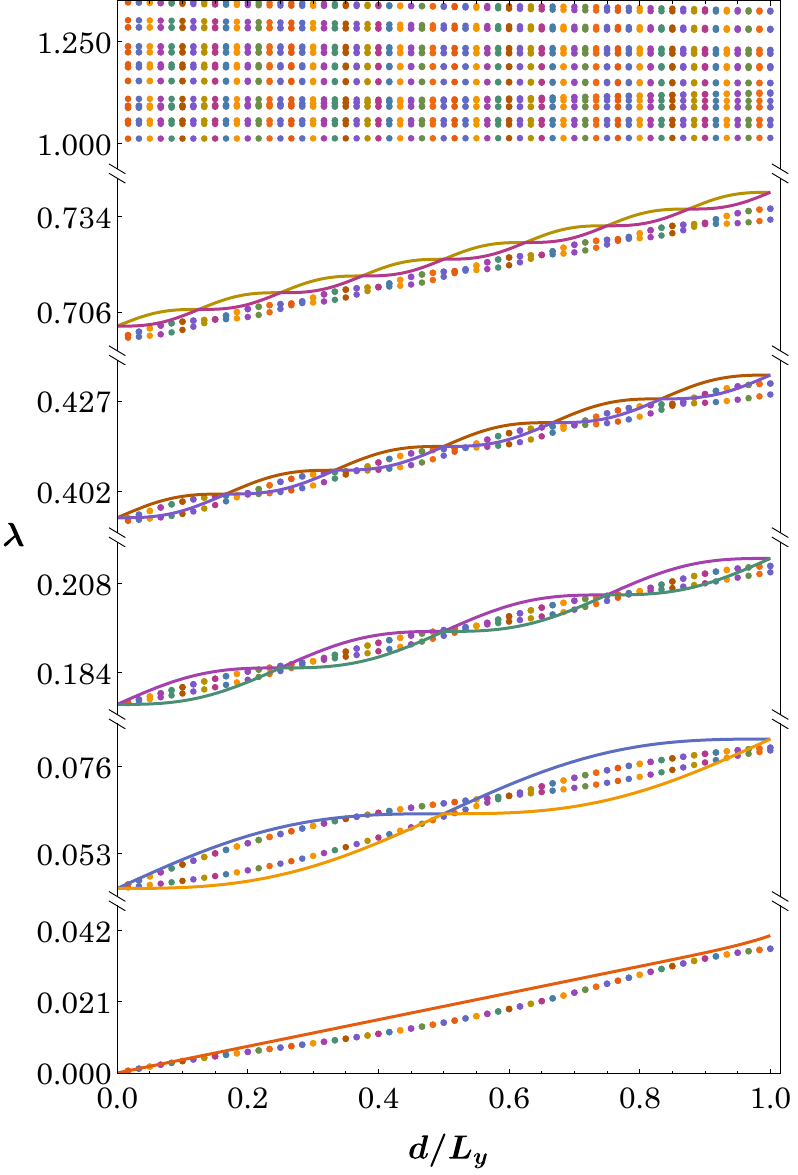}}}
    \caption{Detailed graph of squared eigenfrequencies $\lambda=\omega^2$ calculated for the static kink configuration trapped by a well (without dissipation and bias current) depending on the value of $d/L_y$ for $\varepsilon = 0.01$ on the left and $\varepsilon = 0.1$ on the right. In both cases $\varepsilon = 0.1$, $h=4$. The lines represent the analytical results obtained in Appendix B.}
    \label{fig_15}
\end{figure}

\begin{figure}
    \centering
    \subfloat{{\includegraphics[width=8cm]{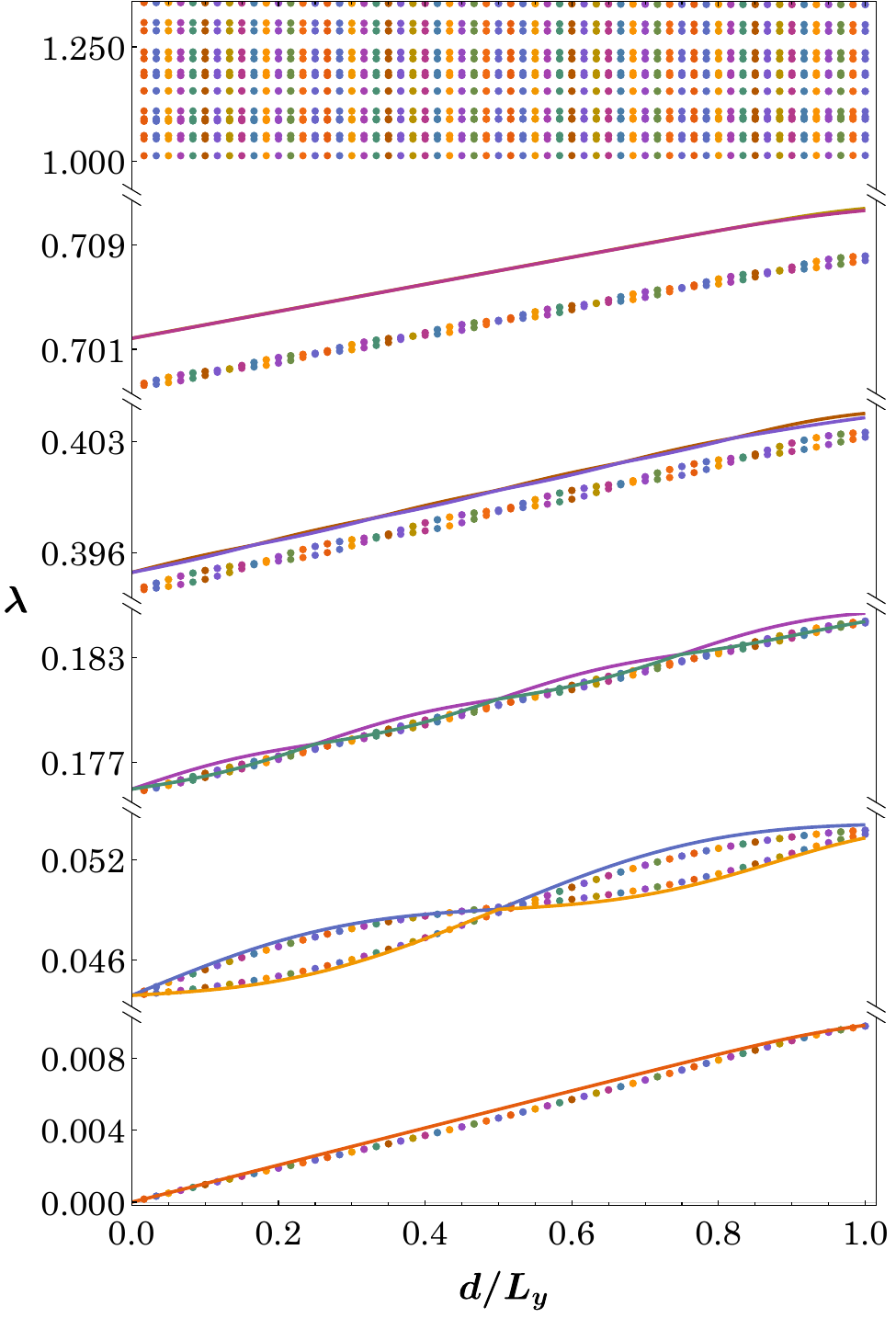}}}
    \qquad
    \subfloat{{\includegraphics[width=8cm]{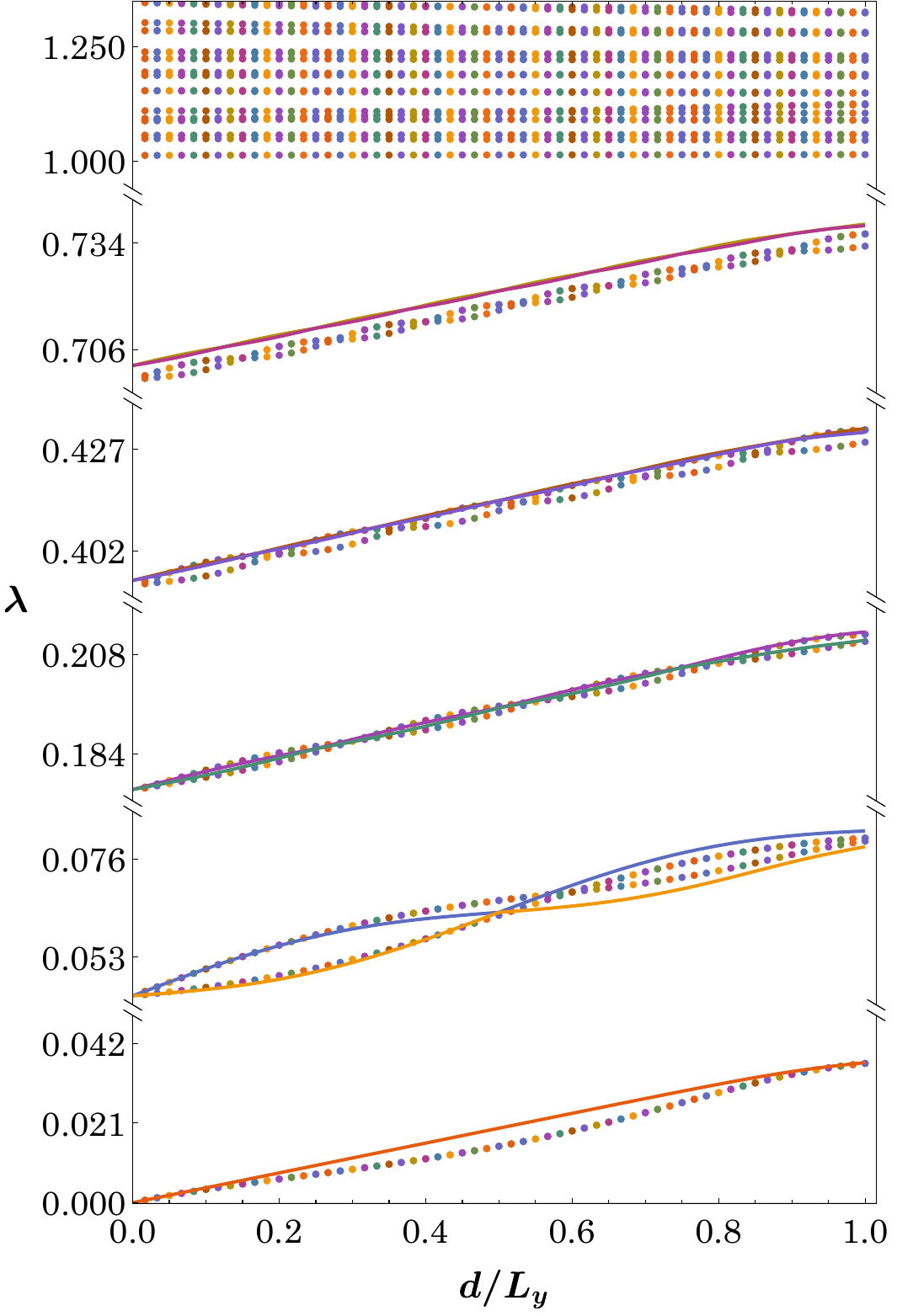}}}
    \caption{Detailed graph of squared eigenfrequencies $\lambda=\omega^2$ calculated for the static kink configuration trapped by a well (without dissipation and bias current) depending on the value of $d/L_y$ for $\varepsilon = 0.01$ on the left and $\varepsilon = 0.1$ on the right. In both cases $h=4$. The lines represent the results of Appendix B with integrals determined numerically for \eqref{chi+}.}
    \label{fig_16}
\end{figure}

 We introduce into equation \eqref{s-1} a configuration $\phi$ consisting of the solution $\phi_0$ and a small correction $\psi$ i.e. $\phi(t,x,y) = \phi_0(x,y) + \psi(t,x,y)$. Moreover, we assume a 
 separation of variables of the perturbation in
 terms of its time and space
 dependence as: $\psi(t,x,y) = e^{i \omega t} v(x,y).$
In a linear approximation with respect to the correction, we obtain 
\begin{equation}
\label{s-2}   - \partial_x \left( {\cal
F}(x,y)\, \partial_x v(x,y)\right) - \partial_y^2 v(x,y)+ ( \cos
\phi_0 ) \, v(x,y) = \lambda v(x,y) \, ,
\end{equation}
where
$\lambda=\omega^2 $.
We can briefly write this equation using the $\hat{{\cal L}}$ operator, which includes a dependence on the analytical form of inhomogeneity 
\begin{equation}
\label{s-3}  \hat{{\cal L}} v + \cos \phi_0  \, v = \lambda v \, .
\end{equation}
{The above equation has the character of a stationary Schr{\"o}dinger equation with a potential defined by the cosine of the straight kink front configuration $\phi_0$ intercepted by the inhomogeneity. An important feature of this configuration, is that, similarly to the $\hat{{\cal L}}$ operator, it depends in part on the form of the inhomogeneity. In the region of heterogeneity, it has an analytical form different from that of the free kink (denoted $\phi_K$ in this work). This modification of the analytical form of the field is a consequence of the interaction of the kink with the inhomogeneity.}
Based on this equation, an analysis of the excitation spectrum of the static kink captured by the inhomogeneity was carried out. The results can be found in Figures \ref{fig_14} and \ref{fig_15}. Figure \ref{fig_14} shows with dotted lines the dependence of the squares of the frequency on the parameter $d$ describing the transverse size of the inhomogeneity. In the figure, the values of the parameters are assumed to be $h=4$ and $\varepsilon=0.1$ (in addition, the size of the system is determined by the values $L_x=30$ and $L_y=30$). The 
lowest energy state in this diagram is the non-degenerate state and it corresponds to the zero mode of the sine-Gordon model without inhomogeneities. 
In addition, the figure includes the fit obtained for this state using an energy landscape study of the one-degree-of-freedom effective model (see Appendix C for a description of this approach). Note that up to a value of about $0.4$ of the $d/L_y$ ratio, this simple model captures the course of the numerical dependence well. Above that
lie the excited states. At the scale adopted in the figure, it is almost imperceptible that each line actually consists of two lines running side by side. Note that the increase in the value of $\lambda$ for the excited states is similar to the increase in the value for the ground state, as indicated by the dashed lines parallel to the red line obtained for the ground state based on the approximate model (Appendix C). Above a value of
unity, we encounter the continuous spectrum of the problem.
A more detailed plot is shown in Figure \ref{fig_15}. In this figure, it is much clearer that the discrete states (except for the ground state) are described by double lines. The spectrum is shown here for two values of $\varepsilon$. The results for $\varepsilon=0.01$ are shown in the left figure, while those for $\varepsilon=0.1$ (as in the previous figure) are shown 
in the right one. The other parameters are identical. The figure also shows the predictions obtained from the degenerate perturbation theory
analysis presented in Appendix B. It can be seen that the analytical result reflects very well the course of the line representing the ground state (especially for small values of $\varepsilon$). The course of the lower excited states is also quite well reproduced. For higher excited states, the similarity of the numerical result to the analytical one is qualitative.
{In order to obtain an analytical estimate of the spectrum of linear excitations of the configuration under study, we need, among other things, the form of deformation $\chi$ of the kink front with respect to the free kink. The method of obtaining the $\chi$-function is presented in Appendix A.}
To check the analytical formulas obtained by approximating, for example, the function $\chi$ in a piecewise form, we performed numerical calculations of the integrals contained in Appendix B based on the approximation \eqref{chi+}. The results are presented in figure \ref{fig_16}, which was made for the same parameters as figure \ref{fig_15}. As can be seen, the improvement in compatibility occurs for the lowest eigenvalues. Specifically, it takes place for the parameter $d/L_y$ close to one. For higher eigenvalues, the situation does not significantly 
improve. It turns out that for higher excited states the analytical formula overestimates the separation of states (corresponding to the degenerate states of the zero approximation), while the result obtained with the fit \eqref{chi+} underestimates this gap.
In any event, given the relatively small size of the
discrepancy, we do not dwell on this further.

\begin{figure}
    \centering
    \subfloat{{\includegraphics[width=8cm]{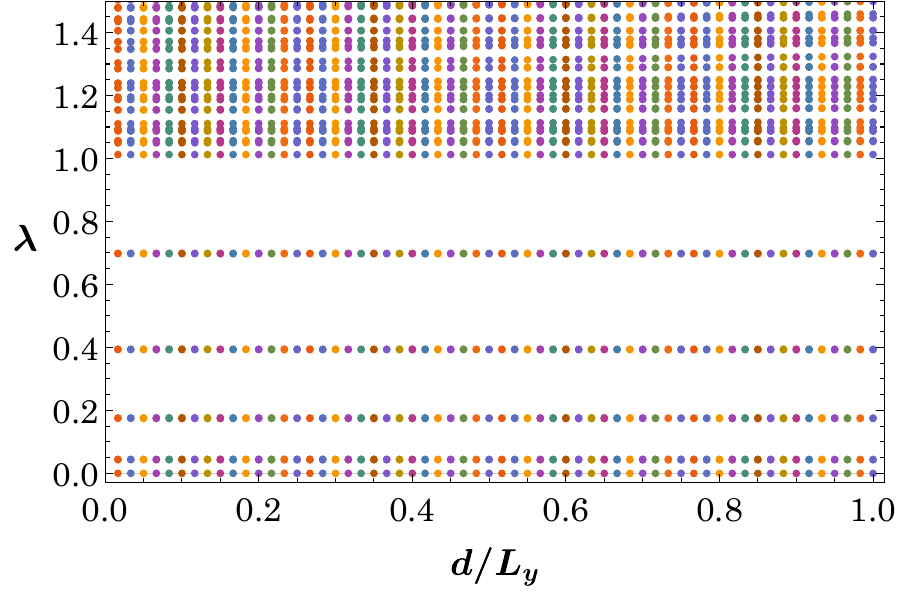}}}
    \qquad
    \subfloat{{\includegraphics[width=8.5cm]{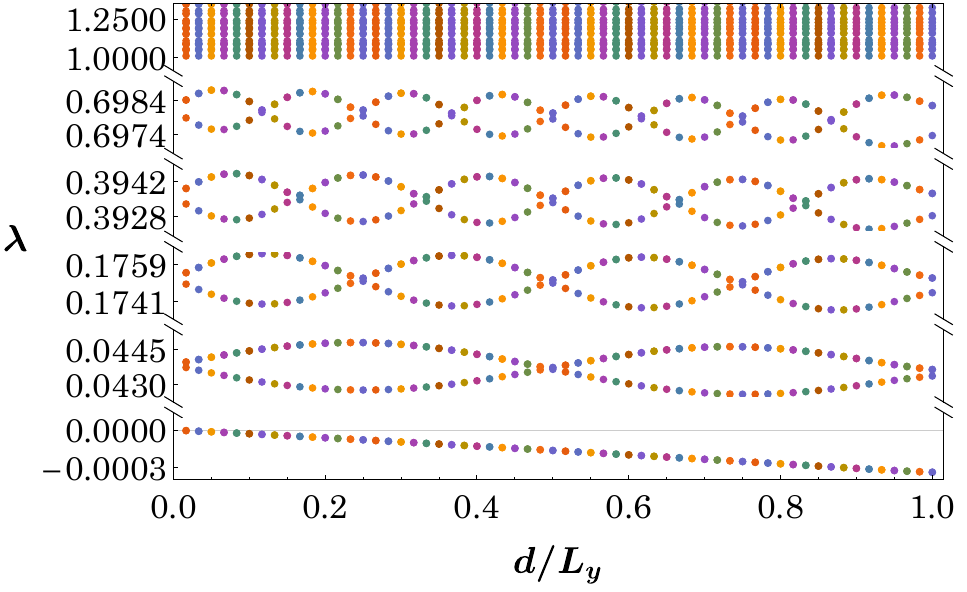}}}
    \caption{Squared eigenfrequencies $\lambda=\omega^2$ calculated for the static configuration on top of inhomogeneity (without dissipation and bias current) depending on the value of $d/L_y$ for $h=4$ and $\varepsilon = 0.1$. The left and right figures differ only in scale.}
    \label{fig_17}
\end{figure}

On the other hand, the results for
a barrier-like inhomogeneity of the form of Figure \ref{fig_08} are presented in Figure \ref{fig_17}. The parameters on the left and right panels of this figure are identical and are $h=4$, $\varepsilon=0.1$, $L_x=30$, $L_y=30$. The figures differ only in scale. 
This time, the configuration of the kink lying
on top of the destabilizing barrier is found to indeed be unstable, which is manifested by the occurrence of a mode with a negative value of $\lambda$ (i.e., an 
imaginary eigenfrequency). This mode corresponds to the translational mode, reflecting in this case the nature
of the effective potential (i.e., a barrier creating
an effective saddle point). Such a value is a manifestation of the kink drifting away  from inhomogeneity. The other modes are quite similar in nature to the excited modes in the case of potential well, which has its origin in the adopted periodic boundary conditions.

\section{Conclusions and Future Challenges}
In the current article we studied the behavior of the kink front in the perturbed $2+1$ dimensional sine-Gordon model. The particular type of perturbation is motivated by the study of the dynamics of gauge-invariant phase difference in one- and quasi-one-dimensional curved Josephson junction \cite{Dobrowolski2012,Gatlik2021,Dobrowolski2009}. We also obtained an effective $1+1$ dimensional model describing the evolution of the kink front based on the non-conservative Lagrangian method \cite{Galley2013,Kevrekidis2014}.
First we tested the usefulness of the approximate model. 
More concretely, we examined the behavior of the kink 
starting from the case
when there are no inhomogeneities in the system.
The agreement between the results of the original  and the effective model turned out to be very satisfactory.
Subsequently, we explored 
the movement of the front in a slightly more complex situation. Namely, we examined inhomogeneities of shape independent of the variable transverse to the direction of movement of the front, i.e., the $y$ variable. The results obtained here are in full analogy with the $1+1$ dimensional model studied earlier \cite{Gatlik2023arXiv}. These studies can be directly applied to the description of quasi-one-dimensional Josephson junctions. 

The most interesting results were obtained
for studies of the behavior of the front in the presence of inhomogeneities with shape genuinely dependent on both spatial variables. This case shows the remarkable richness of the dynamical behaviors of the kink front interacting with heterogeneity.
We studied two types of inhomogeneities. One was in the form of a barrier, while the other was in the form of a well. 

Of particular interest is the process of creating a static final state in the case with dissipation and forcing. We deal with the formation of such a state when a front with too low a velocity is stopped (by a sequence of oscillations) before the peak, and when a front that is too slow is trapped by a well.

We have analyzed the competing factors that 
contribute to the formation of the resulting stationary
states and have shown that our reduced $1+1$-dimensional description can
capture the resulting state very accurately.
It is worth noting that the approximate description in each of the studied cases is also accurate
for long time evolutions for small values of the parameter describing the strength of heterogeneity. 
While deviations might occur in some cases for 
very long times (in Hamiltonian perturbations)
or for sufficiently large perturbations in dissipative
cases, generally, we found that the reduced kink
filament model was very accurate in capturing the 
relevant dynamics.

Finally, we also studied the stability of a straight kink front captured by a single inhomogeneity of the form of a potential well. In this case, the zero mode of the sine-Gordon model without inhomogeneities turns into an oscillating mode in the model with inhomogeneities. 
Indeed, the breaking of translational invariance
leads to either an effective attractive well
or a repulsive barrier (see
also the analytical justification in Appendix C) 
manifested in the presence of an internal oscillation or
a saddle-like departure from the inhomogeneous region. In addition, the periodic boundary conditions we have adopted result in a number of additional discrete modes appearing in the system in addition to the ground state and the continuous spectrum.  
These are effectively the linear modes associated
with the quantized wavenumbers due to the transverse
domain size. In the absence of a genuinely 2d heterogeneity,
this picture can be made precise with the
respective eigenmodes being $k_y=2 n \pi/L_y$.
In the presence of genuinely $2$d heterogeneities,
the picture is still qualitatively valid, but the modes
are locally deformed and then a degenerate perturbation
theory analysis is warranted, as shown in Appendix B, where
we have provided such an analytical description of 
the mode structure
This description matches quite well with the numerical results - especially for the lower states of the spectrum under study.  

Naturally, there are numerous extensions of
the present work that are worth exploring in the future.
More specifically, in the present setting we have
focused on inhomogeneities impacted upon by 
rectilinear kink structures, while numerous earlier
works \cite{Christiansen1981,Geicke1983,Caputo2013,Kevrekidis2018} have considered the interesting additional
effects of curvature in the two-dimensional
setting. In light of the latter, it would be interesting
to examine heterogeneities in such radial cases.
Furthermore, in the sine-Gordon case, the absence
of an internal mode in the quasi-one-dimensional
setting may have a significant bearing of a phenomenology
and the possibility of energy transfer type
effects that occur, e.g., in the $\phi^4$ model \cite{goodman}. It would, thus, be particularly
relevant to explore how the relevant phenomenology
generalizes (or is modified) in the latter setting.
Finally, while two-dimensional settings have yet
to be exhausted (including about the potential
of radial long-lived breathing-like states),
it would naturally also of interest to explore
similar phenomena in the three-dimensional
setting. Such studies are presently under
consideration and will be reported in future publications.

\section{Appendix A}
\subsection{Peak-shaped inhomogeneity}
\begin{figure}
    \centering
    \subfloat{{\includegraphics[width=6cm]{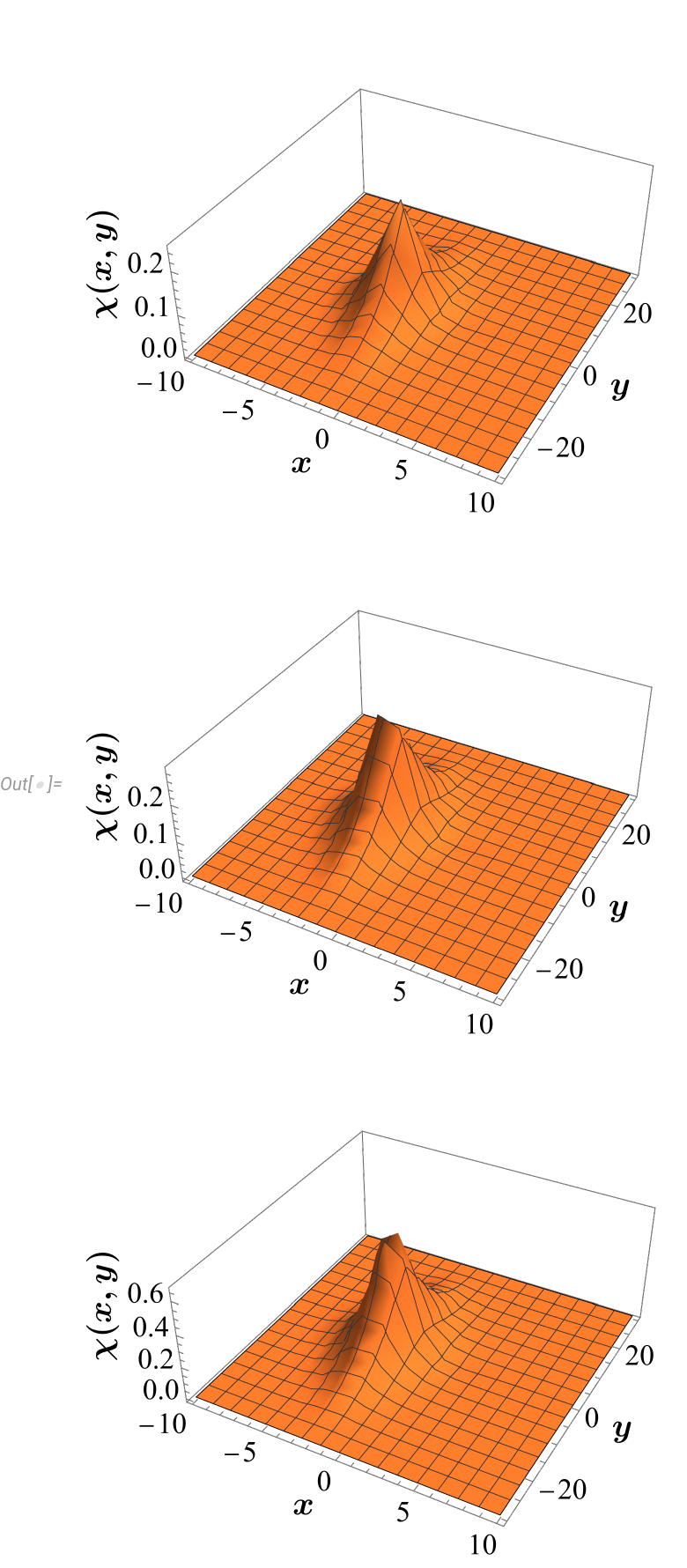}}}
    \qquad
    \subfloat{{\includegraphics[width=5.2cm]{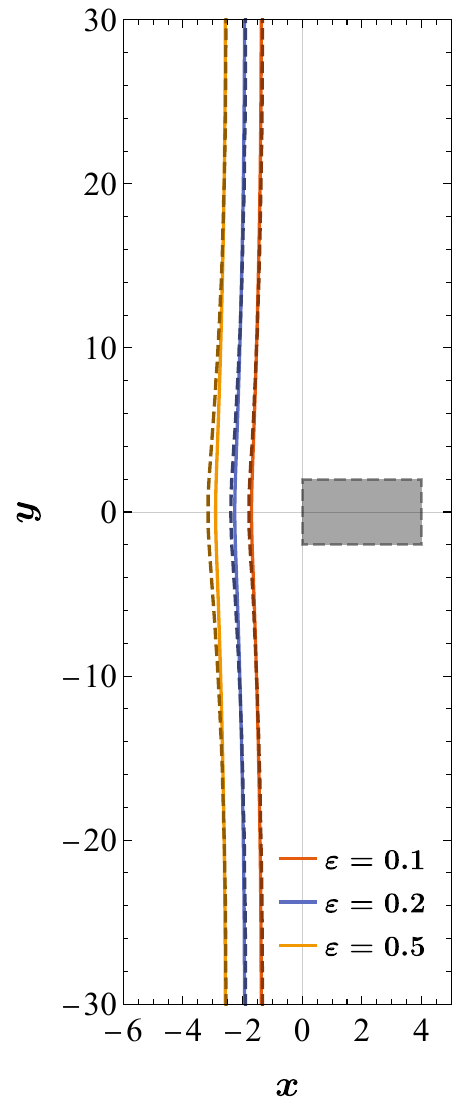}}}
    \caption{The left panel shows the function $\chi(x,y)$ for, starting from top, $\varepsilon = 0.1,0.2$ and $0.5$. The right figure compares the shape of the kink front 
    obtained using the equation \eqref{x-2}
    with the exact results represented by the dashed black line, for the same values of $\varepsilon$.
The forcing is assumed here to be $\Gamma=0.001$. }
    \label{fig_18}
\end{figure}

We will consider the case of a kink front stopped by 
the inhomogeneity (in the form of a barrier; see Fig.~\ref{fig_08}) in the presence of forcing and dissipation.
The static configuration in this case is the solution of the following equation
\begin{equation}
\label{x-1}
 - \partial_x (\mathcal{F}(x,y)\partial_x \phi_0) - \partial_y^2 \phi_0 + \sin \phi_0 = -\Gamma .
\end{equation}
To begin with, we will show that the solution can be represented
(for small perturbations) as the sum of a kink profile $\phi_K = 4
\arctan e^{x-X_0(y)}$ and a  correction that depends
only on the shape of the inhomogeneity and the external forcing i.e.
$\phi_0(x,y) = \phi_K(x-X_0)+\chi(x,y).$ The equation satisfied by the correction $\chi$,
to leading order,is of the form
\begin{equation}
\label{x-2}
    -\partial_x \left( {\cal F}(x,y)\partial_x \chi \right) - \partial_y^2 \chi + ( \cos \phi_K(x-X_0) ) \chi = \varepsilon\partial_x\left( g(x,y)\partial \phi_K(x-X_0)\right) - \Gamma.
\end{equation}
The results of simulations performed on the ground of
approximation \eqref{x-2} and the field model \eqref{x-1} are demonstrated in Figure
\ref{fig_18}.
This figure shows in the left panel the $\chi$ profiles obtained
for different values of the $\varepsilon$ parameter. Starting from
the top, we have $\varepsilon=0.1$, $\varepsilon=0.2$ and
$\varepsilon=0.5$. In all cases, $\Gamma=0.001$. The right panel
shows the profile of the static kink front in the same cases. This
panel, on the one hand, shows the static kink front obtained
from equation  \eqref{x-1} (black dashed line), and on the other
hand, the fronts obtained from the solutions of equation 
\eqref{x-2} for different values of the parameter $\varepsilon$.
The red line corresponds to $\varepsilon=0.1$, the blue line
corresponds to $\varepsilon=0.2$, while the yellow line
corresponds to $\varepsilon=0.5$. These fronts were determined for
the $\phi_K+\chi$ configuration.
 The deformation of the kink center is due to the fact that it is supported by the inhomogeneity in the central part, and on the other hand, at the edges it is stretched by the existing constant forcing. Of course, due to the tension of the kink front, stretching cannot take place unrestrictedly because this would lead to an excessive increase in the total energy stored in the kink configuration. Let us notice that in all cases, qualitatively the shape of the static kink front is correctly reproduced. On the other hand, in the case of $\varepsilon=0.5$ we observe some quantitative deviations in the central part.

\begin{figure}
    \centering
    {\includegraphics[width=8cm]{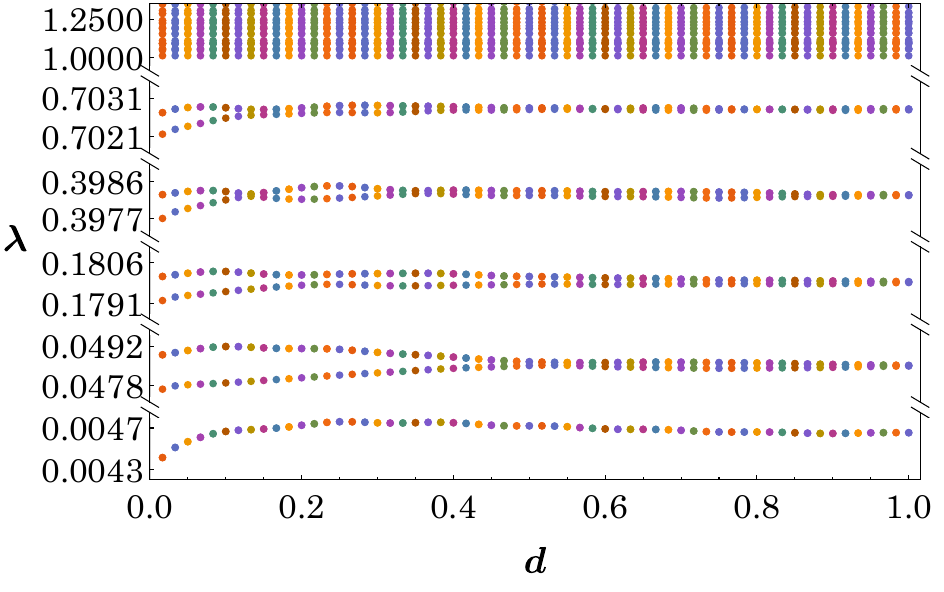}}
    \caption{Graph of squared eigenfrequencies $\omega^2$ calculated for the static kink configuration stopped by a barrier (with dissipation and bias current) depending on the value of $d/L_y$ for $\varepsilon = 0.5$ and $h=4$.}
    \label{fig_19}
\end{figure}

We also test the stability of the above described solution is based on the equation 
which looks identical to the equation \eqref{s-2}, however, the main difference is the relationship of the eigenvalue $\lambda$ to the frequency. In the case considered in this section $\lambda=\omega (\omega-i \alpha) $. 
Figure \ref{fig_19} shows the dependence of the square of the frequency $\omega$ on the parameter $d/L_y$.
It can be seen that the excitation spectrum determined for the configuration shown in Figure \ref{fig_18}, consists of a ground state, excited states and a continuous spectrum. The form of this spectrum is to a significant degree similar to the excitation spectrum of the kink front trapped by the potential well, and shown in Figures \ref{fig_15}, \ref{fig_16}. The main difference from the previous diagrams is that the discrete excited states show less periodicity as in the previous figures.

\subsection{Heterogeneity with a form of well}
\begin{figure}
    \centering
    \includegraphics[width=8cm]{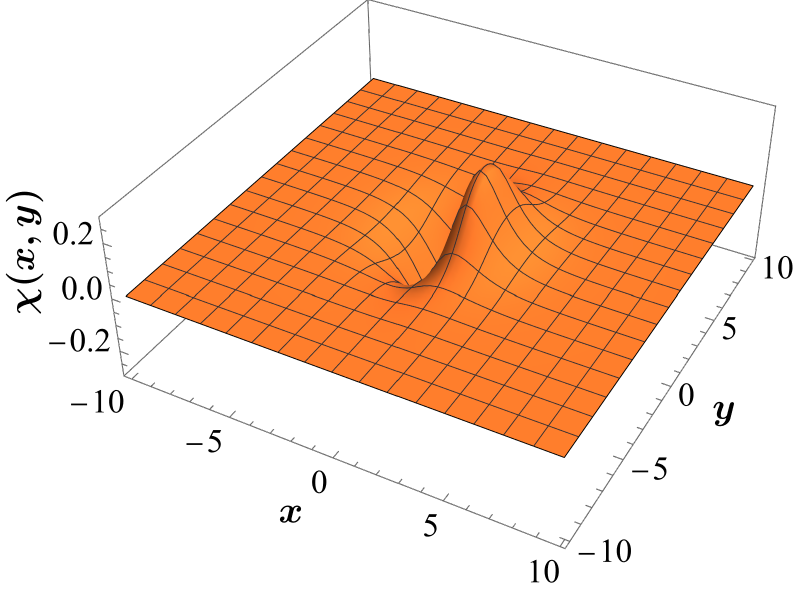}
    \caption{The shape of the function $\chi(x,y)$, for a static kink front trapped by a well-shaped inhomogeneity. The parameters in the figure are as follows: $d=4$, $h=4$, $\varepsilon=0.1$. }
    \label{fig_20}
\end{figure}

\begin{figure}
    \centering
    \subfloat{{\includegraphics[width=8cm]{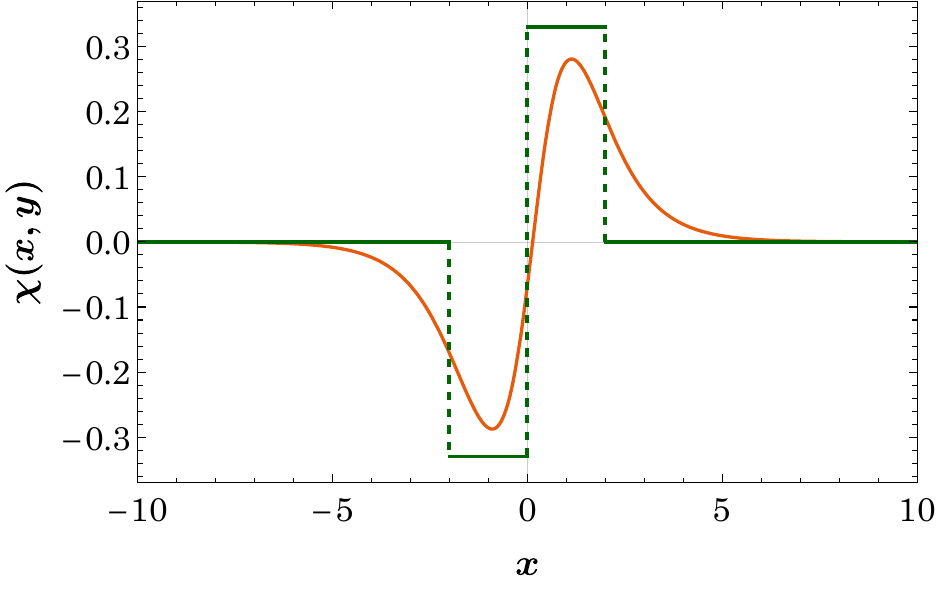}}}
    \qquad
    \subfloat{{\includegraphics[width=8cm]{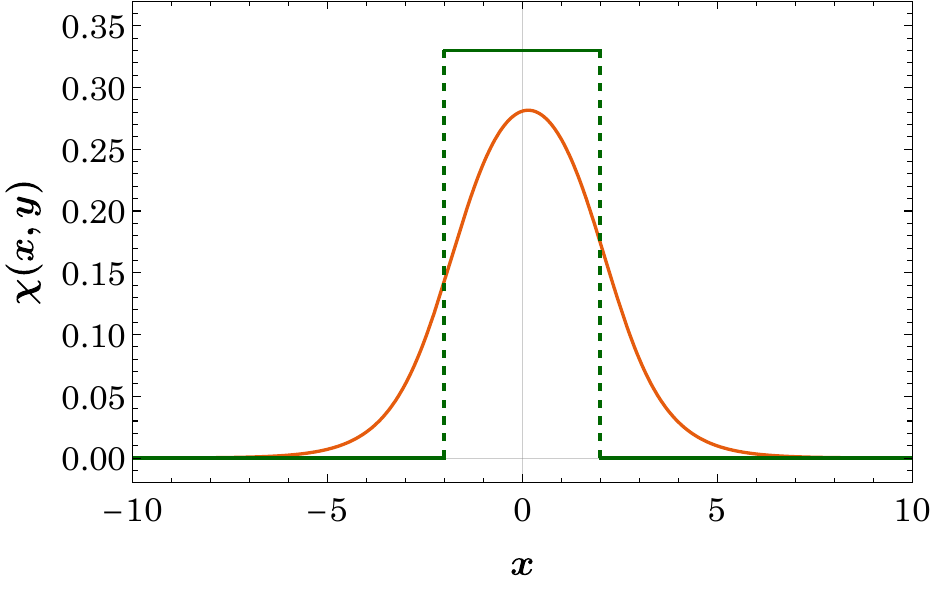}}}
    \caption{Cross sections with fitting for $\chi(x,y)$. The value of $\chi_0=0.67$ was determined by fit. Here $h = 4$, $d=4$ and $\varepsilon = 0.1$.}
    \label{fig_21}
\end{figure}

\begin{figure}
    \centering
    \subfloat{{\includegraphics[width=8cm]{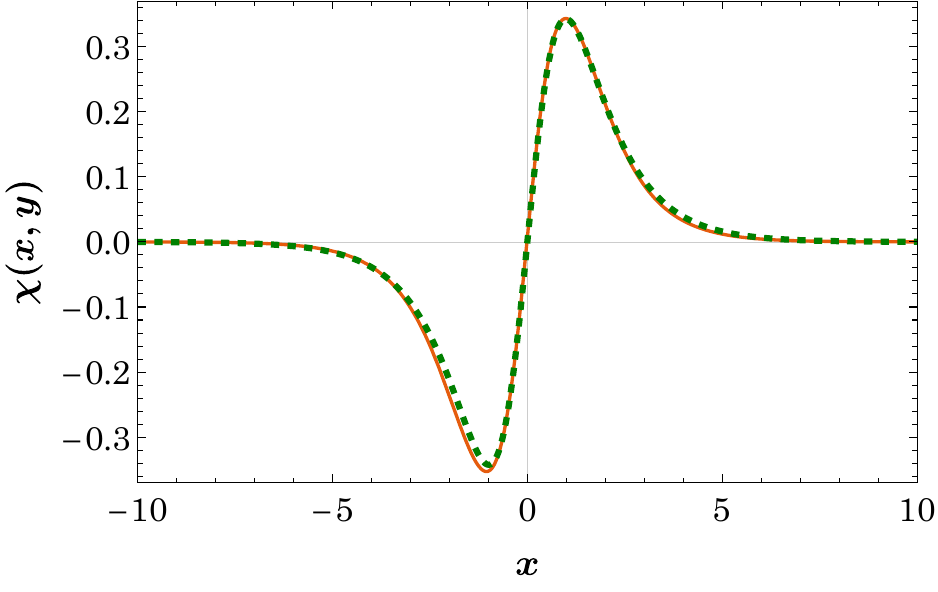}}}
    \qquad
    \subfloat{{\includegraphics[width=8cm]{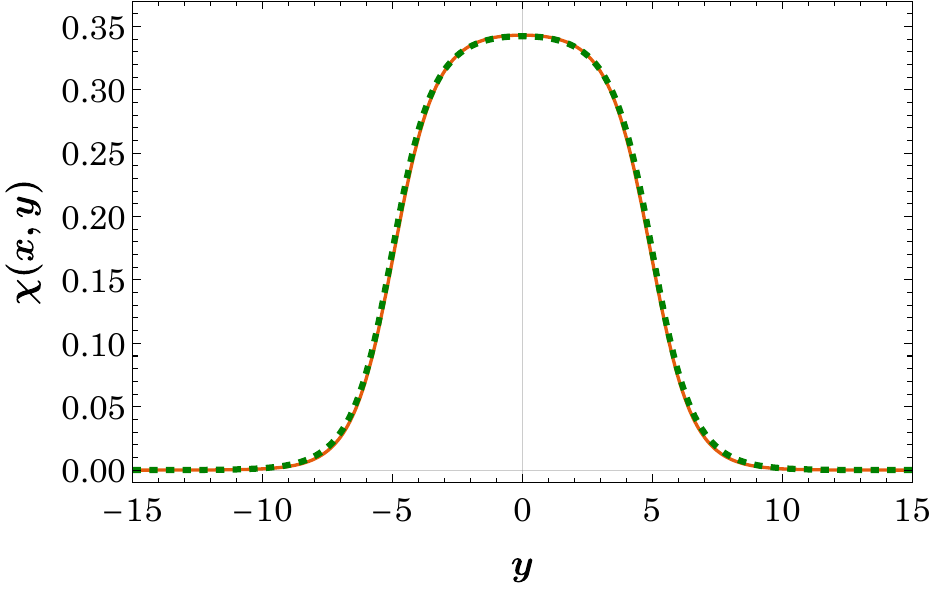}}}
    \caption{Cross sections with fitting for $\chi(x,y)$. The dashed green line represents a fit function of the form $\chi(x,y) = \chi_0 \tanh(a x) \sech(a x) \left(4\arctan e^{y+\frac{d}{2}}-4\arctan e^{y-\frac{d}{2}}\right)$. Here $h = 4$, $d=10$ and $\varepsilon = 0.1$.}
    \label{fig_22}
\end{figure}

In this section, we describe the change in the profile of the static kink that results from the existence of an inhomogeneity in the form of a well. We assume that the well is centrally located and has dimensions defined by the parameters $h$ and $d$ i.e. ${\cal F}(x,y)=1+ \varepsilon g(x,y) = 1 - \varepsilon p(x) q(y)$ and
\begin{equation}
    \label{p}
    p(x) = \frac{1}{2}\left[\tanh\left(x+\frac{h}{2}\right)-\tanh\left(x-\frac{h}{2}\right) \right] \approx
    \begin{cases}
        1, & x \in [-\frac{h}{2},+\frac{h}{2}] \\
        0, & x \notin [-\frac{h}{2},+\frac{h}{2}] 
    \end{cases} ,
\end{equation}
\begin{equation}
    \label{q}
    q(y) = \frac{1}{2} \left[\tanh \left(y+\frac{d}{2}\right)-\tanh\left(y-\frac{d}{2}\right)\right] \approx
    \begin{cases}
        1, & y \in [-\frac{d}{2},+\frac{d}{2}] \\
        0, & y \notin [-\frac{d}{2},+\frac{d}{2}] 
    \end{cases} .
\end{equation}
The approximate form used to calculate some integrals when determining the analytical form of the eigenvalues (see Appendix B) is also given 
in the above expression.
An example profile obtained from equation \eqref{x-2} in the absence of bias current ($\Gamma=0$) is shown in Figure~\ref{fig_20}. The shape of the $\chi$ function, although shown for specific parameter values (i.e., $h = 4$, $d=4$ and $\varepsilon = 0.1$), is characteristic over a wide range of parameters.
The profile  shown in Figures~\ref{fig_21} is an even function in the $y$ variable and an odd function in the $x$ variable.
The panels of figure~\ref{fig_21} also include a simple fit in the form of the step function. The parameter $\chi_0$ was chosen so that the areas under the curves $\alpha = \alpha(x)$, $\beta=\beta(y)$ and the fit were identical.
In the next section (appendix B), we use this form of the $\chi$ function to approximate the eigenvalues when studying the stability of a  static configuration trapped by a well-like inhomogeneity
\begin{equation}
\label{chi}
    \chi(x,y) = \chi_0 \,\, \alpha(x) \beta(y),
\end{equation}
\begin{equation}
    \label{alphabeta}
    \alpha(x) \approx
    \begin{cases}
        -1, & x \in [-\frac{h}{2},0) \\
        +1,  & x \in [0,+\frac{h}{2}] \\
        0, & \text{otherwise} 
    \end{cases},\;\;\;\;
        \beta(y) \approx
    \begin{cases}
        +1,  & x \in [\frac{d}{2},+\frac{d}{2}] \\
        0, & \text{otherwise} 
    \end{cases} \,\,\, .
\end{equation}
In order to validate the analytical expressions \eqref{xx-22} and \eqref{xx-35} for the eigenvalues of the linear excitation operator, we also determined a much better fit for the $\chi$ function. We looked for the fit in the form:
\begin{equation}
\label{chi+}
    \chi(x,y) = \chi_0 \tanh(a x) \sech(a x) \left(4\arctan e^{y+\frac{d}{2}}-4\arctan e^{y-\frac{d}{2}}\right).
\end{equation}
The shape of the fit was compared with the numerical result. The example figure \ref{fig_22} shows a very good convergence between the fit (dashed line) and the numerical result (solid line). The figure was made for parameters equal to $\chi_0= 0.67$, $a=0.85$, $h=4$, respectively. The fit form described by equation \eqref{chi+} was also used to determine the numerical value of the integrals in Appendix B. The results obtained on this basis are presented in Figure \ref{fig_16}. As can be seen for lower eigenvalues, we observe improved agreement with numerical results. Moreover, the improvement is evident for values of $d/L_y$ close to one.

\section{Appendix B - Kink stability in the potential well}
In this section, we will present analytical results on the spectrum of linear excitations of a deformed kink bounded by an inhomogeneity in the form of a potential well. We start with the equation \eqref{s-3} 
\begin{equation}
\label{xx-3}  \hat{{\cal L}} v + \cos \phi_0  \, v = \lambda v \, .
\end{equation}
Since we plan to use perturbation calculus in the parameter $\varepsilon$ determining the magnitude of the inhomogeneity, we separate the operator $\hat{{\cal L}}$ into a part $\hat{{\cal L}_0}$ that does not depend on the perturbation parameter and a part $\hat{W}$ preceded by this parameter. The relationships between operators and the other quantities used in this section are summarized below
\begin{equation}
\label{xx-4}  \hat{{\cal L}} v = \hat{{\cal L}_0} v + \varepsilon \hat{W} v , \,\,\,\,\,\,
\hat{{\cal L}_0}v = -
\partial_x^2 v -
\partial_y^2 v , \,\,\,\,\,\,
\hat{W} v = -  \partial_x \left( g(x,y) \,
\partial_x v \right) , \,\,\,\,\,\, {\cal F}(x,y) = 1 + \varepsilon g(x,y) .
\end{equation}
According to the results presented in appendix A, we can separate the static kink configuration in the presence of inhomogeneity into static free kink $\phi_K$ and deformation associated with the existence of inhomogeneity $\chi$
\begin{equation}
\label{xx-5}   \phi_0(x,y)  = \phi_K(x) + \chi(x,y) .
\end{equation}
Next, we expand the quantities appearing in formula \eqref{xx-3} with respect to the parameter $\varepsilon$
\begin{eqnarray}
\label{xx-6}
  & v   = v^{(0)} +
\varepsilon v^{(1)} + \varepsilon^2 v^{(2)} +
... &  \\
& \lambda = \lambda^{(0)} + \varepsilon \lambda^{(1)} +
\varepsilon^2 \lambda^{(2)} + ... , & \\ \nonumber
& \chi   =
\chi^{(0)} + \varepsilon \chi^{(1)} + \varepsilon^2 \chi^{(2)} +
... .    &  \nonumber
\end{eqnarray}
The function $\chi$ is defined in such a way that it does not appear in the zero order, i.e. $\chi^{(0)}=0$.
In addition, since in the system under consideration we assume periodic boundary conditions in the direction of the $y$ variable we also take $v(x,-\frac{1}{2}L_y) = v(x,+\frac{1}{2} L_y).$ Moreover, it is assumed that the inhomogeneity disappears at the edges of the system (in the direction of the variable $x$), i.e.  $g(x,y)
\rightarrow 0$ for $x \rightarrow \pm \frac{1}{2}L_x . $ Note also that, like $\partial_x \phi (\pm \frac{1}{2} L_x, y)$,  also $\partial_x v(\pm \frac{1}{2} L_x, y)$ disappears at  the $x$ boundaries of the area under consideration.

\subsection{The lowest order of expansion}
In the lowest order, we get the equation
\begin{equation}
\label{xx-7}  \hat{{\cal L}_0} v^{(0)} + \cos \phi_K  \, v^{(0)} =
\lambda^{(0)} v^{(0)} \, ,
\end{equation}
where $\phi_K(x)=4 \arctan (e^x)$ describes the kink front located at $x=0$ and stretched along the $y$-axis. For the function $\phi_K(x)$, the equation can be separated into two equations. One depending on the $x$ variable and the other on $y$. Using periodicity in the $y$ variable, we obtain a series of eigenvalues and eigenfunctions. The ground state in this approximation corresponds to zero eigenvalue
\begin{eqnarray}
\label{xx-8}
& \lambda^{(0)}_{0} = 0 , \,\,\,\,\, v^{(0)}_{0}(x,y)  = A_0 \, \mathrm{sech}(x) , \,\,\,\,\, A_0 = \frac{1}{\sqrt{2 L_y \tanh \frac{L_x}{2}}} .
\end{eqnarray}
The subsequent eigenstates correspond to non-zero eigenvalues

\begin{equation}
    \label{xx-9}
    \lambda^{(0)}_{n \pm} = \left( \frac{2 \pi}{L_y}\right)^2 n^2 , \,\,\,\,\,\,\,
    \begin{cases}
      v^{(0)}_{n+}(x,y)  = A \, \mathrm{sech}(x) \cos(2 \pi n
\frac{y}{L_y})  \\
 v^{(0)}_{n-}(x,y)  = A \,
\mathrm{sech}(x) \sin(2 \pi n \frac{y}{L_y}) 
    \end{cases},  \,\,\,\,\,\,
    A = \frac{1}{\sqrt{L_y \tanh \frac{L_x}{2}}} .
\end{equation}
In the lowest order of the perturbation calculus, all non-zero eigenvalues are degenerate twice.
The normalization coefficients $A$ and $A_0$ were chosen so that the eigenfunctions were normalized to one in the sense of the product 
defined as the integral over the area $[-L_x/2,+L_x/2] \times [-L_y/2,+L_y/2]$, according to the formula
\begin{equation}
    \label{xx-10}
\langle u, v \rangle \equiv \int_{-\frac{L_x}{2}}^{+\frac{L_x}{2}}
\int_{-\frac{L_y}{2}}^{+\frac{L_y}{2}} u(x,y) v(x,y) dx dy ,
\end{equation}
where we assume that functions are periodic with respect to the variable $y$ and their $x$-derivatives disappear at the boundaries $x=\pm \frac{L_x}{2}$ .

\subsection{The first order of expansion}
In the first-order of expansion the equation is of the form
\begin{equation}
\label{xx-11}  \hat{{\cal L}_0} v^{(1)} + \cos \phi_K  \, v^{(1)}  +
\hat{G} v^{(0)} = \lambda^{(0)} v^{(1)} + \lambda^{(1)} v^{(0)} \, .
\end{equation}
In order to shorten the formulas that appear in this section, the operator $\hat{G}$ was introduced
\begin{equation}
\label{xx-12}
\hat{G} v^{(0)} \equiv \hat{W} v^{(0)} - (\sin \phi_K)  \chi^{(1)} v^{(0)} .
\end{equation}

\subsubsection{Correction to the ground state}
We project equation \eqref{xx-11} for the ground state onto the state $v_0^{(0)}$ which leads to the equation
\begin{equation}
\label{xx-13} \langle v_0^{(0)}, ( \hat{{\cal L}_0} + \cos \phi_K ) \,
v_0^{(1)} \rangle + \langle v_0^{(0)} , \hat{G} v_0^{(0)} \rangle =
\lambda^{(0)}_0 \langle v_0^{(0)} , v_0^{(1)}\rangle + \lambda^{(1)}_0
\langle v_0^{(0)}, v_0^{(0)} \rangle\, .
\end{equation}
Due to the normalization of the state $v_0^{(0)}$ and the fact that the operator $\hat{{\cal L}_0} + \cos \phi_K$ is hermitian, i.e.,

\begin{equation}
\label{xx-14} \langle v, ( \hat{{\cal L}_0} + \cos \phi_K ) \,
u \rangle = \langle ( \hat{{\cal L}_0} + \cos
\phi_K ) v, u \rangle\, ,
\end{equation}
equation \eqref{xx-13} can be reduced to the form
\begin{equation}
\label{xx-15}
 \lambda_0^{(1)} = \langle v_0^{(0)} , \hat{G} v_0^{(0)} \rangle  .
\end{equation}
We determine the value of $\lambda_0^{(1)}$ based on equations \eqref{xx-12} and \eqref{xx-4}. In the appendix, we take the following form of $g(x,y) = - p(x) q(y)$. As for the function describing the deformation of the function $\phi_0$ resulting from the existence of inhomogeneities, i.e., $\chi^{(1)}$, we write it as follows $\chi^{(1)} = \chi_0 \,\, \alpha(x) \beta(y)$. Under the above conditions, the correction of first order 
is of the form
\begin{equation}
\label{xx-16}
 \lambda_0^{(1)} = \frac{1}{2 L_y \tanh \frac{L_x}{2}} \left(2 \chi_0 J_{\alpha} I_{\beta} - J_p I_q \right)  .
\end{equation}
The integrals that appear in the above formula are defined below
\begin{equation}
\label{xxx-16}
J_p \equiv \int^{+\frac{L_x}{2}}_{-\frac{L_x}{2}} p(x)~
\mathrm{sech}^2(x) \tanh^2 (x) ~, \,\,\,\,\,\, I_q \equiv
\int^{+\frac{L_y}{2}}_{-\frac{L_y}{2}} q(y) d y ,
\end{equation}
\begin{equation}
\label{xx-17}
J_{\alpha} \equiv \int^{+\frac{L_x}{2}}_{-\frac{L_x}{2}}
\alpha(x)~ \mathrm{sech}^3(x) \tanh (x)~, \,\,\,\,\,\, I_{\beta}
\equiv \int^{+\frac{L_y}{2}}_{-\frac{L_y}{2}} \beta(y) d y .
\end{equation}
The $p(x)$ and $q(x)$ functions appearing in the above integrals, 
in the paper, are taken in the form of \eqref{p} and \eqref{q}.
On the other hand, the form of the function $\chi(x,y) \approx \chi^{(1)}(x,y)$ is approximated, according to considerations contained in appendix  A in formulas \eqref{alphabeta}.
Two of the above integrals approximately describe the width of the inhomogeneity in the direction of the $y$ variable i.e.
$I_q \approx d$ , $I_{\beta} \approx d .$ Consequently, the eigenvalue of the ground state takes the form of
\begin{equation}
\label{xx-22}
\lambda_0 = \lambda_0^{(0)} + \varepsilon \lambda_0^{(1)} + ... \approx
\frac{\varepsilon}{2\tanh \frac{L_x}{2}} ~\frac{d}{L_y}~
\left(2 \chi_0 J_{\alpha} - J_p \right)  .
\end{equation}
To complete the result obtained, we provide the integrals appearing in this formula 
\begin{equation}
\label{xx-23}
J_{\alpha} \approx \frac{2}{3} \left(1 - \mathrm{sech}^3
\left(\frac{h}{2}\right)\right) ,
\end{equation}
\begin{equation}
\label{xx-24}
J_{p} = \coth \left(\frac{h}{2}\right) \left[ \frac{2 \tanh
\left(\frac{L_x}{2}\right) - \coth \left(\frac{h}{2}\right) ~\ln
\left( \frac{\cosh \left(\frac{L_x+h}{2}\right)}{\cosh
\left(\frac{L_x-h}{2}\right)}\right) }{\sinh^2
\left(\frac{h}{2}\right)}+ \frac{2}{3} \tanh^3 \left(
\frac{L_x}{2} \right)\right] .
\end{equation}

\subsubsection{Correction to the degenerate states}
In the case of degenerate states, we perform a projection of equation \eqref{xx-11} into a state that is a combination of zero-order eigenstates
\begin{equation}
\label{xx-25} v_n = \sum_{i=\pm} c_i v^{(0)}_{n i} .
\end{equation}
Projection of the equation of the first order written for the degenerate state $v_{n j}^{0}$ onto the $v$ state gives
\begin{equation}
\label{xx-26} \langle v_n, ( \hat{{\cal L}_0} + \cos \phi_K ) \,
v_{n j}^{(1)} \rangle + \langle v_n , \hat{G} v_{n j}^{(0)} \rangle =
\lambda^{(0)}_n \langle v_n , v_{n j}^{(1)}\rangle + \lambda^{(1)}_n
\langle v_n, v_{n j}^{(0)} \rangle\, .
\end{equation}
Orthonormality of the zero-order states and hermiticity of the operator $\hat{{\cal L}_0} + \cos \phi_K$ leads to a system of equations for the coefficients $c_i$
\begin{equation}
\label{xx-27} \sum_{i=\pm} c_i \langle v^{(0)}_{n i}, \hat{G}
v^{(0)}_{n j} ~ \rangle =  \lambda^{(1)}_n \sum_{i=\pm} c_i ~
\delta_{ij}  .
\end{equation}
Due to the second degree of degeneracy, we can write the last equation in $2 \times 2$ matrix form
\begin{equation}\label{xx-28}
   \left[%
\begin{array}{cc}
  G_{++}-\lambda^{(1)}_n & G_{+-} \\
  G_{-+} & G_{++}-\lambda^{(1)}_n \\
\end{array}%
\right] \Biggl[%
\begin{array}{c}
  c_+ \\
  c_- \\
\end{array}%
\Biggr] = \Biggl[%
\begin{array}{c}
  0 \\
  0 \\
\end{array}%
\Biggr] ,
\end{equation}
where the matrix elements $G_{i j}$ are written in the basis that consists of eigenstates of the zero order approximation
\begin{equation}
\label{xx-29} 
G_{i j} = \langle v^{(0)}_{n i}, \hat{G} v^{(0)}_{n j} ~ \rangle .
\end{equation}
The condition for the existence of non-trivial solutions of the above equation is the zeroing of the determinant (so that nontrivial solutions
of the homogeneous system exist)
\begin{equation}\label{xx-30}
   \left|%
\begin{array}{cc}
  G_{++}-\lambda^{(1)}_n & G_{+-} \\
  G_{-+} & G_{++}-\lambda^{(1)}_n \\
\end{array}%
\right| = 0 .
\end{equation}
According to the above equation, corrections of the first order remove the degeneracy, leading
to the eigenvalue corrections:
\begin{equation}\label{xx-31}
    \lambda^{(1)}_{n \pm} = \frac{1}{2} \left[ ~(G_{++}+G_{--}) \pm \sqrt{(G_{++}-G_{--})^2 + 4 G_{+ -} G_{- +}} ~\right] .
\end{equation}
The expression above is greatly simplified due to the evenness of the $q(-y) = q(y)$ and $\beta(-y) = \beta(y) $ functions in the $y$ variable. This property removes the matrix element $G_{+-}=0$ which leads to a significant simplification of the last formula  
\begin{equation}\label{xx-32}
    \lambda^{(1)}_{n \pm} = \frac{1}{2} \left[ ~(G_{++}+G_{--}) \pm |G_{++}-G_{--}| ~\right].
\end{equation}
Matrix elements that appear in the above expression
\begin{equation}\label{xxx-32}
G_{++} = A^2 \left( 2 \chi_0 ~ J_{\alpha} I^{+}_{\beta} - J_p
I^{+}_{q} ~\right) , \,\,\,\,\,\,  G_{--} = A^2 \left( 2 \chi_0 ~
J_{\alpha} I^{-}_{\beta} - J_p I^{-}_{q} ~\right) ,
\end{equation}
are written using integrals
\begin{equation}\label{xx-33}
I_q^+ = \int_{-\frac{L_y}{2}}^{+\frac{L_y}{2}} q(y) \cos^2 \left(2
\pi n \frac{y}{L_y} \right) dy , \,\,\,\, I_q^- =
\int_{-\frac{L_y}{2}}^{+\frac{L_y}{2}} q(y) \sin^2 \left(2 \pi n
\frac{y}{L_y} \right) dy ,
\end{equation}
\begin{equation}\label{xx-34}
I_{\beta}^+ = \int_{-\frac{L_y}{2}}^{+\frac{L_y}{2}} \beta(y)
\cos^2 \left(2 \pi n \frac{y}{L_y} \right) dy , \,\,\,\,
I_{\beta}^- = \int_{-\frac{L_y}{2}}^{+\frac{L_y}{2}} \beta(y)
\sin^2 \left(2 \pi n \frac{y}{L_y} \right) dy .
\end{equation}
The final result shows the disappearance
of the degeneracy of the higher eigenvalues (the integrals $J_{\alpha}$ and $J_p$ are defined by the formulas \eqref{xx-23} and \eqref{xx-24}) 
\begin{equation}\label{xx-35}
\lambda_{n \pm} = \lambda^{(0)}_{n} + \varepsilon \lambda^{(1)}_{n
\pm} + ... \approx \left( \frac{2 \pi}{L_y}\right)^2 n^2 +
\frac{\varepsilon}{2 \tanh \left( \frac{L_x}{2}\right)} ~ ( 2
\chi_0 ~J_{\alpha} - J_{p} ) \left[ \frac{d}{L_y} \pm \left|
\frac{\sin \left(2 \pi n \frac{d}{L_y} \right)}{ 2 \pi n} \right|~
\right] .
\end{equation}
This result was obtained by means of the approximation:
\begin{equation}\label{xx-36}
I_q^{\pm} \approx \frac{1}{2}~L_y \left(\frac{d}{L_y} \pm
\frac{\sin \left( 2 \pi n \frac{d}{L_y} \right) }{2 \pi n} \right)
, \,\,\,\, I_{\beta}^{\pm} \approx \frac{1}{2}~L_y
\left(\frac{d}{L_y} \pm \frac{\sin \left( 2 \pi n \frac{d}{L_y}
\right) }{2 \pi n} \right) . 
\end{equation}
In addition, the normalization factor $A$ included in formula \eqref{xx-9} was used, while the values of the integrals $J_{\alpha}$ and $J_p$ are defined by the formulas \eqref{xx-23} and \eqref{xx-24}.

\section{Appendix C}
In this section, we will estimate the value of $\lambda=\omega^2$ corresponding to the ground state, based on the shape of the energy landscape of the system under study. We consider the Lagrangian density of the sine-Gordon model in the presence of inhomogeneity
\begin{equation}
\label{C-1}
    {\cal L} = \frac{1}{2} (\partial_t \phi)^2 - \frac{1}{2} {\cal F}(x,y) (\partial_x \phi)^2 - \frac{1}{2} (\partial_y \phi)^2 - V(\phi) .
\end{equation}
The energy density in this model is of the form
\begin{equation}
 \label{C-2}   \rho =
 \frac{1}{2} (\partial_t \phi)^2 + \frac{1}{2} {\cal F}(x,y) (\partial_x \phi)^2 + \frac{1}{2} (\partial_y \phi)^2 + V(\phi) .
\end{equation}
As in previous parts $V(\phi) = 1 - \cos \phi$ and ${\cal F}(x,y) = 1 + \varepsilon g(x,y).$ 
Into the expression for the energy density we insert the kink ansatz $\phi_K(t,x) = 4 \arctan e^{x-x_0(t)}$, where $x_0=x_0(t)$  determines 
the position of the kink. 
Based on expression \eqref{C-2}, we calculate the energy per unit length of the kink front
\begin{equation}
 \label{C-3}   E(x_0) = \frac{1}{L_y} \int_{-\frac{L_x}{2}}^{+\frac{L_x}{2}} \int_{-\frac{L_y}{2}}^{+\frac{L_y}{2}} \rho(x,y,x_0) dx dy 
    =  \frac{1}{2} m \dot{x_0}^2 + \widetilde{V}(x_0) .
\end{equation}
The first term has its origin in the differentiation of the kink ansatz with respect to the time variable $\partial_t \phi_K = - \dot{x_0} \, \partial_x \phi_K $ and $m= 8 \tanh{\frac{L_x}{2}} \approx 8$ is the mass of a free, resting kink (where $L_x=30$). 
The next terms define the potential energy.  Under the assumption as to the form of inhomogeneity $g(x,y)=-p(x) q(y)$, the potential energy can be expressed by two integrals
\begin{equation}
 \label{C-4}   \widetilde{V}(x_0)
    =  8 - 2 \varepsilon I(d) J(x_0,h)   ,
\end{equation}
where we denoted
\begin{equation}
     I(d) = \frac{1}{Ly} \int_{-\frac{L_y}{2}}^{+\frac{L_y}{2}} q(y)dy =\frac{1}{L_y} \ln \left( \frac{ \cosh\left(\frac{L_y+d}{2}\right)}{\cosh \left(\frac{L_y-d}{2}\right)} \right) \approx \frac{d}{L_y} , \,\,\,\,\,\,  J(x_0,h) = \int_{-\frac{L_x}{2}}^{+\frac{L_x}{2}} p(x) \sech^2(x-x_0) d x .
\end{equation}
For a more compact result (and because of the rapid disappearance of the $p$-function when approaching the edge), we approximate the second integral as follows
\begin{equation}
 \label{C-5}   J(x_0,h) \approx \int_{-\infty}^{+\infty} p(x) \sech^2(x-x_0) d x = - \left( \frac{2 x_0 +h - \sinh(2 x_0 +h)}{\cosh (2 x_0 +h) -1} -  \frac{2 x_0 -h - \sinh(2 x_0 -h)}{\cosh (2 x_0 -h) -1} \right) .
\end{equation}
In the vicinity of the center of the well (i.e., for $x_0 = 0$), we can approximate the potential energy \eqref{C-4} to the accuracy of the harmonic term
\begin{equation}
 \label{C-6}
    \widetilde{V}(x_0) \approx A + B x_0^2 ,
\end{equation}
where the expansion coefficients are respectively 
\begin{equation}
     \label{C-7}
    A =8+ 4 \varepsilon \frac{d}{L_y} \, \left( \frac{h - \sinh h}{\cosh h -1}\right) \,\, , \,\,\,\,\, B = 2 \varepsilon \frac{d}{L_y} \, \mathrm{csch}^4 \left(\frac{h}{2}\right) \, \left[ h (2 + \cosh h) - 3 \sinh h \right] .
\end{equation}
We can rescale the original potential  $\widetilde{V}(x_0)$ by a constant getting a new potential  $V(x_0) = \widetilde{V}(x_0) -A$.
The effective Lagrangian for this system is thus of the form 
\begin{equation}
 \label{C-8}
    L = \frac{1}{2} m \dot{x_0}^2 - B x_0^2 .
\end{equation}
The effective equation is that of a harmonic oscillator 
\begin{equation}
    \label{C-9}
    \Ddot{x}_0 + \frac{2 B}{m} x_0 = 0 .
\end{equation}
The eigenfrequency of this oscillator describes, in a manner independent of the perturbation calculus performed in Appendix B (i.e., the latter is
at the level of the equation of motion, while
here we work at the level of the corresponding Lagrangian and energy functionals), the ground state appearing in the description of the linear stability of a kink trapped by a well-shaped inhomogeneity.
\begin{equation}
 \label{C-10}
    \omega^2 = \frac{2 B}{m} = \frac{1}{2} \varepsilon \frac{d}{L_y} \, \mathrm{csch}^4 \left(\frac{h}{2}\right) \, \left[ h (2 + \cosh h) - 3 \sinh h \right] .
\end{equation}
The relevant results is showcased in Fig.~\ref{fig_14}.

\section*{Acknowledgement}
This research has been made possible by the Kosciuszko Foundation The American Centre of Polish Culture (JG). This research was supported in part by PLGrid Infrastructure (TD and JG).
This material is based upon work supported by the U.S. National Science Foundation under the awards PHY-2110030 and DMS-2204702 (PGK).

\printbibliography

\end{document}